\documentclass[%
 reprint,
 amsmath,amssymb,
 aps,
prb,
]{revtex4-2}
\usepackage{graphicx}
\usepackage{dcolumn}
\usepackage{bm}
\usepackage{hyperref}
\usepackage{xcolor} 

\newcommand{\panel}[1]{(\MakeLowercase{#1})}

\usepackage[left]{lineno}
\begin{document}

\preprint{APS/123-QED}
%\linenumbers
\title{Topological flowscape reveals state transitions in nonreciprocal living matter}

\author{Hyunseok~Lee$^{1,6}$}
\author{EliseAnne~Koskelo$^{1,2,6}$}
\author{Shreyas~Gokhale$^{1,6}$}
\author{Junang~Li$^{3,6}$}
\author{Chenyi~Fei$^4$}
\author{Chih-Wei Joshua~Liu$^1$}
\author{Lisa~Lin$^1$}
\author{J\"{o}rn~Dunkel$^4$}
\author{Dominic~J~Skinner$^5$}
\author{Nikta~Fakhri$^{1}$}
\email{Corresponding author: fakhri@mit.edu}

\affiliation{\vspace{0.25cm}\\ $^1$ Department of Physics, Massachusetts Institute of Technology, Cambridge, MA, USA \\ $^2$ Department of Physics, Harvard University, Cambridge, MA, USA \\ $^3$ Center for the Physics of Biological Function, Princeton University, Princeton, NJ, USA \\ $^4$ Department of Mathematics, Massachusetts Institute of Technology, Cambridge, MA, USA \\ $^5$ Center for Computational Biology, Flatiron Institute, New York, NY, USA \\ $^6$ These authors contributed equally and are joint first authors.}

\date{\today}

\begin{abstract}
Nonreciprocal interactions---where forces between entities are asymmetric---govern a wide range of nonequilibrium phenomena, yet their role in structural transitions in living and active systems remains elusive. Here, we demonstrate a transition between nonreciprocal states using starfish embryos at different stages of development, where interactions are inherently asymmetric and tunable. Experiments, interaction inference, and topological analysis yield a nonreciprocal state diagram spanning crystalline,  {flock-like}, and fragmented states, revealing that weak nonreciprocity promotes structural order while stronger asymmetry disrupts it. To capture these transitions, we introduce topological landscapes, mapping the distribution of structural motifs across state space. We further develop topological flowscapes, a dynamic framework that quantifies transitions between collective states and detects an informational rate shift  {and an emergent proofreading} from the experimental state transition. Together, these results establish a general approach for decoding nonequilibrium transitions and uncover how asymmetric interactions sculpt the dynamical and structural architecture of active and living matter.

\end{abstract}

\maketitle

Nature produces diverse structures and behaviors whose complexity greatly surpasses equilibrium constraints~\cite{needleman2017active, marchetti2013hydrodynamics}. From flocks of birds~\cite{vicsek1995novel, ballerini2008interaction} and colonies of microbes {~\cite{drescher2009dancing,hartmann2019emergence,ishikawa2020stability,liu2024emergence,von2025spatiotemporal}} to cellular tissues~\cite{giavazzi2018flocking, tang2022collective} and synthetic active materials~\cite{bricard2013emergence, veenstra2024non}, remarkable forms and functionalities emerge spontaneously from local interactions among constituents, without external coordination. This self-organized emergence, driven by the agency of individual components, not only underpins biological complexity but also inspires new classes of active and adaptive materials, where nonequilibrium dynamics are harnessed for novel functionalities~\cite{hopfield1982neural, paxton2004catalytic, yan2016reconfiguring, mallory2018active, scheibner2020odd, meredith2020predator}.
\\

At the core of these complex phenomena lie nonreciprocal interactions, where action and reaction between components are inherently asymmetric~\cite{soto2014self, meredith2020predator, saha2020scalar, you2020nonreciprocity, fruchart2021non}. Such asymmetry profoundly influences nonequilibrium behavior, driving rich dynamical phenomena including traveling waves~\cite{you2020nonreciprocity, saha2020scalar, pisegna2024emergent, mandal2024robustness}, spontaneous oscillations~\cite{fruchart2021non, chen2024emergent, parkavousi2025enhanced}, and chiral flows~\cite{markovich2024nonreciprocity}. Yet, despite growing recognition of their significance, how nonreciprocity shapes structural transitions and the emergence of stable configurations remains largely unexplored~\cite{dinelli2023non, osat2023non, brauns2024nonreciprocal, kreienkamp2024nonreciprocal, guillet2025melting, kole2025non}. Furthermore, the extent to which nonreciprocity governs information transfer and state evolution in living systems is unclear~\cite{loos2020irreversibility, bowick2022symmetry}.
\\

Here, we address this gap by investigating how nonreciprocity orchestrates collective self-organization and structural transitions in a biologically tunable system: starfish embryos at distinct developmental stages, which interact through inherently asymmetric fluid-mediated forces. Building upon our previous discovery of living chiral crystals composed of homogeneous embryo ensembles~\cite{tan2022odd}, we systematically examine binary mixtures of embryos at different developmental ages. Through experiments integrated with interaction inference  (Section I) and novel topological analyses  (Section II), we identify distinct emergent states, including crystalline, self-propelled crystalline,  flock-like, and fragmented configurations.  We introduce topological landscapes (Section III), revealing symmetry-breaking transitions between these nonreciprocal states, and topological flowscapes (Section IV), quantifying the informational trajectory of the transitions. These conceptual tools provide a versatile framework to uncover structural and dynamic transitions across biological and synthetic active matter.

\begin{figure*}[htbp]
    \centering
    \small
    \includegraphics[width=\textwidth]{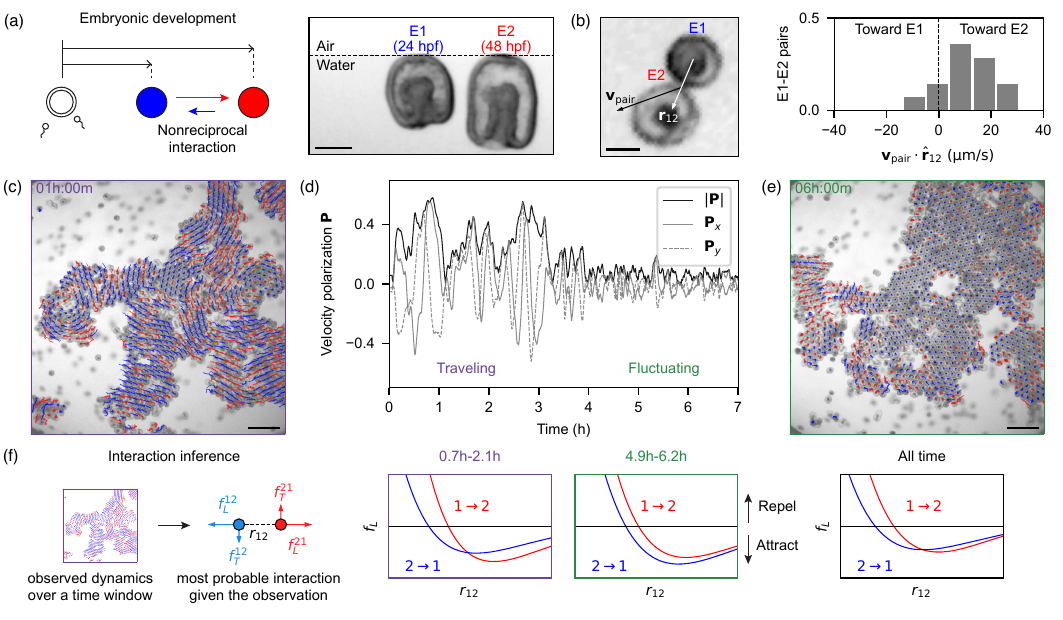}
    \caption{\textbf{A nonreciprocal mixture of starfish embryos transitions from a traveling state to a fluctuating state.} 
    \panel{A}~Nonreciprocity arises from a developmental gap between embryos. An E1 embryo (24 hours post-fertilization) and an E2 embryo (48 hours post-fertilization) form  {a bound pair~\cite{drescher2009dancing}} at the air-water interface. Scale bar: $\mathrm{100\ \mu m}$.
    \panel{B}~The velocity of the embryo pair, $\mathbf{v}_{\mathrm{pair}}$, tends to align with the displacement vector $\mathbf{r}_{12}$ from E1 to E2. The average drift speed toward E2, $\mathbf{v}_{\mathrm{pair}} \cdot \hat{\mathbf{r}}_{12}$, is $\mathrm{12 \pm 3\ \mu m/s}$ (SEM, n = 14). Scale bar: $\mathrm{100\ \mu m}$.
    \panel{C}~Snapshot of an E1--E2 mixture at 1 hour, overlaid with 2-minute trajectories (blue: E1, red: E2). Scale bar: $\mathrm{1\ mm}$.
    \panel{D}~Time series of velocity polarization, $\mathbf{P}(t) \equiv \frac{1}{N(t)} \sum_{i = 1}^{N(t)} \frac{\mathbf{v}_i(t)}{|\mathbf{v}_i(t)|}$, reveals a transition from a traveling state to a fluctuating state. Oscillations and a phase shift between $\mathbf{P}_x$ and $\mathbf{P}_y$ highlight the chirality of the velocity polarization, which rotates clockwise during the traveling state.
    \panel{E}~Snapshot of the same mixture at 6 hours, overlaid with 2-minute trajectories (blue: E1, red: E2). Scale bar: $\mathrm{1\ mm}$.  {\panel{F}~Interaction inference quantifies the nonreciprocal interaction and its time evolution. \textit{Left}: We implement Bayesian inference to quantify pairwise inter-embryo interactions $f_{L/T}^{ij}(r)$, which depend on embryo types $i,j$ and pairwise distance $r$, from embryo trajectories (see Supplementary~Section~II). \textit{Center}: $f_{L}^{21}$ (red) and $f_{L}^{12}$ (blue), the longitudinal interactions by E1 on E2 and by E2 on E1 respectively, are inferred using data from early and late times. As embryonic development continues over the experiment, interactions become less nonreciprocal. \textit{Right}: $f_{L}^{21}$ (red) and $f_{L}^{12}$ (blue) are inferred using data from all time. We use this all-time inference as a basis for our inference-based model (Fig. \ref{fig:fig2}). }
}
    \label{fig:fig1}
\end{figure*}

\section{A nonreciprocal mixture of starfish embryos transitions from a traveling state to a fluctuating state}

To experimentally uncover how nonreciprocity shapes collective dynamics, we developed a living nonreciprocal system composed of embryos from the starfish \textit{Patiria miniata}. Unlike homogeneous embryo populations, which display emergent nonreciprocity through broken chiral symmetries~\cite{tan2022odd, chao2026selective}, our heterogeneous system exhibits tunable interactions between two types of embryos (E1 and E2), controlled by their developmental ages (Fig.~\ref{fig:fig1}a). Embryos act as building blocks in this reconfigurable system, and can interact in both a symmetric (with same-type neighbors) and asymmetric (with different-type neighbors) manner. Quantitative video microscopy reveals a striking run-and-chase dynamic: younger E1 embryos actively pursue older E2 embryos, whereas E2 embryos move preferentially away from E1 (Fig.~\ref{fig:fig1}b, Supplementary~Video~1, and Supplementary Section I.A-I.B~\cite{suppinfo}). As in previous studies of chiral swimmers~\cite{drescher2009dancing, chao2026selective}, the dynamics of anterior-posterior (AP) axes offer mechanistic insights-- further experiments demonstrate that this  {broken symmetry} arises from asymmetric precession of the embryos'  {AP} axes, establishing a clear physical mechanism underlying nonreciprocity~(Supplementary~Section~I.B.3 and Supplementary~Video~2). In addition to this bound pair dynamics \cite{drescher2009dancing}, when embryos are apart, they approach each other with asymmetric speeds, demonstrating nonreciprocity in their hydrodynamic attractions~(Supplementary~Video~1 and Supplementary~Section~I.B.4, I.C). 
\\

When thousands of E1 and E2 embryos are mixed, the system spontaneously exhibits collective translation, a hallmark of nonreciprocal interactions~\cite{you2020nonreciprocity}. Initially, the embryos flock together, forming a dynamically ordered \textit{traveling} state characterized by strong polar alignment, quantified by the velocity polarization order parameter $\mathbf{P}(t) \equiv \frac{1}{N(t)} \sum_{i = 1}^{N(t)} \frac{\mathbf{v}_i(t)}{|\mathbf{v}_i(t)|}$, where $N(t)$ is the number of embryos at time $t$ and $\frac{\mathbf{v}_i(t)}{|\mathbf{v}_i(t)|}$ is the direction of $i^{th}$ particle's instantaneous velocity~(Fig.~\ref{fig:fig1}c--d, Supplementary~Videos~3--4, and Supplementary Section I.D.
Remarkably, this polar order emerges exclusively in mixed-age embryo populations, whereas it is absent in homogeneous populations. This experimentally verifies the theoretical prediction that nonreciprocity is an ingredient that can lead to emergent polar order~\cite{you2020nonreciprocity, saha2020scalar}.
The velocity polarization $\mathbf{P}$ is time-dependent, rotating clockwise with a fluctuating magnitude~(Fig.~\ref{fig:fig1}d and Supplementary~Section~I.D.3). This rotation arises from broken chiral symmetry of individual embryos rather than from emergent chirality via nonreciprocal alignment~\cite{fruchart2021non}. Furthermore, we find that E2 embryos consistently lead E1 in collective translation,  {suggesting} that nonreciprocity drives the flocking of embryos~(Supplementary~Section~I.D.4). 
\\

After hours of collective translation, the embryo mixture transitions into a distinct self-organized state~(Supplementary~Videos~3--4). In this \textit{fluctuating} state, embryos arrange into an ordered lattice, while small fluctuations arise from their persistent velocity moduli that lack alignment
~(Fig.~\ref{fig:fig1}e, Supplementary~Section~I.D.5). As a result, the velocity polarization $\mathbf{P}$ exhibits near-zero fluctuations, demarcating the departure from the traveling state~(Fig.~\ref{fig:fig1}d). 
\\

 To confirm the nonreciprocity, we extracted the underlying interactions from many-body experimental trajectories using a data-driven inference method that reconstructs the most likely longitudinal and transverse forces between embryo pairs as functions of their separation and types (E1 or E2)~\cite{zhang2024nonlinear, zuo2025predicting} (Supplementary Section~II). 
Applied to the full experimental dataset~(Fig.~\ref{fig:fig1}f), this inference yields an E1--E2 longitudinal interaction that is nonreciprocal, in agreement with the isolated run-and-chase dynamics~(Fig.~\ref{fig:fig1}b),  {and evolves to be less asymmetric over time  (Supplementary~Section~II.B)}.
\\

The transition from ordered dynamics to ordered configurations reveals a rich state space of nonequilibrium behaviors that depend on nonreciprocity. By tuning nonreciprocity, the system exhibits a wide variety of emergent collective states, including chiral orbits and demixed regimes. This tunability highlights the versatility of the experimental platform~(Supplementary~Section~I.E and Supplementary~Videos~5--6).
\\

\begin{figure*}[htbp]
    \centering
    \small
    \includegraphics[width=\textwidth]{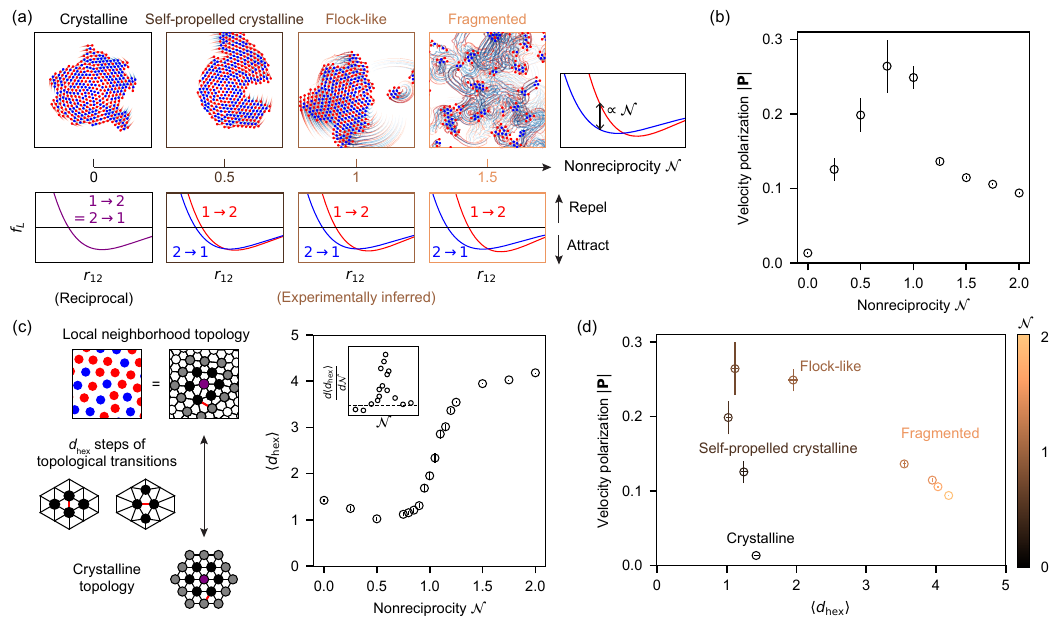}
    \caption{\textbf{Topological order parameter distinguishes emergent states in an inference-based model of nonreciprocal mixtures}. 
    \panel{A}~In the inference-based model, we use experimentally inferred pairwise interactions $f_{L/T}^{ij}$, except for longitudinal inter-type interactions whose nonreciprocity is scaled by $\mathcal{N}$. The model exhibits four distinct states as the nonreciprocity $\mathcal{N}$ increases: crystalline, self-propelled crystalline,  {flock-like}, and fragmented states.
    \panel{B}~Velocity polarization magnitude $|\mathbf{P}|$ as a function of $\mathcal{N}$, averaged over 20 initial configurations (error bars: SEM). $|\mathbf{P}|$ decreases for $\mathcal{N}>1$ .
    \panel{C}~\textit{Left:} The structural metric $d_{\mathrm{hex}}$ quantifies the number of local topological (T1) transitions relative to a perfect hexagonal crystal; red edges highlight topology-changing transitions. \textit{Right:} The structural order parameter $\langle d_{\mathrm{hex}} \rangle$ reaches a minimum at $\mathcal{N} = 0.5$ and  {rises} at $\mathcal{N} = 1$ (error bars: SEM, n=20).  {\textit{Inset:} The slope $ \frac{ d \langle d_{\mathrm{hex}} \rangle}{d\mathcal{N}}$ is negative for small $\mathcal{N}$ and peaks near $\mathcal{N}=1$ (dotted line is at zero).}
    \panel{D}~State diagram of nonreciprocal mixtures, summarizing  emergent states across the pairwise nonreciprocity $\mathcal{N}$.
}
    \label{fig:fig2}
\end{figure*}

 {\section{Topological order parameter distinguishes emergent states in an inference-based model of nonreciprocal mixtures}}

 Motivated by the diversity in self-organization and dynamics of the experimental platform, we systematically mapped the nonequilibrium states of nonreciprocal mixtures.  {We developed a pairwise interaction model that allows us to} tune the nonreciprocity $\mathcal{N}$ \emph{in-silico}. Here, nonreciprocity $\mathcal{N}$  encapsulates the asymmetry of the inter-type longitudinal interactions. Through the model, we identify different states as a function of $\mathcal{N}$ that we characterize with velocity polarization and a novel topological order parameter.
\\

 To construct a tunable model, we take the interaction inference as baseline, and define nonreciprocity parameter $\mathcal{N}$ such that:
\begin{eqnarray} 
    f_L^{12}(r;\mathcal{N}) = f_L^S(r) + \mathcal{N} f_L^A(r), \notag\\
    f_L^{21}(r;\mathcal{N}) = f_L^S(r) - \mathcal{N} f_L^A(r).
\end{eqnarray} where $f_L^S(r) \equiv \frac{f_L^{12}(r) + f_L^{21}(r)}{2}$ and $f_L^A(r) \equiv \frac{f_L^{12}(r) - f_L^{21}(r)}{2}$ are the symmetric and antisymmetric parts of the inferred longitudinal interactions (Fig. \ref{fig:fig1}f and \ref{fig:fig2}a).
\\

 This form ensures that the system is reciprocal when $\mathcal{N} = 0$, and matches the inferred interactions when $\mathcal{N} = 1$. We explore a range of values including weaker ($\mathcal{N} < 1$) and stronger ($\mathcal{N} > 1$) nonreciprocity to elucidate its effect on emergent behaviors (Fig.~\ref{fig:fig2}a, Supplementary Section III and Supplementary~Video~7). \\

At $\mathcal{N}=0$ (reciprocal interactions), we find a \textit{crystalline} state that rotates clockwise, recapitulating homogeneous embryos forming living chiral crystals~\cite{tan2022odd}. For $0 < \mathcal{N} < 1$ (weak nonreciprocity), we observe a \textit{self-propelled crystalline} state, where a crystal undergoes collective translation. At $\mathcal{N}=1$ (nonreciprocity inferred from experiment), the system exhibits \textit{ {flock-like}}  {dynamics}, maintaining collective motion but not a global crystalline arrangement. Instead, the crystals repeatedly merge and fragment, reflecting the early dynamics observed in embryo mixture experiments (Fig. \ref{fig:fig1}d and Supplementary~Section~III.C ). For $\mathcal{N}>1$ (strong nonreciprocity), the system is further \textit{fragmented}, with each fragment propelling itself in different directions.
\\

Although existing theories of nonreciprocal interactions predict a jump in velocity polarization $|\mathbf{P}|$ as the system enters a traveling state~\cite{fruchart2021non,you2020nonreciprocity} (Supplementary Section VII.F), our model reveals a richer complexity (Fig.~\ref{fig:fig2}b). The broken chiral symmetry in transverse interactions causes clusters to rotate rather than move in a straight line, competing with collective translation induced by nonreciprocity. This interplay leads $|\mathbf{P}|$ to rise continuously, rather than abruptly, with increasing $\mathcal{N}$. Surprisingly, we find that $|\mathbf{P}|$ peaks near $\mathcal{N}=1$ and then decreases for $\mathcal{N}>1$, a behavior explained by fragmentation into independently moving clusters whose internal self-propulsion dominates interactions between clusters. Consequently, states with disparate collective dynamics, such as self-propelled crystals and fragmented clusters, exhibit similar velocity polarization magnitudes despite their distinct differences in structure.
\\

To robustly distinguish these structurally diverse states beyond velocity polarization $|\mathbf{P}|$ alone, we need a structural order parameter.
Notably, while nonreciprocity can drive structures far away from translational or orientational order, these complex arrangements can still be systematically compared through their topologies~\cite{skinner2023topological, ballerini2008interaction}. Hence, we define a topological metric $d_{\mathrm{hex}}$ based on a recently developed topological packing statistics framework~\cite{skinner2021topological, skinner2023topological}.
This metric quantifies structural order by counting the number of local T1 topological rearrangements required to reach a perfect crystalline arrangement from a given local neighborhood  {a.k.a. ``motif''} (Fig.~\ref{fig:fig2}c and Supplementary~Section~IV).
Accordingly, the system-average $\langle d_{\mathrm{hex}} \rangle$ serves as a measure of the topological distance between the system's self-organized structure and a perfect crystal.
\\

The structural order parameter $\langle d_{\mathrm{hex}} \rangle$ quantifies the average topological distance from a crystal (Fig.~\ref{fig:fig2}c). At small $\mathcal{N}$, $\langle d_{\mathrm{hex}} \rangle$ indicates crystalline structures whose local neighborhoods are on average within one to two T1 transitions from a hexagonal motif. Interestingly, in this regime, we find that $\langle d_{\mathrm{hex}}\rangle$ decreases with increasing $\mathcal{N}$.
We call this counterintuitive trend nonreciprocal error correction~\cite{harrington2022engineered, lake2025squeezing, pajouheshgar2026exploring}: topological motifs with defects, which are unstable under nonreciprocity-driven stress, are annealed into a crystalline motif which remains stable~(Supplementary~Section~IV.D).   This behavior reflects a selective stabilization of low-defect configurations under nonequilibrium driving. The crystalline motif eventually becomes unstable under a sufficiently strong nonreciprocity, such as the experimentally inferred $\mathcal{N}=1$. This destabilization is captured by the steep increase in $\langle d_{\mathrm{hex}}\rangle$ as an annealed crystal transitions into a dynamic, multi-cluster structure~(Supplementary~Video~8).
\\

Combining the two order parameters, $|\mathbf{P}|$ for dynamics and $\langle d_{\mathrm{hex}}\rangle$ for structure, we construct a state diagram that maps all emergent nonreciprocal states~(Fig.~\ref{fig:fig2}d). 
As nonreciprocity increases from $\mathcal{N}=0$, $\langle d_{\mathrm{hex}}\rangle$ decreases and $|\mathbf{P}|$ increases. 
Once nonreciprocity reaches the experimental $\mathcal{N}=1$, $\langle d_{\mathrm{hex}}\rangle$ jumps while $|\mathbf{P}|$ remains large. 
Finally, for $\mathcal{N}>1$, $\langle d_{\mathrm{hex}}\rangle$ is large while $|\mathbf{P}|$ is small. 
Overall, our  state diagram captures nonreciprocity-driven transitions in both collective translation, which  {onsets at weak nonreciprocity and} peaks at intermediate nonreciprocity, and self-organized crystalline structure, which is enhanced by weak nonreciprocity but destabilized by strong nonreciprocity. \\

\section{Topological landscapes reveal a symmetry-breaking transition in self-organized structures as nonreciprocity increases} 

\begin{figure*}
    \centering
    \small
    \includegraphics[width=\textwidth]{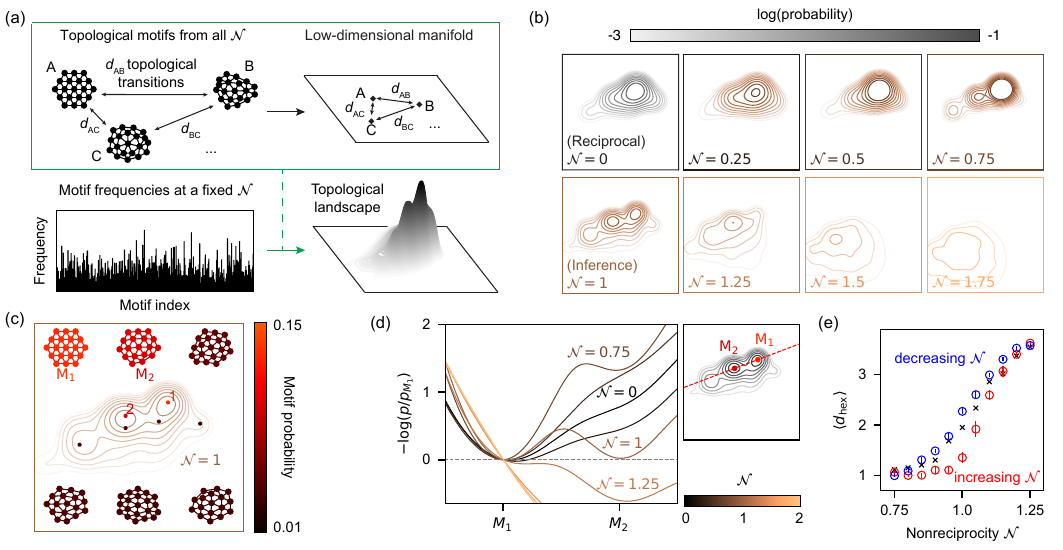}
    \caption{\textbf{Topological landscape reveals a symmetry-breaking transition in self-organized structures as nonreciprocity increases.}
    \panel{A}~We construct a low-dimensional manifold of topological motifs based on their pairwise topological distances.
    The resulting \textit{topological landscape} visualizes the probability distribution of local topologies (from each level of nonreciprocity $\mathcal{N}$) on this manifold.
    \panel{B}~Contour plots of topological landscapes display a transition from single-peaked to multi-peaked distributions as $\mathcal{N}$ increases.
    \panel{C}~At $\mathcal{N} = 1$ (experimentally inferred nonreciprocity), 14 topological motifs exhibit a frequency greater than 1\%. In particular, the top two dominant motifs correspond to the two distinct peaks of the topological landscape.  \textit{Top}: Motifs from $\mathrm{M_1}$ to $\mathrm{M_ {3}}$. $\mathrm{M_1}$ is a perfect crystal, while $\mathrm{M_2}$ contains a 5-defect. \textit{Bottom}: Motifs from $\mathrm{M_ {4}}$ to $\mathrm{M_ {6}}$. See Supplementary~Section~V.C for an enlarged motif atlas with full annotations.
    \panel{D}~Negative log-likelihood ratio relative to $\mathrm{M_1}$, measured along the $\mathrm{M_1}$--$\mathrm{M_2}$ axis. At $\mathcal{N} = 0$, the system shows a single minimum at $\mathrm{M_1}$. Near $\mathcal{N} = 1$, the structures undergo a first-order-like transition from $\mathrm{M_1}$-dominated to $\mathrm{M_2}$-dominated states, with coexistence at $\mathcal{N} = 1$. 
     \panel{E} Nonreciprocal hysteresis emerges upon slowly decreasing or increasing $\mathcal{N}$ about $\mathcal{N} = 1$. Its existence is a testable hypothesis that stems from the analogy to a first-order phase transition and its metastability.
    }
    \label{fig:fig3}
\end{figure*}

Using the structural order parameter $\langle d_{\mathrm{hex}}\rangle$, we observe that weak nonreciprocity stabilizes crystalline structures, whereas strong nonreciprocity leads to structural destabilization. However, this average measure does not provide insight into the precise local structural changes driving these transitions. To address this limitation, we introduce a novel approach to systematically characterize the distribution of local topologies observed in our experiments and simulations, involving approximately 15,000 distinct motifs~(Supplementary~Section~V).
\\

 First, we recall that the topological motifs have a metric structure, determined by the number of T1 transitions between them, and thus that the motifs exist as points in a high-dimensional space \cite{skinner2021topological}. Using these pairwise topological distances, we construct a two-dimensional manifold (Fig.~\ref{fig:fig3}a) which is a low-dimensional embedding of this metric space using classical multidimensional scaling (MDS) (Supplementary Section V.A). This manifold positions motifs  {that} differ by only a few T1 transitions close to each other, allowing an intuitive representation of structural proximity.  We then construct \textit{topological landscapes} by mapping motif frequencies onto this manifold as height values. The landscape $p(x,y)$ is found using kernel density estimation (Supplementary Section V.B). Specifically, $p$ quantifies the probability density of motifs in the vicinity of $(x,y)$ on the low-dimensional manifold. Thus, topological landscapes provide a comprehensive visualization of how local structural motifs distribute and evolve as nonreciprocity $\mathcal{N}$ varies. Similar to an autoencoder, these topological landscapes embed distributions in a lower-dimensional latent space. 
\\

As nonreciprocity increases near $\mathcal{N} = 1$, these landscapes exhibit a clear transition from single-peaked to multi-peaked distributions, indicating increased structural diversity (Fig.~\ref{fig:fig3}b). Specifically, motif probabilities shift systematically towards a region of the manifold associated with less crystalline order. This trend aligns with the observed increase in the structural order parameter $\langle d_{\mathrm{hex}}\rangle$, supporting the association between this manifold region and low-symmetry structures.
\\

Further analysis identifies the atlas of frequent motifs at the experimentally inferred nonreciprocity level $\mathcal{N}=1$ (Fig.~\ref{fig:fig3}c). Among those  {are} the two dominant motifs that contribute to the two peaks of the landscape: motif $\mathrm{M_1}$, corresponding to a perfect crystalline structure, and motif $\mathrm{M_2}$, representing an almost-crystalline structure with a single five-fold defect~(Supplementary~Section~V.D and Supplementary~Video~9). Crucially, nonreciprocity not only breaks the sixfold symmetry inherent to the perfect crystal but also introduces directional anisotropy within the motif space, selecting a preferred structural transition pathway along the $\mathrm{M_1}$--$\mathrm{M_2}$ axis. \textit{A posteriori} analysis of this axis reveals its physical interpretation as an emergent order parameter, the number of five-fold defects contained within a motif (Supplementary Section V.A.III). For example, moving away from $\mathrm{M_1}$, towards $\mathrm{M_2}$, and then to $\mathrm{M_6}$ is associated with an increase in the number of five-fold defects.
\\

To quantitatively capture this symmetry-breaking transition, we examine one-dimensional slices of the topological landscape along the $\mathrm{M_1}$--$\mathrm{M_2}$ axis (Fig.~\ref{fig:fig3}d). The negative log-likelihood ratio $-\log(p/p_{\mathrm{M}_1})$, where $p_{M1}$ is the probability density at the location of motif M$_1$, clearly reveals a global minimum at motif $\mathrm{M_1}$ for low nonreciprocity levels. This minimum deepens as $\mathcal{N}$ increases, indicating enhanced stability of crystalline order through nonreciprocal  {error correction}. A second minimum emerges at motif $\mathrm{M_2}$ as $\mathcal{N}$ approaches 1, resulting in a coexistence at $\mathcal{N} = 1$ and a subsequent dominance of defect-containing motifs at higher nonreciprocity levels. This behavior is reminiscent of a first-order phase transition, where a state transition is mediated by the coexistence of distant configurations. Revealing such hidden analogies provides a concrete framework to analyze nonequilibrium state transitions that otherwise lack a standard thermodynamic description.\\

 The analogy to a first-order phase transition points towards a set of testable hypotheses~(Supplementary~Section~V.E-F). For example, we demonstrate the existence of nonreciprocal hysteresis in the structural order parameter $\langle d_\mathrm{hex} \rangle$ as a function of $\mathcal{N}$ ~(Fig. 3e and Supplementary Video 12). Such hysteresis may arise if the transition between $\mathrm{M_1}$ and $\mathrm{M_2}$ is separated by some kinetic barrier and must proceed via less likely motifs. In this case, near the crossover at $\mathcal{N} = 1$, an $\mathrm{M_1}$-dominated configuration does not immediately break, but instead requires an additional increase in $\mathcal{N}$ to destabilize $\mathrm{M_1}$. This metastability results in a nonreciprocal hysteresis in which $\langle d_{\mathrm{hex}}\rangle$ forms a loop as a function of $\mathcal{N}$~(Fig.~2c), as $\mathcal{N}$ is slowly increased and decreased around the transition. We demonstrate this effect in our model~(Fig.~3e, Supplementary Video 12, Supplementary~Section~V.F), supporting the coexistence of $\mathrm{M}_1$ and $\mathrm{M}_2$ motifs at the $\mathcal{N} = 1$ structural transition and highlighting the power of the analogy from topological landscapes.
\\

Extending this analysis to experimental data, we construct time-resolved experimental topological landscapes by tracking motif frequencies within temporal windows (Fig.~\ref{fig:fig4}a and Supplementary~Videos~10-11).  These experimental landscapes resemble the model predictions at $\mathcal{N} = 1$, displaying identical dual-peaked distributions around motifs $\mathrm{M_1}$ and $\mathrm{M_2}$. Additionally, the experimental progression over time mirrors the model predictions under decreasing nonreciprocity (Fig. \ref{fig:fig4}b), including passage through a coexistence regime at the onset of the transition from traveling to fluctuating states, as quantified by velocity polarization dynamics. This agreement implies that pairwise nonreciprocal interactions at the nearest neighbor level are the dominant driver of self-organized structures and dynamics in the experiment. It also reinforces our   {independent} interaction inference that pairwise nonreciprocity between embryos decreases during the course of the experiment.  All together, these results support a striking correspondence between time-driven transitions in the experiment and those induced by varying nonreciprocity in the theoretical model (Supplementary~Section~II.B). \\

\section{Topological flowscapes reveal information flow and dissipation-driven bias in collective state transitions in the experiment}

\begin{figure*}
    \centering
    \small
    \includegraphics[width=\textwidth]{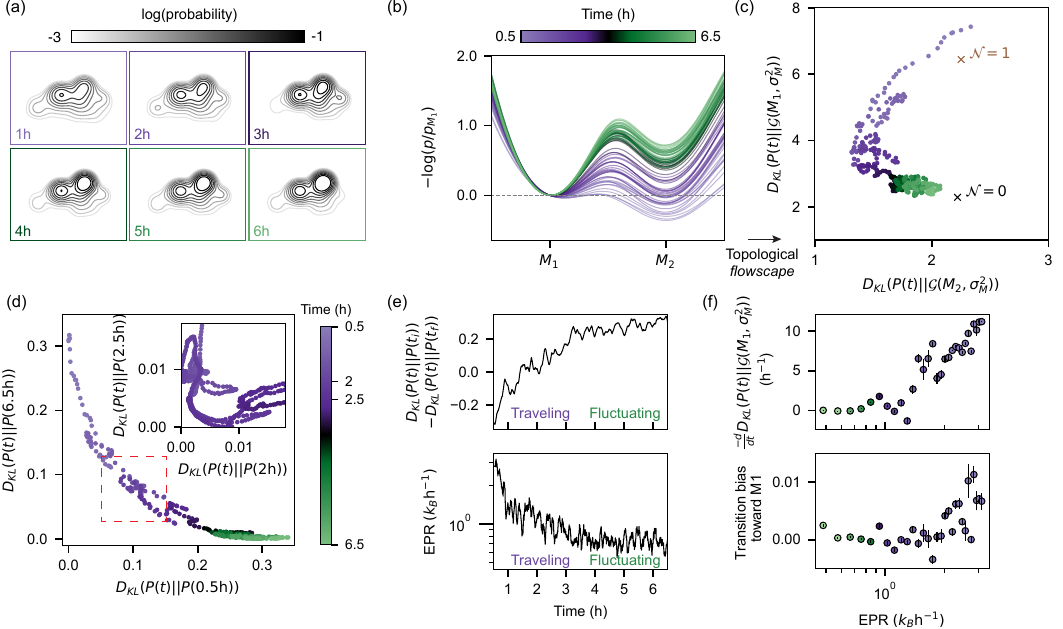}
    \caption{\textbf{ {Topological flowscapes reveal information flow and dissipation-driven bias in collective state transitions in the experiment.}}
    \panel{A}~Contour plots of experimental topological landscapes reveal a shifting balance between two dominant peaks over time. These peaks correspond to motifs $\mathrm{M_1}$ and $\mathrm{M_2}$, previously identified in the model at the experimentally inferred nonreciprocity $\mathcal{N} = 1$.
    \panel{B}~Negative log-likelihood relative to $\mathrm{M_1}$, measured along the $\mathrm{M_1}$--$\mathrm{M_2}$ axis. The self-organized structures transition from an $\mathrm{M_2}$-dominated to an $\mathrm{M_1}$-dominated state as the embryo mixture shifts from a traveling to a fluctuating state~(Fig.~\ref{fig:fig1}d).
    \panel{C}~\textit{Topological flowscape} of the experiment.
    Each landscape is mapped to a point using KL divergence from two synthetic reference distributions centered at $\mathrm{M_1}$ and $\mathrm{M_2}$, with widths $\sigma_M$ matched to the kernel of experimental landscape (Supplementary~Section~VI.B). The trajectory flows from a region near $\mathcal{N} = 1$ toward one resembling $\mathcal{N} = 0$. \panel{D} Flowscape constructed using experimental states at different times as dynamic references. Using 0.5 h and 6.5 h as references captures a long-timescale transition. \emph{Inset:} Using 2 h and 2.5 h as references reveals short-timescale cycle transitions associated with clusters merging and breaking (Supplementary Section VI.E)  {\panel{E}~Informational rate shift between experimental states. \textit{Top:} Diagonal displacement in the flowscape quantifies the log-likelihood ``progress bar'' of the current-time distribution from the initial state to the final state. The progress is rapid during the traveling state and slows significantly in the fluctuating state. \textit{Bottom:} Entropy production rate (EPR), estimated from changes in topological motif frequencies using thermodynamic speed limits,  {correlates with} the rate transition found in the flowscape. }
     \panel{F}~Emergent proofreading in topological motif transitions. \textit{Top:} Rate at which the system approaches the $\mathrm{M}_1$-dominated state, $-\frac{d}{dt}D_{KL}(P(t) \parallel \mathcal{G}(\mathrm{M}_1, \sigma^2_M))$, binned over the EPR estimates (colors indicate the average time of each bin with the same colorbar as (b)). The self-organized structure approaches $\mathrm{M}_1$ faster under a higher dissipation. \textit{Bottom:} Transition bias toward $\mathrm{M}_1$, $\frac{(\sum_{i \neq 1} {\textrm{M}_i\to \textrm{M}_1}) - (\sum_{i \neq 1} {\textrm{M}_1\to \textrm{M}_i})}{\sum_{i \neq j} \textrm{M}_i \to \textrm{M}_j} $, binned over the EPR estimates (Supplementary Section VII.C). The transition bias is enhanced with an increasing EPR estimate.
    }
    \label{fig:fig4}
\end{figure*}

Topological landscapes describe steady states without explicitly capturing the dynamic pathways connecting these states. While landscapes capture the key structural changes between snapshots, we also asked whether the latent representation adequately captures the nonequilibrium dynamic flows. To address this, we introduce the \textit{topological flowscape}, a quantitative framework that assigns coordinates to entire landscapes, analogous to how individual topological motifs are embedded~(Fig.~\ref{fig:fig4}c). 
The coordinates for the flowscape are defined by Kullback--Leibler (KL) divergences from two informed choices of reference distributions, representing $\mathrm{M_1}$ and $\mathrm{M_2}$, respectively (Supplementary~Section~VI). These divergences quantify the distinguishability between observed system states and reference states based on local motif frequencies, providing an information-theoretic interpretation: each axis measures the informational ``cost'' of describing the entire system state in terms of  {either} $\mathrm{M_1}$ or $\mathrm{M_2}$. 
Thus, the flowscape, a low-dimensional projection of the landscapes, visually captures their temporal evolution as a continuous trajectory or ``flow.''
\\

The topological flowscape quantitatively shows that the observed transition from an $\mathrm{M_2}$-dominated (disordered) state toward an $\mathrm{M_1}$-dominated (ordered) state is not gradual (Fig. \ref{fig:fig4}c). Instead, around 3 hours, the system undergoes a sharp directional change, initially moving rapidly toward both motifs to subsequently drifting slowly toward $\mathrm{M_1}$ while moving away from $\mathrm{M_2}$. 
Moreover, while the model predicts that the intermediate states are more crystalline than the final state, the experimental path deviates from this behavior~(Supplementary~Sections~VI.C--D). This discrepancy indicates potential higher-order mechanisms  {in addition to} the development-driven decrease in nonreciprocity~(Supplementary~Section~II.B), highlighting the flowscape's potential as a tool for dissecting complex experimental dynamics.
\\

The flexibility of reference distributions in the flowscape framework further allows us to probe structural dynamics  {and its corresponding information flow} across multiple temporal scales (Fig.~\ref{fig:fig4}d). Large temporal gaps between references elucidate global structural transitions, while shorter intervals reveal rapid cyclic dynamics such as repeated cluster merging and fragmentations (Supplementary Section VI.E).  {These cyclic dynamics indicate broken detailed balance between  {microscopic} topological configurations in the experiment}  which persist at the level of information flow of the macroscopic structural state.  
\\

Beyond the directional shift towards or away from M$_2$, the flowscape trajectory also exhibits a rate shift  {in the flow of structural information} between the traveling and fluctuating states. In particular, the flowscape built with temporal references (Fig.~\ref{fig:fig4}d) provides a log-likelihood-based ``progress bar'' (Fig.~\ref{fig:fig4}e), defined by $D_{KL}\left(P(t)||P(t_i)\right) - D_{KL}\left(P(t)||P(t_f)\right)$, whose rate tracks how  {quickly} the system transitions from initial to final states. This displacement shows a clear  { decrease in the rate of information transfer} that coincides with the independently measured shift in velocity polarization $|\mathbf{P}|$ from the traveling to fluctuating state (Fig.~\ref{fig:fig1}d),  {underlining a possible} connection between macroscopic state transitions and microscopic structural changes. To validate this rate transition, we estimate  {a lower bound for} the entropy production rate (EPR) using motif transition statistics and thermodynamic speed-limit principles (Fig.~\ref{fig:fig4}e and Supplementary~Section~VII.A). The EPR  {estimate}  {exhibits} a transition from high dissipation during the traveling state to lower dissipation in the fluctuating state,  {in concordance with} the rate transition captured by the flowscape. Moreover, the information  {rate} derived from information geometry also reveals a rate shift at the state transition (Supplementary~Section~VII.B). \\

 {In addition, we analyze the rate at which the system approaches the M$_1$-dominated reference state on the flowscape, $-\frac{d}{dt}D_{KL}\left(P(t)||\mathcal{G}(M_1,\sigma_M^2)\right)$, and its relation with the EPR estimate (Fig. \ref{fig:fig4}f). The two rates are positively correlated suggesting that enhanced dissipation, which occurs primarily in the traveling state, speeds up information flow towards the $\mathrm{M}_1$-dominated state.} In a generic nonequilibrium state transition, such a speed-up might simply arise from an overall upscaling of all transition fluxes. However, by normalizing the net influx toward M$_1$ by the total transition rate, we show that an increasing EPR estimate disproportionately enhances the directional bias toward the M$_1$ motif (Fig. \ref{fig:fig4}f, bottom).
 Overall, we find that in the traveling state, higher dissipation biases motif-motif transitions in favor of low-defect configurations, including the M$_2$ motif (Supplementary Section VII.C).  In the fluctuating state when dissipation is lower, this information flow and transition bias towards M$_1$ are likewise reduced. Because this directional bias suppresses structural defects, we interpret these results as an emergent proofreading, in which higher dissipation selectively favors information flow and transitions towards target states. Together, these results demonstrate that flowscapes uncover the experimental signatures of nonequilibrium dynamics underlying state transitions. 

\section{Discussion}

We have investigated the self-organization and state transitions driven by nonreciprocal interactions in active living matter. Through experiments on heterogeneous populations of starfish embryos, whose developmental ages control their nonreciprocal interactions, we  {identify two distinct} nonequilibrium states: a traveling state characterized by dynamic flocking clusters, and a fluctuating state defined by a crystal-like structure and reduced  {collective motion}. By developing a predictive model, we map out a broader spectrum of  {behaviors}, revealing that collective motion peaks at intermediate levels of nonreciprocity, while  {weaker nonreciprocity enhances structural order and} higher nonreciprocity induces structural instability and fragmentation.
\\

To characterize these emergent states and their transitions, we introduced two  {complementary} frameworks: \textit{topological landscapes} and \textit{topological flowscapes}. Topological landscapes provide a quantitative  {description of} local structural distributions and symmetry-breaking phenomena,  {but do not encode how systems move between states. Topological flowscapes address this limitation by representing the evolution of landscapes as trajectories of information flow in a reduced space, capturing both the directions and rates of transitions. This dynamical representation reveals that  nonequilibrium state transitions can be governed by directed probability flows, with measurable rates of progression and dissipation.\\ }

 Our conceptual framework suggests a broader perspective on nonequilibrium organization. Topological landscapes  elucidate the order of a system, while flowscapes reveal how  this order evolves and, further, how irreversibility  affects the dynamics of its evolution. Together, they  comprise a physics-informed, low-dimensional representation of structure and information flow in nonequilibrium systems.\\ 

 This perspective is broadly applicable to nonequilibrium systems whose behavior is  also governed by transition dynamics between states, rather than by steady-state structure alone.  Our framework connects structure, dynamics,  information flow and thermodynamics to provide a general method for  analyzing nonequilibrium transitions,  unpacking how microscopic interactions give rise to macroscopic organization. More broadly, these results highlight the nonequilibrium principle that self-organization depends not only on the states it occupies, but also on the directed flows between these states.

\begin{acknowledgments}
We thank Mehran Kardar, Raymond E. Goldstein, Ryo Hanai, Sarah A.M. Loos, Suraj Shankar, and Pedro E. Harunari for valuable discussions.
This research was supported by an Alfred P. Sloan Foundation Grant (G-2021-16758) to N.F. and J.D., and a National Science Foundation CAREER Award (PHY-1848247) to N.F.. H.L. and S.G. acknowledge the Gordon and Betty Moore Foundation for support as Physics of Living Systems Fellows through Grant No. GBMF4513. E.K. was supported by the National Science Foundation for a Graduate Research Fellowship under Grant No. DGE 2140743.
L.L. was supported by the National Science Foundation for a Graduate Research Fellowship under Grant No. 2141064. J.L. acknowledges the support of the Center for the Physics of Biological Function (PHY-1734030). 
This research received support through Schmidt Sciences, LLC (to J.D.), the MathWorks Professorship Fund (to J.D.), and National Science Foundation Award DMR-2214021 (to J.D.). N.F. and J.D. thank the WPI-SKCM$^2$ Hiroshima University for hospitality and support. The authors acknowledge the MIT SuperCloud and Lincoln Laboratory Supercomputing Center for providing HPC resources that have contributed to the research results reported within this paper. 
\end{acknowledgments}

\begin{appendix}
\section*{Supplementary Material}
Materials and methods are provided in the Supplementary Information \cite{suppinfo}, which includes references~\cite{stringer2021cellpose,pachitariu2022cellpose,allan2024trackpy,crocker1996methods,squires_effective_2001, blake_fundamental_1974, short_flows_2006,vanderwalt2014skimage,sternberg1983biomedical,pizer1987adaptive,breiman1996bagging,pedregosa2011sklearn,breiman1984classification,simpson1949measurement,clopper1934confidence,tipping2001sparse,wipf2004sparse,krogh1991simple,mackay1992practical,mackay1992bayesian,borg_modern_2005,lin_multidimensional_2024,bowman1997applied,shiraishi2018speed,hatano2001steady,crooks2007measuring,kim2021information} . The code and data for main figures is available from Ref.~\cite{lee_2026_20492239}.
\end{appendix}

\appendix

\bibliography{references}

\begin{thebibliography}{32}%
\makeatletter
\providecommand \@ifxundefined [1]{%
 \@ifx{#1\undefined}
}%
\providecommand \@ifnum [1]{%
 \ifnum #1\expandafter \@firstoftwo
 \else \expandafter \@secondoftwo
 \fi
}%
\providecommand \@ifx [1]{%
 \ifx #1\expandafter \@firstoftwo
 \else \expandafter \@secondoftwo
 \fi
}%
\providecommand \natexlab [1]{#1}%
\providecommand \enquote  [1]{``#1''}%
\providecommand \bibnamefont  [1]{#1}%
\providecommand \bibfnamefont [1]{#1}%
\providecommand \citenamefont [1]{#1}%
\providecommand \href@noop [0]{\@secondoftwo}%
\providecommand \href [0]{\begingroup \@sanitize@url \@href}%
\providecommand \@href[1]{\@@startlink{#1}\@@href}%
\providecommand \@@href[1]{\endgroup#1\@@endlink}%
\providecommand \@sanitize@url [0]{\catcode `\\12\catcode `\$12\catcode
  `\&12\catcode `\#12\catcode `\^12\catcode `\_12\catcode `\%12\relax}%
\providecommand \@@startlink[1]{}%
\providecommand \@@endlink[0]{}%
\providecommand \url  [0]{\begingroup\@sanitize@url \@url }%
\providecommand \@url [1]{\endgroup\@href {#1}{\urlprefix }}%
\providecommand \urlprefix  [0]{URL }%
\providecommand \Eprint [0]{\href }%
\providecommand \doibase [0]{https://doi.org/}%
\providecommand \selectlanguage [0]{\@gobble}%
\providecommand \bibinfo  [0]{\@secondoftwo}%
\providecommand \bibfield  [0]{\@secondoftwo}%
\providecommand \translation [1]{[#1]}%
\providecommand \BibitemOpen [0]{}%
\providecommand \bibitemStop [0]{}%
\providecommand \bibitemNoStop [0]{.\EOS\space}%
\providecommand \EOS [0]{\spacefactor3000\relax}%
\providecommand \BibitemShut  [1]{\csname bibitem#1\endcsname}%
\let\auto@bib@innerbib\@empty
%</preamble>
\bibitem [{\citenamefont {Stringer}\ \emph {et~al.}(2021)\citenamefont
  {Stringer}, \citenamefont {Wang}, \citenamefont {Michaelos},\ and\
  \citenamefont {Pachitariu}}]{stringer2021cellpose}%
  \BibitemOpen
  \bibfield  {author} {\bibinfo {author} {\bibfnamefont {C.}~\bibnamefont
  {Stringer}}, \bibinfo {author} {\bibfnamefont {T.}~\bibnamefont {Wang}},
  \bibinfo {author} {\bibfnamefont {M.}~\bibnamefont {Michaelos}},\ and\
  \bibinfo {author} {\bibfnamefont {M.}~\bibnamefont {Pachitariu}},\ }\bibfield
   {title} {\bibinfo {title} {Cellpose: a generalist algorithm for cellular
  segmentation},\ }\href {https://doi.org/10.1038/s41592-020-01018-x}
  {\bibfield  {journal} {\bibinfo  {journal} {Nature Methods}\ }\textbf
  {\bibinfo {volume} {18}},\ \bibinfo {pages} {100} (\bibinfo {year}
  {2021})}\BibitemShut {NoStop}%
\bibitem [{\citenamefont {Pachitariu}\ and\ \citenamefont
  {Stringer}(2022)}]{pachitariu2022cellpose}%
  \BibitemOpen
  \bibfield  {author} {\bibinfo {author} {\bibfnamefont {M.}~\bibnamefont
  {Pachitariu}}\ and\ \bibinfo {author} {\bibfnamefont {C.}~\bibnamefont
  {Stringer}},\ }\bibfield  {title} {\bibinfo {title} {Cellpose 2.0: how to
  train your own model},\ }\href {https://doi.org/10.1038/s41592-022-01663-4}
  {\bibfield  {journal} {\bibinfo  {journal} {Nature Methods}\ }\textbf
  {\bibinfo {volume} {18}},\ \bibinfo {pages} {1634} (\bibinfo {year}
  {2022})}\BibitemShut {NoStop}%
\bibitem [{\citenamefont {Allan}\ \emph {et~al.}(2024)\citenamefont {Allan},
  \citenamefont {Caswell}, \citenamefont {Keim}, \citenamefont {van~der Wel},\
  and\ \citenamefont {Verweij}}]{allan2024trackpy}%
  \BibitemOpen
  \bibfield  {author} {\bibinfo {author} {\bibfnamefont {D.~B.}\ \bibnamefont
  {Allan}}, \bibinfo {author} {\bibfnamefont {T.}~\bibnamefont {Caswell}},
  \bibinfo {author} {\bibfnamefont {N.~C.}\ \bibnamefont {Keim}}, \bibinfo
  {author} {\bibfnamefont {C.~M.}\ \bibnamefont {van~der Wel}},\ and\ \bibinfo
  {author} {\bibfnamefont {R.~W.}\ \bibnamefont {Verweij}},\ }\href
  {https://doi.org/10.5281/zenodo.12708864} {\bibinfo {title}
  {soft-matter/trackpy: v0.6.4}} (\bibinfo {year} {2024})\BibitemShut {NoStop}%
\bibitem [{\citenamefont {Crocker}\ and\ \citenamefont
  {Grier}(1996)}]{crocker1996methods}%
  \BibitemOpen
  \bibfield  {author} {\bibinfo {author} {\bibfnamefont {J.~C.}\ \bibnamefont
  {Crocker}}\ and\ \bibinfo {author} {\bibfnamefont {D.~G.}\ \bibnamefont
  {Grier}},\ }\bibfield  {title} {\bibinfo {title} {Methods of digital video
  microscopy for colloidal studies},\ }\href
  {https://doi.org/10.1006/jcis.1996.0217} {\bibfield  {journal} {\bibinfo
  {journal} {Journal of Colloid and Interface Science}\ }\textbf {\bibinfo
  {volume} {179}},\ \bibinfo {pages} {298} (\bibinfo {year}
  {1996})}\BibitemShut {NoStop}%
\bibitem [{\citenamefont {Tan}\ \emph {et~al.}(2022)\citenamefont {Tan},
  \citenamefont {Mietke}, \citenamefont {Li}, \citenamefont {Chen},
  \citenamefont {Higinbotham}, \citenamefont {Foster}, \citenamefont {Gokhale},
  \citenamefont {Dunkel},\ and\ \citenamefont {Fakhri}}]{tan2022odd}%
  \BibitemOpen
  \bibfield  {author} {\bibinfo {author} {\bibfnamefont {T.~H.}\ \bibnamefont
  {Tan}}, \bibinfo {author} {\bibfnamefont {A.}~\bibnamefont {Mietke}},
  \bibinfo {author} {\bibfnamefont {J.}~\bibnamefont {Li}}, \bibinfo {author}
  {\bibfnamefont {Y.}~\bibnamefont {Chen}}, \bibinfo {author} {\bibfnamefont
  {H.}~\bibnamefont {Higinbotham}}, \bibinfo {author} {\bibfnamefont {P.~J.}\
  \bibnamefont {Foster}}, \bibinfo {author} {\bibfnamefont {S.}~\bibnamefont
  {Gokhale}}, \bibinfo {author} {\bibfnamefont {J.}~\bibnamefont {Dunkel}},\
  and\ \bibinfo {author} {\bibfnamefont {N.}~\bibnamefont {Fakhri}},\
  }\bibfield  {title} {\bibinfo {title} {Odd dynamics of living chiral
  crystals},\ }\href@noop {} {\bibfield  {journal} {\bibinfo  {journal}
  {Nature}\ }\textbf {\bibinfo {volume} {607}},\ \bibinfo {pages} {287}
  (\bibinfo {year} {2022})}\BibitemShut {NoStop}%
\bibitem [{\citenamefont {Drescher}\ \emph {et~al.}(2009)\citenamefont
  {Drescher}, \citenamefont {Leptos}, \citenamefont {Tuval}, \citenamefont
  {Ishikawa}, \citenamefont {Pedley},\ and\ \citenamefont
  {Goldstein}}]{drescher2009dancing}%
  \BibitemOpen
  \bibfield  {author} {\bibinfo {author} {\bibfnamefont {K.}~\bibnamefont
  {Drescher}}, \bibinfo {author} {\bibfnamefont {K.~C.}\ \bibnamefont
  {Leptos}}, \bibinfo {author} {\bibfnamefont {I.}~\bibnamefont {Tuval}},
  \bibinfo {author} {\bibfnamefont {T.}~\bibnamefont {Ishikawa}}, \bibinfo
  {author} {\bibfnamefont {T.~J.}\ \bibnamefont {Pedley}},\ and\ \bibinfo
  {author} {\bibfnamefont {R.~E.}\ \bibnamefont {Goldstein}},\ }\bibfield
  {title} {\bibinfo {title} {Dancing volvox: hydrodynamic bound states of
  swimming algae},\ }\href@noop {} {\bibfield  {journal} {\bibinfo  {journal}
  {Physical Review Letters}\ }\textbf {\bibinfo {volume} {102}},\ \bibinfo
  {pages} {168101} (\bibinfo {year} {2009})}\BibitemShut {NoStop}%
\bibitem [{\citenamefont {Squires}(2001)}]{squires_effective_2001}%
  \BibitemOpen
  \bibfield  {author} {\bibinfo {author} {\bibfnamefont {T.~M.}\ \bibnamefont
  {Squires}},\ }\bibfield  {title} {\bibinfo {title} {Effective
  pseudo-potentials of hydrodynamic origin},\ }\href
  {https://doi.org/10.1017/S0022112001005432} {\bibfield  {journal} {\bibinfo
  {journal} {Journal of Fluid Mechanics}\ }\textbf {\bibinfo {volume} {443}},\
  \bibinfo {pages} {403} (\bibinfo {year} {2001})}\BibitemShut {NoStop}%
\bibitem [{\citenamefont {Blake}\ and\ \citenamefont
  {Chwang}(1974)}]{blake_fundamental_1974}%
  \BibitemOpen
  \bibfield  {author} {\bibinfo {author} {\bibfnamefont {J.~R.}\ \bibnamefont
  {Blake}}\ and\ \bibinfo {author} {\bibfnamefont {A.~T.}\ \bibnamefont
  {Chwang}},\ }\bibfield  {title} {\bibinfo {title} {Fundamental singularities
  of viscous flow: {Part} {I}: {The} image systems in the vicinity of a
  stationary no-slip boundary},\ }\href {https://doi.org/10.1007/BF02353701}
  {\bibfield  {journal} {\bibinfo  {journal} {Journal of Engineering
  Mathematics}\ }\textbf {\bibinfo {volume} {8}},\ \bibinfo {pages} {23}
  (\bibinfo {year} {1974})}\BibitemShut {NoStop}%
\bibitem [{\citenamefont {Short}\ \emph {et~al.}(2006)\citenamefont {Short},
  \citenamefont {Solari}, \citenamefont {Ganguly}, \citenamefont {Powers},
  \citenamefont {Kessler},\ and\ \citenamefont {Goldstein}}]{short_flows_2006}%
  \BibitemOpen
  \bibfield  {author} {\bibinfo {author} {\bibfnamefont {M.~B.}\ \bibnamefont
  {Short}}, \bibinfo {author} {\bibfnamefont {C.~A.}\ \bibnamefont {Solari}},
  \bibinfo {author} {\bibfnamefont {S.}~\bibnamefont {Ganguly}}, \bibinfo
  {author} {\bibfnamefont {T.~R.}\ \bibnamefont {Powers}}, \bibinfo {author}
  {\bibfnamefont {J.~O.}\ \bibnamefont {Kessler}},\ and\ \bibinfo {author}
  {\bibfnamefont {R.~E.}\ \bibnamefont {Goldstein}},\ }\bibfield  {title}
  {\bibinfo {title} {Flows driven by flagella of multicellular organisms
  enhance long-range molecular transport},\ }\href@noop {} {\bibfield
  {journal} {\bibinfo  {journal} {Proceedings of the National Academy of
  Sciences}\ }\textbf {\bibinfo {volume} {103}},\ \bibinfo {pages} {8315}
  (\bibinfo {year} {2006})}\BibitemShut {NoStop}%
\bibitem [{\citenamefont {van~der Walt}\ \emph {et~al.}(2014)\citenamefont
  {van~der Walt}, \citenamefont {Sch{\"o}nberger}, \citenamefont
  {Nunez-Iglesias}, \citenamefont {Boulogne}, \citenamefont {Warner},
  \citenamefont {Yager}, \citenamefont {Guillart}, \citenamefont {Yu},\ and\
  \citenamefont {the scikit-image contributers}}]{vanderwalt2014skimage}%
  \BibitemOpen
  \bibfield  {author} {\bibinfo {author} {\bibfnamefont {S.}~\bibnamefont
  {van~der Walt}}, \bibinfo {author} {\bibfnamefont {J.~L.}\ \bibnamefont
  {Sch{\"o}nberger}}, \bibinfo {author} {\bibfnamefont {J.}~\bibnamefont
  {Nunez-Iglesias}}, \bibinfo {author} {\bibfnamefont {F.}~\bibnamefont
  {Boulogne}}, \bibinfo {author} {\bibfnamefont {J.~D.}\ \bibnamefont
  {Warner}}, \bibinfo {author} {\bibfnamefont {N.}~\bibnamefont {Yager}},
  \bibinfo {author} {\bibfnamefont {E.}~\bibnamefont {Guillart}}, \bibinfo
  {author} {\bibfnamefont {T.}~\bibnamefont {Yu}},\ and\ \bibinfo {author}
  {\bibnamefont {the scikit-image contributers}},\ }\bibfield  {title}
  {\bibinfo {title} {{S}cikit-image: Image processing in python},\ }\href
  {https://doi.org/10.7717/peerj.453} {\bibfield  {journal} {\bibinfo
  {journal} {PeerJ}\ }\textbf {\bibinfo {volume} {2}},\ \bibinfo {pages} {e453}
  (\bibinfo {year} {2014})}\BibitemShut {NoStop}%
\bibitem [{\citenamefont {Sternberg}(1983)}]{sternberg1983biomedical}%
  \BibitemOpen
  \bibfield  {author} {\bibinfo {author} {\bibfnamefont {S.~R.}\ \bibnamefont
  {Sternberg}},\ }\bibfield  {title} {\bibinfo {title} {Biomedical image
  processing},\ }\href {https://doi.org/10.1109/MC.1983.1654163} {\bibfield
  {journal} {\bibinfo  {journal} {Computer}\ }\textbf {\bibinfo {volume}
  {16}},\ \bibinfo {pages} {22} (\bibinfo {year} {1983})}\BibitemShut {NoStop}%
\bibitem [{\citenamefont {Pizer}\ \emph {et~al.}(1987)\citenamefont {Pizer},
  \citenamefont {Amburn}, \citenamefont {Austin}, \citenamefont {Cromartie},
  \citenamefont {Geselowitz}, \citenamefont {Greer}, \citenamefont {ter
  Haar~Romeny}, \citenamefont {Zimmerman},\ and\ \citenamefont
  {Zuiderveld}}]{pizer1987adaptive}%
  \BibitemOpen
  \bibfield  {author} {\bibinfo {author} {\bibfnamefont {S.~M.}\ \bibnamefont
  {Pizer}}, \bibinfo {author} {\bibfnamefont {E.~P.}\ \bibnamefont {Amburn}},
  \bibinfo {author} {\bibfnamefont {J.~D.}\ \bibnamefont {Austin}}, \bibinfo
  {author} {\bibfnamefont {R.}~\bibnamefont {Cromartie}}, \bibinfo {author}
  {\bibfnamefont {A.}~\bibnamefont {Geselowitz}}, \bibinfo {author}
  {\bibfnamefont {T.}~\bibnamefont {Greer}}, \bibinfo {author} {\bibfnamefont
  {B.}~\bibnamefont {ter Haar~Romeny}}, \bibinfo {author} {\bibfnamefont
  {J.~B.}\ \bibnamefont {Zimmerman}},\ and\ \bibinfo {author} {\bibfnamefont
  {K.}~\bibnamefont {Zuiderveld}},\ }\bibfield  {title} {\bibinfo {title}
  {Adaptive histogram equalization and its variations},\ }\href
  {https://doi.org/10.1016/S0734-189X(87)80186-X} {\bibfield  {journal}
  {\bibinfo  {journal} {Computer Vision, Graphics, and Image Processing}\
  }\textbf {\bibinfo {volume} {39}},\ \bibinfo {pages} {355} (\bibinfo {year}
  {1987})}\BibitemShut {NoStop}%
\bibitem [{\citenamefont {Breiman}(1996)}]{breiman1996bagging}%
  \BibitemOpen
  \bibfield  {author} {\bibinfo {author} {\bibfnamefont {L.}~\bibnamefont
  {Breiman}},\ }\bibfield  {title} {\bibinfo {title} {Bagging predictors},\
  }\href {https://doi.org/10.1007/BF00058655} {\bibfield  {journal} {\bibinfo
  {journal} {Machine Learning}\ }\textbf {\bibinfo {volume} {24}},\ \bibinfo
  {pages} {123} (\bibinfo {year} {1996})}\BibitemShut {NoStop}%
\bibitem [{\citenamefont {Pedregosa}\ \emph {et~al.}(2011)\citenamefont
  {Pedregosa}, \citenamefont {Varoquaux}, \citenamefont {Gramfort},
  \citenamefont {Michel}, \citenamefont {Thirion}, \citenamefont {Grisel},
  \citenamefont {Blondel}, \citenamefont {Prettenhofer}, \citenamefont {Weiss},
  \citenamefont {Dubourg}, \citenamefont {Vanderplas}, \citenamefont {Passos},
  \citenamefont {Cournapeau}, \citenamefont {Brucher}, \citenamefont {Perrot},\
  and\ \citenamefont {Duchesnay}}]{pedregosa2011sklearn}%
  \BibitemOpen
  \bibfield  {author} {\bibinfo {author} {\bibfnamefont {F.}~\bibnamefont
  {Pedregosa}}, \bibinfo {author} {\bibfnamefont {G.}~\bibnamefont
  {Varoquaux}}, \bibinfo {author} {\bibfnamefont {A.}~\bibnamefont {Gramfort}},
  \bibinfo {author} {\bibfnamefont {V.}~\bibnamefont {Michel}}, \bibinfo
  {author} {\bibfnamefont {B.}~\bibnamefont {Thirion}}, \bibinfo {author}
  {\bibfnamefont {O.}~\bibnamefont {Grisel}}, \bibinfo {author} {\bibfnamefont
  {M.}~\bibnamefont {Blondel}}, \bibinfo {author} {\bibfnamefont
  {P.}~\bibnamefont {Prettenhofer}}, \bibinfo {author} {\bibfnamefont
  {R.}~\bibnamefont {Weiss}}, \bibinfo {author} {\bibfnamefont
  {V.}~\bibnamefont {Dubourg}}, \bibinfo {author} {\bibfnamefont
  {J.}~\bibnamefont {Vanderplas}}, \bibinfo {author} {\bibfnamefont
  {A.}~\bibnamefont {Passos}}, \bibinfo {author} {\bibfnamefont
  {D.}~\bibnamefont {Cournapeau}}, \bibinfo {author} {\bibfnamefont
  {M.}~\bibnamefont {Brucher}}, \bibinfo {author} {\bibfnamefont
  {M.}~\bibnamefont {Perrot}},\ and\ \bibinfo {author} {\bibfnamefont
  {E.}~\bibnamefont {Duchesnay}},\ }\bibfield  {title} {\bibinfo {title}
  {Scikit-learn: Machine learning in {P}ython},\ }\href
  {https://doi.org/10.48550/arXiv.1201.0490} {\bibfield  {journal} {\bibinfo
  {journal} {J. Mach. Learn. Res.}\ }\textbf {\bibinfo {volume} {12}},\
  \bibinfo {pages} {2825} (\bibinfo {year} {2011})}\BibitemShut {NoStop}%
\bibitem [{\citenamefont {Breiman}\ \emph {et~al.}(1984)\citenamefont
  {Breiman}, \citenamefont {Friedman}, \citenamefont {Olshen},\ and\
  \citenamefont {Stone}}]{breiman1984classification}%
  \BibitemOpen
  \bibfield  {author} {\bibinfo {author} {\bibfnamefont {L.}~\bibnamefont
  {Breiman}}, \bibinfo {author} {\bibfnamefont {J.}~\bibnamefont {Friedman}},
  \bibinfo {author} {\bibfnamefont {R.~A.}\ \bibnamefont {Olshen}},\ and\
  \bibinfo {author} {\bibfnamefont {C.~J.}\ \bibnamefont {Stone}},\ }\href
  {https://doi.org/10.1201/9781315139470} {\emph {\bibinfo {title}
  {Classification and Regression Trees}}}\ (\bibinfo  {publisher} {Taylor \&
  Francis},\ \bibinfo {address} {New York, NY, USA},\ \bibinfo {year}
  {1984})\BibitemShut {NoStop}%
\bibitem [{\citenamefont {Simpson}(1949)}]{simpson1949measurement}%
  \BibitemOpen
  \bibfield  {author} {\bibinfo {author} {\bibfnamefont {E.~H.}\ \bibnamefont
  {Simpson}},\ }\bibfield  {title} {\bibinfo {title} {Measurement of
  diversity},\ }\bibfield  {journal} {\bibinfo  {journal} {Nature}\ }\textbf
  {\bibinfo {volume} {163}},\ \href {https://doi.org/10.1038/163688a0}
  {10.1038/163688a0} (\bibinfo {year} {1949})\BibitemShut {NoStop}%
\bibitem [{\citenamefont {Clopper}\ and\ \citenamefont
  {Pearson}(1934)}]{clopper1934confidence}%
  \BibitemOpen
  \bibfield  {author} {\bibinfo {author} {\bibfnamefont {C.~J.}\ \bibnamefont
  {Clopper}}\ and\ \bibinfo {author} {\bibfnamefont {E.~S.}\ \bibnamefont
  {Pearson}},\ }\bibfield  {title} {\bibinfo {title} {On the use of confidence
  or fiducial limits illustrated in the case of the binomial},\ }\href
  {https://doi.org/10.1093/biomet/26.4.404} {\bibfield  {journal} {\bibinfo
  {journal} {Biometrika}\ }\textbf {\bibinfo {volume} {26}},\ \bibinfo {pages}
  {404} (\bibinfo {year} {1934})}\BibitemShut {NoStop}%
\bibitem [{\citenamefont {Tipping}(2001)}]{tipping2001sparse}%
  \BibitemOpen
  \bibfield  {author} {\bibinfo {author} {\bibfnamefont {M.~E.}\ \bibnamefont
  {Tipping}},\ }\bibfield  {title} {\bibinfo {title} {Sparse {Bayesian}
  learning and the relevance vector machine},\ }\href@noop {} {\bibfield
  {journal} {\bibinfo  {journal} {J. Mach. Learn. Res.}\ }\textbf {\bibinfo
  {volume} {1}},\ \bibinfo {pages} {211} (\bibinfo {year} {2001})}\BibitemShut
  {NoStop}%
\bibitem [{\citenamefont {Wipf}\ and\ \citenamefont
  {Rao}(2004)}]{wipf2004sparse}%
  \BibitemOpen
  \bibfield  {author} {\bibinfo {author} {\bibfnamefont {D.~P.}\ \bibnamefont
  {Wipf}}\ and\ \bibinfo {author} {\bibfnamefont {B.~D.}\ \bibnamefont {Rao}},\
  }\bibfield  {title} {\bibinfo {title} {Sparse {B}ayesian learning for basis
  selection},\ }\href@noop {} {\bibfield  {journal} {\bibinfo  {journal} {IEEE
  Trans. Signal Process.}\ }\textbf {\bibinfo {volume} {52}},\ \bibinfo {pages}
  {2153} (\bibinfo {year} {2004})}\BibitemShut {NoStop}%
\bibitem [{\citenamefont {Krogh}\ and\ \citenamefont
  {Hertz}(1991)}]{krogh1991simple}%
  \BibitemOpen
  \bibfield  {author} {\bibinfo {author} {\bibfnamefont {A.}~\bibnamefont
  {Krogh}}\ and\ \bibinfo {author} {\bibfnamefont {J.}~\bibnamefont {Hertz}},\
  }\bibfield  {title} {\bibinfo {title} {{A Simple Weight Decay Can Improve
  Generalization}},\ }in\ \href
  {https://proceedings.neurips.cc/paper_files/paper/1991/file/8eefcfdf5990e441f0fb6f3fad709e21-Paper.pdf}
  {\emph {\bibinfo {booktitle} {Advances in Neural Information Processing
  Systems}}},\ Vol.~\bibinfo {volume} {4},\ \bibinfo {editor} {edited by\
  \bibinfo {editor} {\bibfnamefont {J.}~\bibnamefont {Moody}}, \bibinfo
  {editor} {\bibfnamefont {S.}~\bibnamefont {Hanson}},\ and\ \bibinfo {editor}
  {\bibfnamefont {R.}~\bibnamefont {Lippmann}}}\ (\bibinfo  {publisher}
  {Morgan-Kaufmann},\ \bibinfo {address} {Burlington, MA},\ \bibinfo {year}
  {1991})\BibitemShut {NoStop}%
\bibitem [{\citenamefont {MacKay}(1992{\natexlab{a}})}]{mackay1992practical}%
  \BibitemOpen
  \bibfield  {author} {\bibinfo {author} {\bibfnamefont {D.~J.}\ \bibnamefont
  {MacKay}},\ }\bibfield  {title} {\bibinfo {title} {A practical {B}ayesian
  framework for backpropagation networks},\ }\href@noop {} {\bibfield
  {journal} {\bibinfo  {journal} {Neural Comput.}\ }\textbf {\bibinfo {volume}
  {4}},\ \bibinfo {pages} {448} (\bibinfo {year}
  {1992}{\natexlab{a}})}\BibitemShut {NoStop}%
\bibitem [{\citenamefont {MacKay}(1992{\natexlab{b}})}]{mackay1992bayesian}%
  \BibitemOpen
  \bibfield  {author} {\bibinfo {author} {\bibfnamefont {D.~J.}\ \bibnamefont
  {MacKay}},\ }\bibfield  {title} {\bibinfo {title} {Bayesian interpolation},\
  }\href@noop {} {\bibfield  {journal} {\bibinfo  {journal} {Neural Comput.}\
  }\textbf {\bibinfo {volume} {4}},\ \bibinfo {pages} {415} (\bibinfo {year}
  {1992}{\natexlab{b}})}\BibitemShut {NoStop}%
\bibitem [{\citenamefont {Skinner}\ \emph {et~al.}(2021)\citenamefont
  {Skinner}, \citenamefont {Song}, \citenamefont {Jeckel}, \citenamefont
  {Jelli}, \citenamefont {Drescher},\ and\ \citenamefont
  {Dunkel}}]{skinner2021topological}%
  \BibitemOpen
  \bibfield  {author} {\bibinfo {author} {\bibfnamefont {D.~J.}\ \bibnamefont
  {Skinner}}, \bibinfo {author} {\bibfnamefont {B.}~\bibnamefont {Song}},
  \bibinfo {author} {\bibfnamefont {H.}~\bibnamefont {Jeckel}}, \bibinfo
  {author} {\bibfnamefont {E.}~\bibnamefont {Jelli}}, \bibinfo {author}
  {\bibfnamefont {K.}~\bibnamefont {Drescher}},\ and\ \bibinfo {author}
  {\bibfnamefont {J.}~\bibnamefont {Dunkel}},\ }\bibfield  {title} {\bibinfo
  {title} {Topological metric detects hidden order in disordered media},\
  }\href@noop {} {\bibfield  {journal} {\bibinfo  {journal} {Physical Review
  Letters}\ }\textbf {\bibinfo {volume} {126}},\ \bibinfo {pages} {048101}
  (\bibinfo {year} {2021})}\BibitemShut {NoStop}%
\bibitem [{\citenamefont {Borg}\ and\ \citenamefont
  {Groenen}(2005)}]{borg_modern_2005}%
  \BibitemOpen
  \bibfield  {author} {\bibinfo {author} {\bibfnamefont {I.}~\bibnamefont
  {Borg}}\ and\ \bibinfo {author} {\bibfnamefont {P.~J.~F.}\ \bibnamefont
  {Groenen}},\ }\href@noop {} {\emph {\bibinfo {title} {Modern multidimensional
  scaling: theory and applications}}},\ \bibinfo {edition} {2nd}\ ed.,\
  Springer {Series} in {Statistics}\ (\bibinfo  {publisher} {Springer},\
  \bibinfo {address} {New York},\ \bibinfo {year} {2005})\BibitemShut {NoStop}%
\bibitem [{\citenamefont {Lin}(2024)}]{lin_multidimensional_2024}%
  \BibitemOpen
  \bibfield  {author} {\bibinfo {author} {\bibfnamefont {D.}~\bibnamefont
  {Lin}},\ }\href {https://juliastats.org/MultivariateStats.jl/dev/mds/}
  {\bibinfo {title} {Multidimensional {Scaling}: {MultivariateStats}.jl}}
  (\bibinfo {year} {2024})\BibitemShut {NoStop}%
\bibitem [{\citenamefont {Bowman}\ and\ \citenamefont
  {Azzalini}(1997)}]{bowman1997applied}%
  \BibitemOpen
  \bibfield  {author} {\bibinfo {author} {\bibfnamefont {A.~W.}\ \bibnamefont
  {Bowman}}\ and\ \bibinfo {author} {\bibfnamefont {A.}~\bibnamefont
  {Azzalini}},\ }\href@noop {} {\emph {\bibinfo {title} {Applied smoothing
  techniques for data analysis: the kernel approach with S-Plus
  illustrations}}},\ Vol.~\bibinfo {volume} {18}\ (\bibinfo  {publisher} {OUP
  Oxford},\ \bibinfo {year} {1997})\BibitemShut {NoStop}%
\bibitem [{\citenamefont {Shiraishi}\ \emph {et~al.}(2018)\citenamefont
  {Shiraishi}, \citenamefont {Funo},\ and\ \citenamefont
  {Saito}}]{shiraishi2018speed}%
  \BibitemOpen
  \bibfield  {author} {\bibinfo {author} {\bibfnamefont {N.}~\bibnamefont
  {Shiraishi}}, \bibinfo {author} {\bibfnamefont {K.}~\bibnamefont {Funo}},\
  and\ \bibinfo {author} {\bibfnamefont {K.}~\bibnamefont {Saito}},\ }\bibfield
   {title} {\bibinfo {title} {Speed limit for classical stochastic processes},\
  }\href@noop {} {\bibfield  {journal} {\bibinfo  {journal} {Physical Review
  Letters}\ }\textbf {\bibinfo {volume} {121}},\ \bibinfo {pages} {070601}
  (\bibinfo {year} {2018})}\BibitemShut {NoStop}%
\bibitem [{\citenamefont {Hatano}\ and\ \citenamefont
  {Sasa}(2001)}]{hatano2001steady}%
  \BibitemOpen
  \bibfield  {author} {\bibinfo {author} {\bibfnamefont {T.}~\bibnamefont
  {Hatano}}\ and\ \bibinfo {author} {\bibfnamefont {S.-i.}\ \bibnamefont
  {Sasa}},\ }\bibfield  {title} {\bibinfo {title} {Steady-state thermodynamics
  of langevin systems},\ }\href@noop {} {\bibfield  {journal} {\bibinfo
  {journal} {Physical Review Letters}\ }\textbf {\bibinfo {volume} {86}},\
  \bibinfo {pages} {3463} (\bibinfo {year} {2001})}\BibitemShut {NoStop}%
\bibitem [{\citenamefont {Crooks}(2007)}]{crooks2007measuring}%
  \BibitemOpen
  \bibfield  {author} {\bibinfo {author} {\bibfnamefont {G.~E.}\ \bibnamefont
  {Crooks}},\ }\bibfield  {title} {\bibinfo {title} {Measuring thermodynamic
  length},\ }\href@noop {} {\bibfield  {journal} {\bibinfo  {journal} {Physical
  Review Letters}\ }\textbf {\bibinfo {volume} {99}},\ \bibinfo {pages}
  {100602} (\bibinfo {year} {2007})}\BibitemShut {NoStop}%
\bibitem [{\citenamefont {Kim}(2021)}]{kim2021information}%
  \BibitemOpen
  \bibfield  {author} {\bibinfo {author} {\bibfnamefont {E.-j.}\ \bibnamefont
  {Kim}},\ }\bibfield  {title} {\bibinfo {title} {Information geometry,
  fluctuations, non-equilibrium thermodynamics, and geodesics in complex
  systems},\ }\href@noop {} {\bibfield  {journal} {\bibinfo  {journal}
  {Entropy}\ }\textbf {\bibinfo {volume} {23}},\ \bibinfo {pages} {1393}
  (\bibinfo {year} {2021})}\BibitemShut {NoStop}%
\bibitem [{\citenamefont {Marchetti}\ \emph {et~al.}(2013)\citenamefont
  {Marchetti}, \citenamefont {Joanny}, \citenamefont {Ramaswamy}, \citenamefont
  {Liverpool}, \citenamefont {Prost}, \citenamefont {Rao},\ and\ \citenamefont
  {Simha}}]{marchetti2013hydrodynamics}%
  \BibitemOpen
  \bibfield  {author} {\bibinfo {author} {\bibfnamefont {M.~C.}\ \bibnamefont
  {Marchetti}}, \bibinfo {author} {\bibfnamefont {J.-F.}\ \bibnamefont
  {Joanny}}, \bibinfo {author} {\bibfnamefont {S.}~\bibnamefont {Ramaswamy}},
  \bibinfo {author} {\bibfnamefont {T.~B.}\ \bibnamefont {Liverpool}}, \bibinfo
  {author} {\bibfnamefont {J.}~\bibnamefont {Prost}}, \bibinfo {author}
  {\bibfnamefont {M.}~\bibnamefont {Rao}},\ and\ \bibinfo {author}
  {\bibfnamefont {R.~A.}\ \bibnamefont {Simha}},\ }\bibfield  {title} {\bibinfo
  {title} {Hydrodynamics of soft active matter},\ }\href@noop {} {\bibfield
  {journal} {\bibinfo  {journal} {Reviews of modern physics}\ }\textbf
  {\bibinfo {volume} {85}},\ \bibinfo {pages} {1143} (\bibinfo {year}
  {2013})}\BibitemShut {NoStop}%
\bibitem [{\citenamefont {Scheibner}\ \emph {et~al.}(2020)\citenamefont
  {Scheibner}, \citenamefont {Souslov}, \citenamefont {Banerjee}, \citenamefont
  {Sur{\'o}wka}, \citenamefont {Irvine},\ and\ \citenamefont
  {Vitelli}}]{scheibner2020odd}%
  \BibitemOpen
  \bibfield  {author} {\bibinfo {author} {\bibfnamefont {C.}~\bibnamefont
  {Scheibner}}, \bibinfo {author} {\bibfnamefont {A.}~\bibnamefont {Souslov}},
  \bibinfo {author} {\bibfnamefont {D.}~\bibnamefont {Banerjee}}, \bibinfo
  {author} {\bibfnamefont {P.}~\bibnamefont {Sur{\'o}wka}}, \bibinfo {author}
  {\bibfnamefont {W.~T.}\ \bibnamefont {Irvine}},\ and\ \bibinfo {author}
  {\bibfnamefont {V.}~\bibnamefont {Vitelli}},\ }\bibfield  {title} {\bibinfo
  {title} {Odd elasticity},\ }\href@noop {} {\bibfield  {journal} {\bibinfo
  {journal} {Nature Physics}\ }\textbf {\bibinfo {volume} {16}},\ \bibinfo
  {pages} {475} (\bibinfo {year} {2020})}\BibitemShut {NoStop}%
\end{thebibliography}%


\providecommand{\noopsort}[1]{}\providecommand{\singleletter}[1]{#1}%
\begin{thebibliography}{74}%
\makeatletter
\providecommand \@ifxundefined [1]{%
 \@ifx{#1\undefined}
}%
\providecommand \@ifnum [1]{%
 \ifnum #1\expandafter \@firstoftwo
 \else \expandafter \@secondoftwo
 \fi
}%
\providecommand \@ifx [1]{%
 \ifx #1\expandafter \@firstoftwo
 \else \expandafter \@secondoftwo
 \fi
}%
\providecommand \natexlab [1]{#1}%
\providecommand \enquote  [1]{``#1''}%
\providecommand \bibnamefont  [1]{#1}%
\providecommand \bibfnamefont [1]{#1}%
\providecommand \citenamefont [1]{#1}%
\providecommand \href@noop [0]{\@secondoftwo}%
\providecommand \href [0]{\begingroup \@sanitize@url \@href}%
\providecommand \@href[1]{\@@startlink{#1}\@@href}%
\providecommand \@@href[1]{\endgroup#1\@@endlink}%
\providecommand \@sanitize@url [0]{\catcode `\\12\catcode `\$12\catcode
  `\&12\catcode `\#12\catcode `\^12\catcode `\_12\catcode `\%12\relax}%
\providecommand \@@startlink[1]{}%
\providecommand \@@endlink[0]{}%
\providecommand \url  [0]{\begingroup\@sanitize@url \@url }%
\providecommand \@url [1]{\endgroup\@href {#1}{\urlprefix }}%
\providecommand \urlprefix  [0]{URL }%
\providecommand \Eprint [0]{\href }%
\providecommand \doibase [0]{https://doi.org/}%
\providecommand \selectlanguage [0]{\@gobble}%
\providecommand \bibinfo  [0]{\@secondoftwo}%
\providecommand \bibfield  [0]{\@secondoftwo}%
\providecommand \translation [1]{[#1]}%
\providecommand \BibitemOpen [0]{}%
\providecommand \bibitemStop [0]{}%
\providecommand \bibitemNoStop [0]{.\EOS\space}%
\providecommand \EOS [0]{\spacefactor3000\relax}%
\providecommand \BibitemShut  [1]{\csname bibitem#1\endcsname}%
\let\auto@bib@innerbib\@empty
%</preamble>
\bibitem [{\citenamefont {Needleman}\ and\ \citenamefont
  {Dogic}(2017)}]{needleman2017active}%
  \BibitemOpen
  \bibfield  {author} {\bibinfo {author} {\bibfnamefont {D.}~\bibnamefont
  {Needleman}}\ and\ \bibinfo {author} {\bibfnamefont {Z.}~\bibnamefont
  {Dogic}},\ }\bibfield  {title} {\bibinfo {title} {Active matter at the
  interface between materials science and cell biology},\ }\href@noop {}
  {\bibfield  {journal} {\bibinfo  {journal} {Nature Reviews Materials}\
  }\textbf {\bibinfo {volume} {2}},\ \bibinfo {pages} {1} (\bibinfo {year}
  {2017})}\BibitemShut {NoStop}%
\bibitem [{\citenamefont {Marchetti}\ \emph {et~al.}(2013)\citenamefont
  {Marchetti}, \citenamefont {Joanny}, \citenamefont {Ramaswamy}, \citenamefont
  {Liverpool}, \citenamefont {Prost}, \citenamefont {Rao},\ and\ \citenamefont
  {Simha}}]{marchetti2013hydrodynamics}%
  \BibitemOpen
  \bibfield  {author} {\bibinfo {author} {\bibfnamefont {M.~C.}\ \bibnamefont
  {Marchetti}}, \bibinfo {author} {\bibfnamefont {J.-F.}\ \bibnamefont
  {Joanny}}, \bibinfo {author} {\bibfnamefont {S.}~\bibnamefont {Ramaswamy}},
  \bibinfo {author} {\bibfnamefont {T.~B.}\ \bibnamefont {Liverpool}}, \bibinfo
  {author} {\bibfnamefont {J.}~\bibnamefont {Prost}}, \bibinfo {author}
  {\bibfnamefont {M.}~\bibnamefont {Rao}},\ and\ \bibinfo {author}
  {\bibfnamefont {R.~A.}\ \bibnamefont {Simha}},\ }\bibfield  {title} {\bibinfo
  {title} {Hydrodynamics of soft active matter},\ }\href@noop {} {\bibfield
  {journal} {\bibinfo  {journal} {Reviews of Modern Physics}\ }\textbf
  {\bibinfo {volume} {85}},\ \bibinfo {pages} {1143} (\bibinfo {year}
  {2013})}\BibitemShut {NoStop}%
\bibitem [{\citenamefont {Vicsek}\ \emph {et~al.}(1995)\citenamefont {Vicsek},
  \citenamefont {Czir{\'o}k}, \citenamefont {Ben-Jacob}, \citenamefont
  {Cohen},\ and\ \citenamefont {Shochet}}]{vicsek1995novel}%
  \BibitemOpen
  \bibfield  {author} {\bibinfo {author} {\bibfnamefont {T.}~\bibnamefont
  {Vicsek}}, \bibinfo {author} {\bibfnamefont {A.}~\bibnamefont {Czir{\'o}k}},
  \bibinfo {author} {\bibfnamefont {E.}~\bibnamefont {Ben-Jacob}}, \bibinfo
  {author} {\bibfnamefont {I.}~\bibnamefont {Cohen}},\ and\ \bibinfo {author}
  {\bibfnamefont {O.}~\bibnamefont {Shochet}},\ }\bibfield  {title} {\bibinfo
  {title} {Novel type of phase transition in a system of self-driven
  particles},\ }\href@noop {} {\bibfield  {journal} {\bibinfo  {journal}
  {Physical Review Letters}\ }\textbf {\bibinfo {volume} {75}},\ \bibinfo
  {pages} {1226} (\bibinfo {year} {1995})}\BibitemShut {NoStop}%
\bibitem [{\citenamefont {Ballerini}\ \emph {et~al.}(2008)\citenamefont
  {Ballerini}, \citenamefont {Cabibbo}, \citenamefont {Candelier},
  \citenamefont {Cavagna}, \citenamefont {Cisbani}, \citenamefont {Giardina},
  \citenamefont {Lecomte}, \citenamefont {Orlandi}, \citenamefont {Parisi},
  \citenamefont {Procaccini} \emph {et~al.}}]{ballerini2008interaction}%
  \BibitemOpen
  \bibfield  {author} {\bibinfo {author} {\bibfnamefont {M.}~\bibnamefont
  {Ballerini}}, \bibinfo {author} {\bibfnamefont {N.}~\bibnamefont {Cabibbo}},
  \bibinfo {author} {\bibfnamefont {R.}~\bibnamefont {Candelier}}, \bibinfo
  {author} {\bibfnamefont {A.}~\bibnamefont {Cavagna}}, \bibinfo {author}
  {\bibfnamefont {E.}~\bibnamefont {Cisbani}}, \bibinfo {author} {\bibfnamefont
  {I.}~\bibnamefont {Giardina}}, \bibinfo {author} {\bibfnamefont
  {V.}~\bibnamefont {Lecomte}}, \bibinfo {author} {\bibfnamefont
  {A.}~\bibnamefont {Orlandi}}, \bibinfo {author} {\bibfnamefont
  {G.}~\bibnamefont {Parisi}}, \bibinfo {author} {\bibfnamefont
  {A.}~\bibnamefont {Procaccini}}, \emph {et~al.},\ }\bibfield  {title}
  {\bibinfo {title} {Interaction ruling animal collective behavior depends on
  topological rather than metric distance: Evidence from a field study},\
  }\href@noop {} {\bibfield  {journal} {\bibinfo  {journal} {Proceedings of the
  National Academy of Sciences}\ }\textbf {\bibinfo {volume} {105}},\ \bibinfo
  {pages} {1232} (\bibinfo {year} {2008})}\BibitemShut {NoStop}%
\bibitem [{\citenamefont {Drescher}\ \emph {et~al.}(2009)\citenamefont
  {Drescher}, \citenamefont {Leptos}, \citenamefont {Tuval}, \citenamefont
  {Ishikawa}, \citenamefont {Pedley},\ and\ \citenamefont
  {Goldstein}}]{drescher2009dancing}%
  \BibitemOpen
  \bibfield  {author} {\bibinfo {author} {\bibfnamefont {K.}~\bibnamefont
  {Drescher}}, \bibinfo {author} {\bibfnamefont {K.~C.}\ \bibnamefont
  {Leptos}}, \bibinfo {author} {\bibfnamefont {I.}~\bibnamefont {Tuval}},
  \bibinfo {author} {\bibfnamefont {T.}~\bibnamefont {Ishikawa}}, \bibinfo
  {author} {\bibfnamefont {T.~J.}\ \bibnamefont {Pedley}},\ and\ \bibinfo
  {author} {\bibfnamefont {R.~E.}\ \bibnamefont {Goldstein}},\ }\bibfield
  {title} {\bibinfo {title} {Dancing volvox: hydrodynamic bound states of
  swimming algae},\ }\href@noop {} {\bibfield  {journal} {\bibinfo  {journal}
  {Physical Review Letters}\ }\textbf {\bibinfo {volume} {102}},\ \bibinfo
  {pages} {168101} (\bibinfo {year} {2009})}\BibitemShut {NoStop}%
\bibitem [{\citenamefont {Hartmann}\ \emph {et~al.}(2019)\citenamefont
  {Hartmann}, \citenamefont {Singh}, \citenamefont {Pearce}, \citenamefont
  {Mok}, \citenamefont {Song}, \citenamefont {D{\'\i}az-Pascual}, \citenamefont
  {Dunkel},\ and\ \citenamefont {Drescher}}]{hartmann2019emergence}%
  \BibitemOpen
  \bibfield  {author} {\bibinfo {author} {\bibfnamefont {R.}~\bibnamefont
  {Hartmann}}, \bibinfo {author} {\bibfnamefont {P.~K.}\ \bibnamefont {Singh}},
  \bibinfo {author} {\bibfnamefont {P.}~\bibnamefont {Pearce}}, \bibinfo
  {author} {\bibfnamefont {R.}~\bibnamefont {Mok}}, \bibinfo {author}
  {\bibfnamefont {B.}~\bibnamefont {Song}}, \bibinfo {author} {\bibfnamefont
  {F.}~\bibnamefont {D{\'\i}az-Pascual}}, \bibinfo {author} {\bibfnamefont
  {J.}~\bibnamefont {Dunkel}},\ and\ \bibinfo {author} {\bibfnamefont
  {K.}~\bibnamefont {Drescher}},\ }\bibfield  {title} {\bibinfo {title}
  {Emergence of three-dimensional order and structure in growing biofilms},\
  }\href@noop {} {\bibfield  {journal} {\bibinfo  {journal} {Nature Physics}\
  }\textbf {\bibinfo {volume} {15}},\ \bibinfo {pages} {251} (\bibinfo {year}
  {2019})}\BibitemShut {NoStop}%
\bibitem [{\citenamefont {Ishikawa}\ \emph {et~al.}(2020)\citenamefont
  {Ishikawa}, \citenamefont {Pedley}, \citenamefont {Drescher},\ and\
  \citenamefont {Goldstein}}]{ishikawa2020stability}%
  \BibitemOpen
  \bibfield  {author} {\bibinfo {author} {\bibfnamefont {T.}~\bibnamefont
  {Ishikawa}}, \bibinfo {author} {\bibfnamefont {T.}~\bibnamefont {Pedley}},
  \bibinfo {author} {\bibfnamefont {K.}~\bibnamefont {Drescher}},\ and\
  \bibinfo {author} {\bibfnamefont {R.~E.}\ \bibnamefont {Goldstein}},\
  }\bibfield  {title} {\bibinfo {title} {Stability of dancing volvox},\
  }\href@noop {} {\bibfield  {journal} {\bibinfo  {journal} {Journal of Fluid
  Mechanics}\ }\textbf {\bibinfo {volume} {903}},\ \bibinfo {pages} {A11}
  (\bibinfo {year} {2020})}\BibitemShut {NoStop}%
\bibitem [{\citenamefont {Liu}\ \emph {et~al.}(2024)\citenamefont {Liu},
  \citenamefont {Li}, \citenamefont {Wang},\ and\ \citenamefont
  {Wu}}]{liu2024emergence}%
  \BibitemOpen
  \bibfield  {author} {\bibinfo {author} {\bibfnamefont {S.}~\bibnamefont
  {Liu}}, \bibinfo {author} {\bibfnamefont {Y.}~\bibnamefont {Li}}, \bibinfo
  {author} {\bibfnamefont {Y.}~\bibnamefont {Wang}},\ and\ \bibinfo {author}
  {\bibfnamefont {Y.}~\bibnamefont {Wu}},\ }\bibfield  {title} {\bibinfo
  {title} {Emergence of large-scale mechanical spiral waves in bacterial living
  matter},\ }\href@noop {} {\bibfield  {journal} {\bibinfo  {journal} {Nature
  Physics}\ }\textbf {\bibinfo {volume} {20}},\ \bibinfo {pages} {1015}
  (\bibinfo {year} {2024})}\BibitemShut {NoStop}%
\bibitem [{\citenamefont {von~der Heyde}\ \emph {et~al.}(2025)\citenamefont
  {von~der Heyde}, \citenamefont {Srinivasan}, \citenamefont {Birwa},
  \citenamefont {von~der Heyde}, \citenamefont {H{\"o}hn}, \citenamefont
  {Goldstein},\ and\ \citenamefont {Hallmann}}]{von2025spatiotemporal}%
  \BibitemOpen
  \bibfield  {author} {\bibinfo {author} {\bibfnamefont {B.}~\bibnamefont
  {von~der Heyde}}, \bibinfo {author} {\bibfnamefont {A.}~\bibnamefont
  {Srinivasan}}, \bibinfo {author} {\bibfnamefont {S.~K.}\ \bibnamefont
  {Birwa}}, \bibinfo {author} {\bibfnamefont {E.~L.}\ \bibnamefont {von~der
  Heyde}}, \bibinfo {author} {\bibfnamefont {S.~S.}\ \bibnamefont {H{\"o}hn}},
  \bibinfo {author} {\bibfnamefont {R.~E.}\ \bibnamefont {Goldstein}},\ and\
  \bibinfo {author} {\bibfnamefont {A.}~\bibnamefont {Hallmann}},\ }\bibfield
  {title} {\bibinfo {title} {Spatiotemporal distribution of the glycoprotein
  pherophorin ii reveals stochastic geometry of the growing ecm of volvox
  carteri},\ }\href@noop {} {\bibfield  {journal} {\bibinfo  {journal}
  {Proceedings of the National Academy of Sciences}\ }\textbf {\bibinfo
  {volume} {122}},\ \bibinfo {pages} {e2425759122} (\bibinfo {year}
  {2025})}\BibitemShut {NoStop}%
\bibitem [{\citenamefont {Giavazzi}\ \emph {et~al.}(2018)\citenamefont
  {Giavazzi}, \citenamefont {Paoluzzi}, \citenamefont {Macchi}, \citenamefont
  {Bi}, \citenamefont {Scita}, \citenamefont {Manning}, \citenamefont
  {Cerbino},\ and\ \citenamefont {Marchetti}}]{giavazzi2018flocking}%
  \BibitemOpen
  \bibfield  {author} {\bibinfo {author} {\bibfnamefont {F.}~\bibnamefont
  {Giavazzi}}, \bibinfo {author} {\bibfnamefont {M.}~\bibnamefont {Paoluzzi}},
  \bibinfo {author} {\bibfnamefont {M.}~\bibnamefont {Macchi}}, \bibinfo
  {author} {\bibfnamefont {D.}~\bibnamefont {Bi}}, \bibinfo {author}
  {\bibfnamefont {G.}~\bibnamefont {Scita}}, \bibinfo {author} {\bibfnamefont
  {M.~L.}\ \bibnamefont {Manning}}, \bibinfo {author} {\bibfnamefont
  {R.}~\bibnamefont {Cerbino}},\ and\ \bibinfo {author} {\bibfnamefont {M.~C.}\
  \bibnamefont {Marchetti}},\ }\bibfield  {title} {\bibinfo {title} {Flocking
  transitions in confluent tissues},\ }\href@noop {} {\bibfield  {journal}
  {\bibinfo  {journal} {Soft Matter}\ }\textbf {\bibinfo {volume} {14}},\
  \bibinfo {pages} {3471} (\bibinfo {year} {2018})}\BibitemShut {NoStop}%
\bibitem [{\citenamefont {Tang}\ \emph {et~al.}(2022)\citenamefont {Tang},
  \citenamefont {Das}, \citenamefont {Pegoraro}, \citenamefont {Han},
  \citenamefont {Huang}, \citenamefont {Roberts}, \citenamefont {Yang},
  \citenamefont {Fredberg}, \citenamefont {Kotton}, \citenamefont {Bi} \emph
  {et~al.}}]{tang2022collective}%
  \BibitemOpen
  \bibfield  {author} {\bibinfo {author} {\bibfnamefont {W.}~\bibnamefont
  {Tang}}, \bibinfo {author} {\bibfnamefont {A.}~\bibnamefont {Das}}, \bibinfo
  {author} {\bibfnamefont {A.~F.}\ \bibnamefont {Pegoraro}}, \bibinfo {author}
  {\bibfnamefont {Y.~L.}\ \bibnamefont {Han}}, \bibinfo {author} {\bibfnamefont
  {J.}~\bibnamefont {Huang}}, \bibinfo {author} {\bibfnamefont {D.~A.}\
  \bibnamefont {Roberts}}, \bibinfo {author} {\bibfnamefont {H.}~\bibnamefont
  {Yang}}, \bibinfo {author} {\bibfnamefont {J.~J.}\ \bibnamefont {Fredberg}},
  \bibinfo {author} {\bibfnamefont {D.~N.}\ \bibnamefont {Kotton}}, \bibinfo
  {author} {\bibfnamefont {D.}~\bibnamefont {Bi}}, \emph {et~al.},\ }\bibfield
  {title} {\bibinfo {title} {Collective curvature sensing and fluidity in
  three-dimensional multicellular systems},\ }\href@noop {} {\bibfield
  {journal} {\bibinfo  {journal} {Nature Physics}\ }\textbf {\bibinfo {volume}
  {18}},\ \bibinfo {pages} {1371} (\bibinfo {year} {2022})}\BibitemShut
  {NoStop}%
\bibitem [{\citenamefont {Bricard}\ \emph {et~al.}(2013)\citenamefont
  {Bricard}, \citenamefont {Caussin}, \citenamefont {Desreumaux}, \citenamefont
  {Dauchot},\ and\ \citenamefont {Bartolo}}]{bricard2013emergence}%
  \BibitemOpen
  \bibfield  {author} {\bibinfo {author} {\bibfnamefont {A.}~\bibnamefont
  {Bricard}}, \bibinfo {author} {\bibfnamefont {J.-B.}\ \bibnamefont
  {Caussin}}, \bibinfo {author} {\bibfnamefont {N.}~\bibnamefont {Desreumaux}},
  \bibinfo {author} {\bibfnamefont {O.}~\bibnamefont {Dauchot}},\ and\ \bibinfo
  {author} {\bibfnamefont {D.}~\bibnamefont {Bartolo}},\ }\bibfield  {title}
  {\bibinfo {title} {Emergence of macroscopic directed motion in populations of
  motile colloids},\ }\href@noop {} {\bibfield  {journal} {\bibinfo  {journal}
  {Nature}\ }\textbf {\bibinfo {volume} {503}},\ \bibinfo {pages} {95}
  (\bibinfo {year} {2013})}\BibitemShut {NoStop}%
\bibitem [{\citenamefont {Veenstra}\ \emph {et~al.}(2024)\citenamefont
  {Veenstra}, \citenamefont {Gamayun}, \citenamefont {Guo}, \citenamefont
  {Sarvi}, \citenamefont {Meinersen},\ and\ \citenamefont
  {Coulais}}]{veenstra2024non}%
  \BibitemOpen
  \bibfield  {author} {\bibinfo {author} {\bibfnamefont {J.}~\bibnamefont
  {Veenstra}}, \bibinfo {author} {\bibfnamefont {O.}~\bibnamefont {Gamayun}},
  \bibinfo {author} {\bibfnamefont {X.}~\bibnamefont {Guo}}, \bibinfo {author}
  {\bibfnamefont {A.}~\bibnamefont {Sarvi}}, \bibinfo {author} {\bibfnamefont
  {C.~V.}\ \bibnamefont {Meinersen}},\ and\ \bibinfo {author} {\bibfnamefont
  {C.}~\bibnamefont {Coulais}},\ }\bibfield  {title} {\bibinfo {title}
  {Non-reciprocal topological solitons in active metamaterials},\ }\href@noop
  {} {\bibfield  {journal} {\bibinfo  {journal} {Nature}\ }\textbf {\bibinfo
  {volume} {627}},\ \bibinfo {pages} {528} (\bibinfo {year}
  {2024})}\BibitemShut {NoStop}%
\bibitem [{\citenamefont {Hopfield}(1982)}]{hopfield1982neural}%
  \BibitemOpen
  \bibfield  {author} {\bibinfo {author} {\bibfnamefont {J.~J.}\ \bibnamefont
  {Hopfield}},\ }\bibfield  {title} {\bibinfo {title} {Neural networks and
  physical systems with emergent collective computational abilities.},\
  }\href@noop {} {\bibfield  {journal} {\bibinfo  {journal} {Proceedings of the
  National Academy of Sciences}\ }\textbf {\bibinfo {volume} {79}},\ \bibinfo
  {pages} {2554} (\bibinfo {year} {1982})}\BibitemShut {NoStop}%
\bibitem [{\citenamefont {Paxton}\ \emph {et~al.}(2004)\citenamefont {Paxton},
  \citenamefont {Kistler}, \citenamefont {Olmeda}, \citenamefont {Sen},
  \citenamefont {St.~Angelo}, \citenamefont {Cao}, \citenamefont {Mallouk},
  \citenamefont {Lammert},\ and\ \citenamefont {Crespi}}]{paxton2004catalytic}%
  \BibitemOpen
  \bibfield  {author} {\bibinfo {author} {\bibfnamefont {W.~F.}\ \bibnamefont
  {Paxton}}, \bibinfo {author} {\bibfnamefont {K.~C.}\ \bibnamefont {Kistler}},
  \bibinfo {author} {\bibfnamefont {C.~C.}\ \bibnamefont {Olmeda}}, \bibinfo
  {author} {\bibfnamefont {A.}~\bibnamefont {Sen}}, \bibinfo {author}
  {\bibfnamefont {S.~K.}\ \bibnamefont {St.~Angelo}}, \bibinfo {author}
  {\bibfnamefont {Y.}~\bibnamefont {Cao}}, \bibinfo {author} {\bibfnamefont
  {T.~E.}\ \bibnamefont {Mallouk}}, \bibinfo {author} {\bibfnamefont {P.~E.}\
  \bibnamefont {Lammert}},\ and\ \bibinfo {author} {\bibfnamefont {V.~H.}\
  \bibnamefont {Crespi}},\ }\bibfield  {title} {\bibinfo {title} {Catalytic
  nanomotors: autonomous movement of striped nanorods},\ }\href@noop {}
  {\bibfield  {journal} {\bibinfo  {journal} {Journal of the American Chemical
  Society}\ }\textbf {\bibinfo {volume} {126}},\ \bibinfo {pages} {13424}
  (\bibinfo {year} {2004})}\BibitemShut {NoStop}%
\bibitem [{\citenamefont {Yan}\ \emph {et~al.}(2016)\citenamefont {Yan},
  \citenamefont {Han}, \citenamefont {Zhang}, \citenamefont {Xu}, \citenamefont
  {Luijten},\ and\ \citenamefont {Granick}}]{yan2016reconfiguring}%
  \BibitemOpen
  \bibfield  {author} {\bibinfo {author} {\bibfnamefont {J.}~\bibnamefont
  {Yan}}, \bibinfo {author} {\bibfnamefont {M.}~\bibnamefont {Han}}, \bibinfo
  {author} {\bibfnamefont {J.}~\bibnamefont {Zhang}}, \bibinfo {author}
  {\bibfnamefont {C.}~\bibnamefont {Xu}}, \bibinfo {author} {\bibfnamefont
  {E.}~\bibnamefont {Luijten}},\ and\ \bibinfo {author} {\bibfnamefont
  {S.}~\bibnamefont {Granick}},\ }\bibfield  {title} {\bibinfo {title}
  {Reconfiguring active particles by electrostatic imbalance},\ }\href@noop {}
  {\bibfield  {journal} {\bibinfo  {journal} {Nature Materials}\ }\textbf
  {\bibinfo {volume} {15}},\ \bibinfo {pages} {1095} (\bibinfo {year}
  {2016})}\BibitemShut {NoStop}%
\bibitem [{\citenamefont {Mallory}\ \emph {et~al.}(2018)\citenamefont
  {Mallory}, \citenamefont {Valeriani},\ and\ \citenamefont
  {Cacciuto}}]{mallory2018active}%
  \BibitemOpen
  \bibfield  {author} {\bibinfo {author} {\bibfnamefont {S.~A.}\ \bibnamefont
  {Mallory}}, \bibinfo {author} {\bibfnamefont {C.}~\bibnamefont {Valeriani}},\
  and\ \bibinfo {author} {\bibfnamefont {A.}~\bibnamefont {Cacciuto}},\
  }\bibfield  {title} {\bibinfo {title} {An active approach to colloidal
  self-assembly},\ }\href@noop {} {\bibfield  {journal} {\bibinfo  {journal}
  {Annual Review of Physical Chemistry}\ }\textbf {\bibinfo {volume} {69}},\
  \bibinfo {pages} {59} (\bibinfo {year} {2018})}\BibitemShut {NoStop}%
\bibitem [{\citenamefont {Scheibner}\ \emph {et~al.}(2020)\citenamefont
  {Scheibner}, \citenamefont {Souslov}, \citenamefont {Banerjee}, \citenamefont
  {Sur{\'o}wka}, \citenamefont {Irvine},\ and\ \citenamefont
  {Vitelli}}]{scheibner2020odd}%
  \BibitemOpen
  \bibfield  {author} {\bibinfo {author} {\bibfnamefont {C.}~\bibnamefont
  {Scheibner}}, \bibinfo {author} {\bibfnamefont {A.}~\bibnamefont {Souslov}},
  \bibinfo {author} {\bibfnamefont {D.}~\bibnamefont {Banerjee}}, \bibinfo
  {author} {\bibfnamefont {P.}~\bibnamefont {Sur{\'o}wka}}, \bibinfo {author}
  {\bibfnamefont {W.~T.}\ \bibnamefont {Irvine}},\ and\ \bibinfo {author}
  {\bibfnamefont {V.}~\bibnamefont {Vitelli}},\ }\bibfield  {title} {\bibinfo
  {title} {Odd elasticity},\ }\href@noop {} {\bibfield  {journal} {\bibinfo
  {journal} {Nature Physics}\ }\textbf {\bibinfo {volume} {16}},\ \bibinfo
  {pages} {475} (\bibinfo {year} {2020})}\BibitemShut {NoStop}%
\bibitem [{\citenamefont {Meredith}\ \emph {et~al.}(2020)\citenamefont
  {Meredith}, \citenamefont {Moerman}, \citenamefont {Groenewold},
  \citenamefont {Chiu}, \citenamefont {Kegel}, \citenamefont {van Blaaderen},\
  and\ \citenamefont {Zarzar}}]{meredith2020predator}%
  \BibitemOpen
  \bibfield  {author} {\bibinfo {author} {\bibfnamefont {C.~H.}\ \bibnamefont
  {Meredith}}, \bibinfo {author} {\bibfnamefont {P.~G.}\ \bibnamefont
  {Moerman}}, \bibinfo {author} {\bibfnamefont {J.}~\bibnamefont {Groenewold}},
  \bibinfo {author} {\bibfnamefont {Y.-J.}\ \bibnamefont {Chiu}}, \bibinfo
  {author} {\bibfnamefont {W.~K.}\ \bibnamefont {Kegel}}, \bibinfo {author}
  {\bibfnamefont {A.}~\bibnamefont {van Blaaderen}},\ and\ \bibinfo {author}
  {\bibfnamefont {L.~D.}\ \bibnamefont {Zarzar}},\ }\bibfield  {title}
  {\bibinfo {title} {Predator--prey interactions between droplets driven by
  non-reciprocal oil exchange},\ }\href@noop {} {\bibfield  {journal} {\bibinfo
   {journal} {Nature Chemistry}\ }\textbf {\bibinfo {volume} {12}},\ \bibinfo
  {pages} {1136} (\bibinfo {year} {2020})}\BibitemShut {NoStop}%
\bibitem [{\citenamefont {Soto}\ and\ \citenamefont
  {Golestanian}(2014)}]{soto2014self}%
  \BibitemOpen
  \bibfield  {author} {\bibinfo {author} {\bibfnamefont {R.}~\bibnamefont
  {Soto}}\ and\ \bibinfo {author} {\bibfnamefont {R.}~\bibnamefont
  {Golestanian}},\ }\bibfield  {title} {\bibinfo {title} {Self-assembly of
  catalytically active colloidal molecules: tailoring activity through surface
  chemistry},\ }\href@noop {} {\bibfield  {journal} {\bibinfo  {journal}
  {Physical Review Letters}\ }\textbf {\bibinfo {volume} {112}},\ \bibinfo
  {pages} {068301} (\bibinfo {year} {2014})}\BibitemShut {NoStop}%
\bibitem [{\citenamefont {Saha}\ \emph {et~al.}(2020)\citenamefont {Saha},
  \citenamefont {Agudo-Canalejo},\ and\ \citenamefont
  {Golestanian}}]{saha2020scalar}%
  \BibitemOpen
  \bibfield  {author} {\bibinfo {author} {\bibfnamefont {S.}~\bibnamefont
  {Saha}}, \bibinfo {author} {\bibfnamefont {J.}~\bibnamefont
  {Agudo-Canalejo}},\ and\ \bibinfo {author} {\bibfnamefont {R.}~\bibnamefont
  {Golestanian}},\ }\bibfield  {title} {\bibinfo {title} {Scalar active
  mixtures: The nonreciprocal cahn-hilliard model},\ }\href@noop {} {\bibfield
  {journal} {\bibinfo  {journal} {Physical Review X}\ }\textbf {\bibinfo
  {volume} {10}},\ \bibinfo {pages} {041009} (\bibinfo {year}
  {2020})}\BibitemShut {NoStop}%
\bibitem [{\citenamefont {You}\ \emph {et~al.}(2020)\citenamefont {You},
  \citenamefont {Baskaran},\ and\ \citenamefont
  {Marchetti}}]{you2020nonreciprocity}%
  \BibitemOpen
  \bibfield  {author} {\bibinfo {author} {\bibfnamefont {Z.}~\bibnamefont
  {You}}, \bibinfo {author} {\bibfnamefont {A.}~\bibnamefont {Baskaran}},\ and\
  \bibinfo {author} {\bibfnamefont {M.~C.}\ \bibnamefont {Marchetti}},\
  }\bibfield  {title} {\bibinfo {title} {Nonreciprocity as a generic route to
  traveling states},\ }\href@noop {} {\bibfield  {journal} {\bibinfo  {journal}
  {Proceedings of the National Academy of Sciences}\ }\textbf {\bibinfo
  {volume} {117}},\ \bibinfo {pages} {19767} (\bibinfo {year}
  {2020})}\BibitemShut {NoStop}%
\bibitem [{\citenamefont {Fruchart}\ \emph {et~al.}(2021)\citenamefont
  {Fruchart}, \citenamefont {Hanai}, \citenamefont {Littlewood},\ and\
  \citenamefont {Vitelli}}]{fruchart2021non}%
  \BibitemOpen
  \bibfield  {author} {\bibinfo {author} {\bibfnamefont {M.}~\bibnamefont
  {Fruchart}}, \bibinfo {author} {\bibfnamefont {R.}~\bibnamefont {Hanai}},
  \bibinfo {author} {\bibfnamefont {P.~B.}\ \bibnamefont {Littlewood}},\ and\
  \bibinfo {author} {\bibfnamefont {V.}~\bibnamefont {Vitelli}},\ }\bibfield
  {title} {\bibinfo {title} {Non-reciprocal phase transitions},\ }\href@noop {}
  {\bibfield  {journal} {\bibinfo  {journal} {Nature}\ }\textbf {\bibinfo
  {volume} {592}},\ \bibinfo {pages} {363} (\bibinfo {year}
  {2021})}\BibitemShut {NoStop}%
\bibitem [{\citenamefont {Pisegna}\ \emph {et~al.}(2024)\citenamefont
  {Pisegna}, \citenamefont {Saha},\ and\ \citenamefont
  {Golestanian}}]{pisegna2024emergent}%
  \BibitemOpen
  \bibfield  {author} {\bibinfo {author} {\bibfnamefont {G.}~\bibnamefont
  {Pisegna}}, \bibinfo {author} {\bibfnamefont {S.}~\bibnamefont {Saha}},\ and\
  \bibinfo {author} {\bibfnamefont {R.}~\bibnamefont {Golestanian}},\
  }\bibfield  {title} {\bibinfo {title} {Emergent polar order in nonpolar
  mixtures with nonreciprocal interactions},\ }\href@noop {} {\bibfield
  {journal} {\bibinfo  {journal} {Proceedings of the National Academy of
  Sciences}\ }\textbf {\bibinfo {volume} {121}},\ \bibinfo {pages}
  {e2407705121} (\bibinfo {year} {2024})}\BibitemShut {NoStop}%
\bibitem [{\citenamefont {Mandal}\ \emph {et~al.}(2024)\citenamefont {Mandal},
  \citenamefont {Jaramillo},\ and\ \citenamefont
  {Sollich}}]{mandal2024robustness}%
  \BibitemOpen
  \bibfield  {author} {\bibinfo {author} {\bibfnamefont {R.}~\bibnamefont
  {Mandal}}, \bibinfo {author} {\bibfnamefont {S.~S.}\ \bibnamefont
  {Jaramillo}},\ and\ \bibinfo {author} {\bibfnamefont {P.}~\bibnamefont
  {Sollich}},\ }\bibfield  {title} {\bibinfo {title} {Robustness of traveling
  states in generic nonreciprocal mixtures},\ }\href@noop {} {\bibfield
  {journal} {\bibinfo  {journal} {Physical Review E}\ }\textbf {\bibinfo
  {volume} {109}},\ \bibinfo {pages} {L062602} (\bibinfo {year}
  {2024})}\BibitemShut {NoStop}%
\bibitem [{\citenamefont {Chen}\ \emph {et~al.}(2024)\citenamefont {Chen},
  \citenamefont {Lei}, \citenamefont {Xiang}, \citenamefont {Duan},
  \citenamefont {Peng},\ and\ \citenamefont {Zhang}}]{chen2024emergent}%
  \BibitemOpen
  \bibfield  {author} {\bibinfo {author} {\bibfnamefont {J.}~\bibnamefont
  {Chen}}, \bibinfo {author} {\bibfnamefont {X.}~\bibnamefont {Lei}}, \bibinfo
  {author} {\bibfnamefont {Y.}~\bibnamefont {Xiang}}, \bibinfo {author}
  {\bibfnamefont {M.}~\bibnamefont {Duan}}, \bibinfo {author} {\bibfnamefont
  {X.}~\bibnamefont {Peng}},\ and\ \bibinfo {author} {\bibfnamefont
  {H.}~\bibnamefont {Zhang}},\ }\bibfield  {title} {\bibinfo {title} {Emergent
  chirality and hyperuniformity in an active mixture with nonreciprocal
  interactions},\ }\href@noop {} {\bibfield  {journal} {\bibinfo  {journal}
  {Physical Review Letters}\ }\textbf {\bibinfo {volume} {132}},\ \bibinfo
  {pages} {118301} (\bibinfo {year} {2024})}\BibitemShut {NoStop}%
\bibitem [{\citenamefont {Parkavousi}\ \emph {et~al.}(2025)\citenamefont
  {Parkavousi}, \citenamefont {Rana}, \citenamefont {Golestanian},\ and\
  \citenamefont {Saha}}]{parkavousi2025enhanced}%
  \BibitemOpen
  \bibfield  {author} {\bibinfo {author} {\bibfnamefont {L.}~\bibnamefont
  {Parkavousi}}, \bibinfo {author} {\bibfnamefont {N.}~\bibnamefont {Rana}},
  \bibinfo {author} {\bibfnamefont {R.}~\bibnamefont {Golestanian}},\ and\
  \bibinfo {author} {\bibfnamefont {S.}~\bibnamefont {Saha}},\ }\bibfield
  {title} {\bibinfo {title} {Enhanced stability and chaotic condensates in
  multispecies nonreciprocal mixtures},\ }\href@noop {} {\bibfield  {journal}
  {\bibinfo  {journal} {Physical Review Letters}\ }\textbf {\bibinfo {volume}
  {134}},\ \bibinfo {pages} {148301} (\bibinfo {year} {2025})}\BibitemShut
  {NoStop}%
\bibitem [{\citenamefont {Markovich}\ and\ \citenamefont
  {Lubensky}(2024)}]{markovich2024nonreciprocity}%
  \BibitemOpen
  \bibfield  {author} {\bibinfo {author} {\bibfnamefont {T.}~\bibnamefont
  {Markovich}}\ and\ \bibinfo {author} {\bibfnamefont {T.~C.}\ \bibnamefont
  {Lubensky}},\ }\bibfield  {title} {\bibinfo {title} {Nonreciprocity and odd
  viscosity in chiral active fluids},\ }\href@noop {} {\bibfield  {journal}
  {\bibinfo  {journal} {Proceedings of the National Academy of Sciences}\
  }\textbf {\bibinfo {volume} {121}},\ \bibinfo {pages} {e2219385121} (\bibinfo
  {year} {2024})}\BibitemShut {NoStop}%
\bibitem [{\citenamefont {Dinelli}\ \emph {et~al.}(2023)\citenamefont
  {Dinelli}, \citenamefont {O’Byrne}, \citenamefont {Curatolo}, \citenamefont
  {Zhao}, \citenamefont {Sollich},\ and\ \citenamefont
  {Tailleur}}]{dinelli2023non}%
  \BibitemOpen
  \bibfield  {author} {\bibinfo {author} {\bibfnamefont {A.}~\bibnamefont
  {Dinelli}}, \bibinfo {author} {\bibfnamefont {J.}~\bibnamefont {O’Byrne}},
  \bibinfo {author} {\bibfnamefont {A.}~\bibnamefont {Curatolo}}, \bibinfo
  {author} {\bibfnamefont {Y.}~\bibnamefont {Zhao}}, \bibinfo {author}
  {\bibfnamefont {P.}~\bibnamefont {Sollich}},\ and\ \bibinfo {author}
  {\bibfnamefont {J.}~\bibnamefont {Tailleur}},\ }\bibfield  {title} {\bibinfo
  {title} {Non-reciprocity across scales in active mixtures},\ }\href@noop {}
  {\bibfield  {journal} {\bibinfo  {journal} {Nature Communications}\ }\textbf
  {\bibinfo {volume} {14}},\ \bibinfo {pages} {7035} (\bibinfo {year}
  {2023})}\BibitemShut {NoStop}%
\bibitem [{\citenamefont {Osat}\ and\ \citenamefont
  {Golestanian}(2023)}]{osat2023non}%
  \BibitemOpen
  \bibfield  {author} {\bibinfo {author} {\bibfnamefont {S.}~\bibnamefont
  {Osat}}\ and\ \bibinfo {author} {\bibfnamefont {R.}~\bibnamefont
  {Golestanian}},\ }\bibfield  {title} {\bibinfo {title} {Non-reciprocal
  multifarious self-organization},\ }\href@noop {} {\bibfield  {journal}
  {\bibinfo  {journal} {Nature Nanotechnology}\ }\textbf {\bibinfo {volume}
  {18}},\ \bibinfo {pages} {79} (\bibinfo {year} {2023})}\BibitemShut {NoStop}%
\bibitem [{\citenamefont {Brauns}\ and\ \citenamefont
  {Marchetti}(2024)}]{brauns2024nonreciprocal}%
  \BibitemOpen
  \bibfield  {author} {\bibinfo {author} {\bibfnamefont {F.}~\bibnamefont
  {Brauns}}\ and\ \bibinfo {author} {\bibfnamefont {M.~C.}\ \bibnamefont
  {Marchetti}},\ }\bibfield  {title} {\bibinfo {title} {Nonreciprocal pattern
  formation of conserved fields},\ }\href@noop {} {\bibfield  {journal}
  {\bibinfo  {journal} {Physical Review X}\ }\textbf {\bibinfo {volume} {14}},\
  \bibinfo {pages} {021014} (\bibinfo {year} {2024})}\BibitemShut {NoStop}%
\bibitem [{\citenamefont {Kreienkamp}\ and\ \citenamefont
  {Klapp}(2024)}]{kreienkamp2024nonreciprocal}%
  \BibitemOpen
  \bibfield  {author} {\bibinfo {author} {\bibfnamefont {K.~L.}\ \bibnamefont
  {Kreienkamp}}\ and\ \bibinfo {author} {\bibfnamefont {S.~H.}\ \bibnamefont
  {Klapp}},\ }\bibfield  {title} {\bibinfo {title} {Nonreciprocal alignment
  induces asymmetric clustering in active mixtures},\ }\href@noop {} {\bibfield
   {journal} {\bibinfo  {journal} {Physical Review Letters}\ }\textbf {\bibinfo
  {volume} {133}},\ \bibinfo {pages} {258303} (\bibinfo {year}
  {2024})}\BibitemShut {NoStop}%
\bibitem [{\citenamefont {Guillet}\ \emph {et~al.}(2025)\citenamefont
  {Guillet}, \citenamefont {Poncet}, \citenamefont {Le~Blay}, \citenamefont
  {Irvine}, \citenamefont {Vitelli},\ and\ \citenamefont
  {Bartolo}}]{guillet2025melting}%
  \BibitemOpen
  \bibfield  {author} {\bibinfo {author} {\bibfnamefont {S.}~\bibnamefont
  {Guillet}}, \bibinfo {author} {\bibfnamefont {A.}~\bibnamefont {Poncet}},
  \bibinfo {author} {\bibfnamefont {M.}~\bibnamefont {Le~Blay}}, \bibinfo
  {author} {\bibfnamefont {W.~T.}\ \bibnamefont {Irvine}}, \bibinfo {author}
  {\bibfnamefont {V.}~\bibnamefont {Vitelli}},\ and\ \bibinfo {author}
  {\bibfnamefont {D.}~\bibnamefont {Bartolo}},\ }\bibfield  {title} {\bibinfo
  {title} {Melting of nonreciprocal solids: How dislocations propel and fission
  in flowing crystals},\ }\href@noop {} {\bibfield  {journal} {\bibinfo
  {journal} {Proceedings of the National Academy of Sciences}\ }\textbf
  {\bibinfo {volume} {122}},\ \bibinfo {pages} {e2412993122} (\bibinfo {year}
  {2025})}\BibitemShut {NoStop}%
\bibitem [{\citenamefont {Kole}\ \emph {et~al.}(2025)\citenamefont {Kole},
  \citenamefont {Chao}, \citenamefont {Mauleon-Amieva}, \citenamefont {Hanai},
  \citenamefont {Royall},\ and\ \citenamefont {Liverpool}}]{kole2025non}%
  \BibitemOpen
  \bibfield  {author} {\bibinfo {author} {\bibfnamefont {S.}~\bibnamefont
  {Kole}}, \bibinfo {author} {\bibfnamefont {X.}~\bibnamefont {Chao}}, \bibinfo
  {author} {\bibfnamefont {A.}~\bibnamefont {Mauleon-Amieva}}, \bibinfo
  {author} {\bibfnamefont {R.}~\bibnamefont {Hanai}}, \bibinfo {author}
  {\bibfnamefont {C.~P.}\ \bibnamefont {Royall}},\ and\ \bibinfo {author}
  {\bibfnamefont {T.~B.}\ \bibnamefont {Liverpool}},\ }\bibfield  {title}
  {\bibinfo {title} {Non-reciprocal interactions drive emergent chiral
  crystallites},\ }\href@noop {} {\bibfield  {journal} {\bibinfo  {journal}
  {arXiv preprint arXiv:2501.15996}\ } (\bibinfo {year} {2025})}\BibitemShut
  {NoStop}%
\bibitem [{\citenamefont {Loos}\ and\ \citenamefont
  {Klapp}(2020)}]{loos2020irreversibility}%
  \BibitemOpen
  \bibfield  {author} {\bibinfo {author} {\bibfnamefont {S.~A.}\ \bibnamefont
  {Loos}}\ and\ \bibinfo {author} {\bibfnamefont {S.~H.}\ \bibnamefont
  {Klapp}},\ }\bibfield  {title} {\bibinfo {title} {Irreversibility, heat and
  information flows induced by non-reciprocal interactions},\ }\href@noop {}
  {\bibfield  {journal} {\bibinfo  {journal} {New Journal of Physics}\ }\textbf
  {\bibinfo {volume} {22}},\ \bibinfo {pages} {123051} (\bibinfo {year}
  {2020})}\BibitemShut {NoStop}%
\bibitem [{\citenamefont {Bowick}\ \emph {et~al.}(2022)\citenamefont {Bowick},
  \citenamefont {Fakhri}, \citenamefont {Marchetti},\ and\ \citenamefont
  {Ramaswamy}}]{bowick2022symmetry}%
  \BibitemOpen
  \bibfield  {author} {\bibinfo {author} {\bibfnamefont {M.~J.}\ \bibnamefont
  {Bowick}}, \bibinfo {author} {\bibfnamefont {N.}~\bibnamefont {Fakhri}},
  \bibinfo {author} {\bibfnamefont {M.~C.}\ \bibnamefont {Marchetti}},\ and\
  \bibinfo {author} {\bibfnamefont {S.}~\bibnamefont {Ramaswamy}},\ }\bibfield
  {title} {\bibinfo {title} {Symmetry, thermodynamics, and topology in active
  matter},\ }\href@noop {} {\bibfield  {journal} {\bibinfo  {journal} {Physical
  Review X}\ }\textbf {\bibinfo {volume} {12}},\ \bibinfo {pages} {010501}
  (\bibinfo {year} {2022})}\BibitemShut {NoStop}%
\bibitem [{\citenamefont {Tan}\ \emph {et~al.}(2022)\citenamefont {Tan},
  \citenamefont {Mietke}, \citenamefont {Li}, \citenamefont {Chen},
  \citenamefont {Higinbotham}, \citenamefont {Foster}, \citenamefont {Gokhale},
  \citenamefont {Dunkel},\ and\ \citenamefont {Fakhri}}]{tan2022odd}%
  \BibitemOpen
  \bibfield  {author} {\bibinfo {author} {\bibfnamefont {T.~H.}\ \bibnamefont
  {Tan}}, \bibinfo {author} {\bibfnamefont {A.}~\bibnamefont {Mietke}},
  \bibinfo {author} {\bibfnamefont {J.}~\bibnamefont {Li}}, \bibinfo {author}
  {\bibfnamefont {Y.}~\bibnamefont {Chen}}, \bibinfo {author} {\bibfnamefont
  {H.}~\bibnamefont {Higinbotham}}, \bibinfo {author} {\bibfnamefont {P.~J.}\
  \bibnamefont {Foster}}, \bibinfo {author} {\bibfnamefont {S.}~\bibnamefont
  {Gokhale}}, \bibinfo {author} {\bibfnamefont {J.}~\bibnamefont {Dunkel}},\
  and\ \bibinfo {author} {\bibfnamefont {N.}~\bibnamefont {Fakhri}},\
  }\bibfield  {title} {\bibinfo {title} {Odd dynamics of living chiral
  crystals},\ }\href@noop {} {\bibfield  {journal} {\bibinfo  {journal}
  {Nature}\ }\textbf {\bibinfo {volume} {607}},\ \bibinfo {pages} {287}
  (\bibinfo {year} {2022})}\BibitemShut {NoStop}%
\bibitem [{\citenamefont {Chao}\ \emph {et~al.}(2026)\citenamefont {Chao},
  \citenamefont {Gokhale}, \citenamefont {Lin}, \citenamefont {Hastewell},
  \citenamefont {Bacanu}, \citenamefont {Chen}, \citenamefont {Li},
  \citenamefont {Liu}, \citenamefont {Lee}, \citenamefont {Dunkel},\ and\
  \citenamefont {Fakhri}}]{chao2026selective}%
  \BibitemOpen
  \bibfield  {author} {\bibinfo {author} {\bibfnamefont {Y.-C.}\ \bibnamefont
  {Chao}}, \bibinfo {author} {\bibfnamefont {S.}~\bibnamefont {Gokhale}},
  \bibinfo {author} {\bibfnamefont {L.}~\bibnamefont {Lin}}, \bibinfo {author}
  {\bibfnamefont {A.}~\bibnamefont {Hastewell}}, \bibinfo {author}
  {\bibfnamefont {A.}~\bibnamefont {Bacanu}}, \bibinfo {author} {\bibfnamefont
  {Y.}~\bibnamefont {Chen}}, \bibinfo {author} {\bibfnamefont {J.}~\bibnamefont
  {Li}}, \bibinfo {author} {\bibfnamefont {J.}~\bibnamefont {Liu}}, \bibinfo
  {author} {\bibfnamefont {H.}~\bibnamefont {Lee}}, \bibinfo {author}
  {\bibfnamefont {J.}~\bibnamefont {Dunkel}},\ and\ \bibinfo {author}
  {\bibfnamefont {N.}~\bibnamefont {Fakhri}},\ }\bibfield  {title} {\bibinfo
  {title} {Selective excitation of work-generating cycles in non-reciprocal
  living solids},\ }\href@noop {} {\bibfield  {journal} {\bibinfo  {journal}
  {Nature Physics}\ }\textbf {\bibinfo {volume} {22}},\ \bibinfo {pages} {474}
  (\bibinfo {year} {2026})}\BibitemShut {NoStop}%
\bibitem [{sup()}]{suppinfo}%
  \BibitemOpen
  \href@noop {} {}\bibinfo {note} {See Supplemental Information for details of
  experimental method, interaction inference, model analysis, topological
  framework, and additional analyses}\BibitemShut {NoStop}%
\bibitem [{\citenamefont {Zhang}\ \emph {et~al.}(2024)\citenamefont {Zhang},
  \citenamefont {Fei},\ and\ \citenamefont {Dunkel}}]{zhang2024nonlinear}%
  \BibitemOpen
  \bibfield  {author} {\bibinfo {author} {\bibfnamefont {S.}~\bibnamefont
  {Zhang}}, \bibinfo {author} {\bibfnamefont {C.}~\bibnamefont {Fei}},\ and\
  \bibinfo {author} {\bibfnamefont {J.}~\bibnamefont {Dunkel}},\ }\bibfield
  {title} {\bibinfo {title} {Nonlinear memory in cell division dynamics across
  species},\ }\href@noop {} {\bibfield  {journal} {\bibinfo  {journal} {arXiv
  preprint arXiv:2408.14564}\ } (\bibinfo {year} {2024})}\BibitemShut {NoStop}%
\bibitem [{\citenamefont {Zuo}\ \emph {et~al.}(2025)\citenamefont {Zuo},
  \citenamefont {Fei}, \citenamefont {Cohen}, \citenamefont {Kim},
  \citenamefont {Carde}, \citenamefont {Dunkel},\ and\ \citenamefont
  {Hu}}]{zuo2025predicting}%
  \BibitemOpen
  \bibfield  {author} {\bibinfo {author} {\bibfnamefont {C.}~\bibnamefont
  {Zuo}}, \bibinfo {author} {\bibfnamefont {C.}~\bibnamefont {Fei}}, \bibinfo
  {author} {\bibfnamefont {A.~E.}\ \bibnamefont {Cohen}}, \bibinfo {author}
  {\bibfnamefont {S.}~\bibnamefont {Kim}}, \bibinfo {author} {\bibfnamefont
  {R.~T.}\ \bibnamefont {Carde}}, \bibinfo {author} {\bibfnamefont
  {J.}~\bibnamefont {Dunkel}},\ and\ \bibinfo {author} {\bibfnamefont {D.~L.}\
  \bibnamefont {Hu}},\ }\bibfield  {title} {\bibinfo {title} {Predicting
  mosquito flight behavior using bayesian dynamical systems learning},\
  }\href@noop {} {\bibfield  {journal} {\bibinfo  {journal} {arXiv preprint
  arXiv:2505.13615}\ } (\bibinfo {year} {2025})}\BibitemShut {NoStop}%
\bibitem [{\citenamefont {Skinner}\ \emph {et~al.}(2023)\citenamefont
  {Skinner}, \citenamefont {Jeckel}, \citenamefont {Martin}, \citenamefont
  {Drescher},\ and\ \citenamefont {Dunkel}}]{skinner2023topological}%
  \BibitemOpen
  \bibfield  {author} {\bibinfo {author} {\bibfnamefont {D.~J.}\ \bibnamefont
  {Skinner}}, \bibinfo {author} {\bibfnamefont {H.}~\bibnamefont {Jeckel}},
  \bibinfo {author} {\bibfnamefont {A.~C.}\ \bibnamefont {Martin}}, \bibinfo
  {author} {\bibfnamefont {K.}~\bibnamefont {Drescher}},\ and\ \bibinfo
  {author} {\bibfnamefont {J.}~\bibnamefont {Dunkel}},\ }\bibfield  {title}
  {\bibinfo {title} {Topological packing statistics of living and nonliving
  matter},\ }\href@noop {} {\bibfield  {journal} {\bibinfo  {journal} {Science
  Advances}\ }\textbf {\bibinfo {volume} {9}},\ \bibinfo {pages} {eadg1261}
  (\bibinfo {year} {2023})}\BibitemShut {NoStop}%
\bibitem [{\citenamefont {Skinner}\ \emph {et~al.}(2021)\citenamefont
  {Skinner}, \citenamefont {Song}, \citenamefont {Jeckel}, \citenamefont
  {Jelli}, \citenamefont {Drescher},\ and\ \citenamefont
  {Dunkel}}]{skinner2021topological}%
  \BibitemOpen
  \bibfield  {author} {\bibinfo {author} {\bibfnamefont {D.~J.}\ \bibnamefont
  {Skinner}}, \bibinfo {author} {\bibfnamefont {B.}~\bibnamefont {Song}},
  \bibinfo {author} {\bibfnamefont {H.}~\bibnamefont {Jeckel}}, \bibinfo
  {author} {\bibfnamefont {E.}~\bibnamefont {Jelli}}, \bibinfo {author}
  {\bibfnamefont {K.}~\bibnamefont {Drescher}},\ and\ \bibinfo {author}
  {\bibfnamefont {J.}~\bibnamefont {Dunkel}},\ }\bibfield  {title} {\bibinfo
  {title} {Topological metric detects hidden order in disordered media},\
  }\href@noop {} {\bibfield  {journal} {\bibinfo  {journal} {Physical Review
  Letters}\ }\textbf {\bibinfo {volume} {126}},\ \bibinfo {pages} {048101}
  (\bibinfo {year} {2021})}\BibitemShut {NoStop}%
\bibitem [{\citenamefont {Harrington}\ \emph {et~al.}(2022)\citenamefont
  {Harrington}, \citenamefont {Mueller},\ and\ \citenamefont
  {Murch}}]{harrington2022engineered}%
  \BibitemOpen
  \bibfield  {author} {\bibinfo {author} {\bibfnamefont {P.~M.}\ \bibnamefont
  {Harrington}}, \bibinfo {author} {\bibfnamefont {E.~J.}\ \bibnamefont
  {Mueller}},\ and\ \bibinfo {author} {\bibfnamefont {K.~W.}\ \bibnamefont
  {Murch}},\ }\bibfield  {title} {\bibinfo {title} {Engineered dissipation for
  quantum information science},\ }\href@noop {} {\bibfield  {journal} {\bibinfo
   {journal} {Nature Reviews Physics}\ }\textbf {\bibinfo {volume} {4}},\
  \bibinfo {pages} {660} (\bibinfo {year} {2022})}\BibitemShut {NoStop}%
\bibitem [{\citenamefont {Lake}\ and\ \citenamefont
  {Ro}(2025)}]{lake2025squeezing}%
  \BibitemOpen
  \bibfield  {author} {\bibinfo {author} {\bibfnamefont {E.}~\bibnamefont
  {Lake}}\ and\ \bibinfo {author} {\bibfnamefont {S.}~\bibnamefont {Ro}},\
  }\bibfield  {title} {\bibinfo {title} {Squeezing codes: robust
  fluctuation-stabilized memories},\ }\href@noop {} {\bibfield  {journal}
  {\bibinfo  {journal} {arXiv preprint arXiv:2509.20730}\ } (\bibinfo {year}
  {2025})}\BibitemShut {NoStop}%
\bibitem [{\citenamefont {Pajouheshgar}\ \emph {et~al.}(2026)\citenamefont
  {Pajouheshgar}, \citenamefont {Bhardwaj}, \citenamefont {Selub},\ and\
  \citenamefont {Lake}}]{pajouheshgar2026exploring}%
  \BibitemOpen
  \bibfield  {author} {\bibinfo {author} {\bibfnamefont {E.}~\bibnamefont
  {Pajouheshgar}}, \bibinfo {author} {\bibfnamefont {A.}~\bibnamefont
  {Bhardwaj}}, \bibinfo {author} {\bibfnamefont {N.}~\bibnamefont {Selub}},\
  and\ \bibinfo {author} {\bibfnamefont {E.}~\bibnamefont {Lake}},\ }\bibfield
  {title} {\bibinfo {title} {Exploring the landscape of nonequilibrium memories
  with neural cellular automata},\ }\href@noop {} {\bibfield  {journal}
  {\bibinfo  {journal} {Physical Review Letters}\ }\textbf {\bibinfo {volume}
  {136}},\ \bibinfo {pages} {037102} (\bibinfo {year} {2026})}\BibitemShut
  {NoStop}%
\bibitem [{\citenamefont {Stringer}\ \emph {et~al.}(2021)\citenamefont
  {Stringer}, \citenamefont {Wang}, \citenamefont {Michaelos},\ and\
  \citenamefont {Pachitariu}}]{stringer2021cellpose}%
  \BibitemOpen
  \bibfield  {author} {\bibinfo {author} {\bibfnamefont {C.}~\bibnamefont
  {Stringer}}, \bibinfo {author} {\bibfnamefont {T.}~\bibnamefont {Wang}},
  \bibinfo {author} {\bibfnamefont {M.}~\bibnamefont {Michaelos}},\ and\
  \bibinfo {author} {\bibfnamefont {M.}~\bibnamefont {Pachitariu}},\ }\bibfield
   {title} {\bibinfo {title} {Cellpose: a generalist algorithm for cellular
  segmentation},\ }\href {https://doi.org/10.1038/s41592-020-01018-x}
  {\bibfield  {journal} {\bibinfo  {journal} {Nature Methods}\ }\textbf
  {\bibinfo {volume} {18}},\ \bibinfo {pages} {100} (\bibinfo {year}
  {2021})}\BibitemShut {NoStop}%
\bibitem [{\citenamefont {Pachitariu}\ and\ \citenamefont
  {Stringer}(2022)}]{pachitariu2022cellpose}%
  \BibitemOpen
  \bibfield  {author} {\bibinfo {author} {\bibfnamefont {M.}~\bibnamefont
  {Pachitariu}}\ and\ \bibinfo {author} {\bibfnamefont {C.}~\bibnamefont
  {Stringer}},\ }\bibfield  {title} {\bibinfo {title} {Cellpose 2.0: how to
  train your own model},\ }\href {https://doi.org/10.1038/s41592-022-01663-4}
  {\bibfield  {journal} {\bibinfo  {journal} {Nature Methods}\ }\textbf
  {\bibinfo {volume} {18}},\ \bibinfo {pages} {1634} (\bibinfo {year}
  {2022})}\BibitemShut {NoStop}%
\bibitem [{\citenamefont {Allan}\ \emph {et~al.}(2024)\citenamefont {Allan},
  \citenamefont {Caswell}, \citenamefont {Keim}, \citenamefont {van~der Wel},\
  and\ \citenamefont {Verweij}}]{allan2024trackpy}%
  \BibitemOpen
  \bibfield  {author} {\bibinfo {author} {\bibfnamefont {D.~B.}\ \bibnamefont
  {Allan}}, \bibinfo {author} {\bibfnamefont {T.}~\bibnamefont {Caswell}},
  \bibinfo {author} {\bibfnamefont {N.~C.}\ \bibnamefont {Keim}}, \bibinfo
  {author} {\bibfnamefont {C.~M.}\ \bibnamefont {van~der Wel}},\ and\ \bibinfo
  {author} {\bibfnamefont {R.~W.}\ \bibnamefont {Verweij}},\ }\href
  {https://doi.org/10.5281/zenodo.12708864} {\bibinfo {title}
  {soft-matter/trackpy: v0.6.4}} (\bibinfo {year} {2024})\BibitemShut {NoStop}%
\bibitem [{\citenamefont {Crocker}\ and\ \citenamefont
  {Grier}(1996)}]{crocker1996methods}%
  \BibitemOpen
  \bibfield  {author} {\bibinfo {author} {\bibfnamefont {J.~C.}\ \bibnamefont
  {Crocker}}\ and\ \bibinfo {author} {\bibfnamefont {D.~G.}\ \bibnamefont
  {Grier}},\ }\bibfield  {title} {\bibinfo {title} {Methods of digital video
  microscopy for colloidal studies},\ }\href
  {https://doi.org/10.1006/jcis.1996.0217} {\bibfield  {journal} {\bibinfo
  {journal} {Journal of Colloid and Interface Science}\ }\textbf {\bibinfo
  {volume} {179}},\ \bibinfo {pages} {298} (\bibinfo {year}
  {1996})}\BibitemShut {NoStop}%
\bibitem [{\citenamefont {Squires}(2001)}]{squires_effective_2001}%
  \BibitemOpen
  \bibfield  {author} {\bibinfo {author} {\bibfnamefont {T.~M.}\ \bibnamefont
  {Squires}},\ }\bibfield  {title} {\bibinfo {title} {Effective
  pseudo-potentials of hydrodynamic origin},\ }\href
  {https://doi.org/10.1017/S0022112001005432} {\bibfield  {journal} {\bibinfo
  {journal} {Journal of Fluid Mechanics}\ }\textbf {\bibinfo {volume} {443}},\
  \bibinfo {pages} {403} (\bibinfo {year} {2001})}\BibitemShut {NoStop}%
\bibitem [{\citenamefont {Blake}\ and\ \citenamefont
  {Chwang}(1974)}]{blake_fundamental_1974}%
  \BibitemOpen
  \bibfield  {author} {\bibinfo {author} {\bibfnamefont {J.~R.}\ \bibnamefont
  {Blake}}\ and\ \bibinfo {author} {\bibfnamefont {A.~T.}\ \bibnamefont
  {Chwang}},\ }\bibfield  {title} {\bibinfo {title} {Fundamental singularities
  of viscous flow: {Part} {I}: {The} image systems in the vicinity of a
  stationary no-slip boundary},\ }\href {https://doi.org/10.1007/BF02353701}
  {\bibfield  {journal} {\bibinfo  {journal} {Journal of Engineering
  Mathematics}\ }\textbf {\bibinfo {volume} {8}},\ \bibinfo {pages} {23}
  (\bibinfo {year} {1974})}\BibitemShut {NoStop}%
\bibitem [{\citenamefont {Short}\ \emph {et~al.}(2006)\citenamefont {Short},
  \citenamefont {Solari}, \citenamefont {Ganguly}, \citenamefont {Powers},
  \citenamefont {Kessler},\ and\ \citenamefont {Goldstein}}]{short_flows_2006}%
  \BibitemOpen
  \bibfield  {author} {\bibinfo {author} {\bibfnamefont {M.~B.}\ \bibnamefont
  {Short}}, \bibinfo {author} {\bibfnamefont {C.~A.}\ \bibnamefont {Solari}},
  \bibinfo {author} {\bibfnamefont {S.}~\bibnamefont {Ganguly}}, \bibinfo
  {author} {\bibfnamefont {T.~R.}\ \bibnamefont {Powers}}, \bibinfo {author}
  {\bibfnamefont {J.~O.}\ \bibnamefont {Kessler}},\ and\ \bibinfo {author}
  {\bibfnamefont {R.~E.}\ \bibnamefont {Goldstein}},\ }\bibfield  {title}
  {\bibinfo {title} {Flows driven by flagella of multicellular organisms
  enhance long-range molecular transport},\ }\href@noop {} {\bibfield
  {journal} {\bibinfo  {journal} {Proceedings of the National Academy of
  Sciences}\ }\textbf {\bibinfo {volume} {103}},\ \bibinfo {pages} {8315}
  (\bibinfo {year} {2006})}\BibitemShut {NoStop}%
\bibitem [{\citenamefont {van~der Walt}\ \emph {et~al.}(2014)\citenamefont
  {van~der Walt}, \citenamefont {Sch{\"o}nberger}, \citenamefont
  {Nunez-Iglesias}, \citenamefont {Boulogne}, \citenamefont {Warner},
  \citenamefont {Yager}, \citenamefont {Guillart}, \citenamefont {Yu},\ and\
  \citenamefont {the scikit-image contributers}}]{vanderwalt2014skimage}%
  \BibitemOpen
  \bibfield  {author} {\bibinfo {author} {\bibfnamefont {S.}~\bibnamefont
  {van~der Walt}}, \bibinfo {author} {\bibfnamefont {J.~L.}\ \bibnamefont
  {Sch{\"o}nberger}}, \bibinfo {author} {\bibfnamefont {J.}~\bibnamefont
  {Nunez-Iglesias}}, \bibinfo {author} {\bibfnamefont {F.}~\bibnamefont
  {Boulogne}}, \bibinfo {author} {\bibfnamefont {J.~D.}\ \bibnamefont
  {Warner}}, \bibinfo {author} {\bibfnamefont {N.}~\bibnamefont {Yager}},
  \bibinfo {author} {\bibfnamefont {E.}~\bibnamefont {Guillart}}, \bibinfo
  {author} {\bibfnamefont {T.}~\bibnamefont {Yu}},\ and\ \bibinfo {author}
  {\bibnamefont {the scikit-image contributers}},\ }\bibfield  {title}
  {\bibinfo {title} {{S}cikit-image: Image processing in python},\ }\href
  {https://doi.org/10.7717/peerj.453} {\bibfield  {journal} {\bibinfo
  {journal} {PeerJ}\ }\textbf {\bibinfo {volume} {2}},\ \bibinfo {pages} {e453}
  (\bibinfo {year} {2014})}\BibitemShut {NoStop}%
\bibitem [{\citenamefont {Sternberg}(1983)}]{sternberg1983biomedical}%
  \BibitemOpen
  \bibfield  {author} {\bibinfo {author} {\bibfnamefont {S.~R.}\ \bibnamefont
  {Sternberg}},\ }\bibfield  {title} {\bibinfo {title} {Biomedical image
  processing},\ }\href {https://doi.org/10.1109/MC.1983.1654163} {\bibfield
  {journal} {\bibinfo  {journal} {Computer}\ }\textbf {\bibinfo {volume}
  {16}},\ \bibinfo {pages} {22} (\bibinfo {year} {1983})}\BibitemShut {NoStop}%
\bibitem [{\citenamefont {Pizer}\ \emph {et~al.}(1987)\citenamefont {Pizer},
  \citenamefont {Amburn}, \citenamefont {Austin}, \citenamefont {Cromartie},
  \citenamefont {Geselowitz}, \citenamefont {Greer}, \citenamefont {ter
  Haar~Romeny}, \citenamefont {Zimmerman},\ and\ \citenamefont
  {Zuiderveld}}]{pizer1987adaptive}%
  \BibitemOpen
  \bibfield  {author} {\bibinfo {author} {\bibfnamefont {S.~M.}\ \bibnamefont
  {Pizer}}, \bibinfo {author} {\bibfnamefont {E.~P.}\ \bibnamefont {Amburn}},
  \bibinfo {author} {\bibfnamefont {J.~D.}\ \bibnamefont {Austin}}, \bibinfo
  {author} {\bibfnamefont {R.}~\bibnamefont {Cromartie}}, \bibinfo {author}
  {\bibfnamefont {A.}~\bibnamefont {Geselowitz}}, \bibinfo {author}
  {\bibfnamefont {T.}~\bibnamefont {Greer}}, \bibinfo {author} {\bibfnamefont
  {B.}~\bibnamefont {ter Haar~Romeny}}, \bibinfo {author} {\bibfnamefont
  {J.~B.}\ \bibnamefont {Zimmerman}},\ and\ \bibinfo {author} {\bibfnamefont
  {K.}~\bibnamefont {Zuiderveld}},\ }\bibfield  {title} {\bibinfo {title}
  {Adaptive histogram equalization and its variations},\ }\href
  {https://doi.org/10.1016/S0734-189X(87)80186-X} {\bibfield  {journal}
  {\bibinfo  {journal} {Computer Vision, Graphics, and Image Processing}\
  }\textbf {\bibinfo {volume} {39}},\ \bibinfo {pages} {355} (\bibinfo {year}
  {1987})}\BibitemShut {NoStop}%
\bibitem [{\citenamefont {Breiman}(1996)}]{breiman1996bagging}%
  \BibitemOpen
  \bibfield  {author} {\bibinfo {author} {\bibfnamefont {L.}~\bibnamefont
  {Breiman}},\ }\bibfield  {title} {\bibinfo {title} {Bagging predictors},\
  }\href {https://doi.org/10.1007/BF00058655} {\bibfield  {journal} {\bibinfo
  {journal} {Machine Learning}\ }\textbf {\bibinfo {volume} {24}},\ \bibinfo
  {pages} {123} (\bibinfo {year} {1996})}\BibitemShut {NoStop}%
\bibitem [{\citenamefont {Pedregosa}\ \emph {et~al.}(2011)\citenamefont
  {Pedregosa}, \citenamefont {Varoquaux}, \citenamefont {Gramfort},
  \citenamefont {Michel}, \citenamefont {Thirion}, \citenamefont {Grisel},
  \citenamefont {Blondel}, \citenamefont {Prettenhofer}, \citenamefont {Weiss},
  \citenamefont {Dubourg}, \citenamefont {Vanderplas}, \citenamefont {Passos},
  \citenamefont {Cournapeau}, \citenamefont {Brucher}, \citenamefont {Perrot},\
  and\ \citenamefont {Duchesnay}}]{pedregosa2011sklearn}%
  \BibitemOpen
  \bibfield  {author} {\bibinfo {author} {\bibfnamefont {F.}~\bibnamefont
  {Pedregosa}}, \bibinfo {author} {\bibfnamefont {G.}~\bibnamefont
  {Varoquaux}}, \bibinfo {author} {\bibfnamefont {A.}~\bibnamefont {Gramfort}},
  \bibinfo {author} {\bibfnamefont {V.}~\bibnamefont {Michel}}, \bibinfo
  {author} {\bibfnamefont {B.}~\bibnamefont {Thirion}}, \bibinfo {author}
  {\bibfnamefont {O.}~\bibnamefont {Grisel}}, \bibinfo {author} {\bibfnamefont
  {M.}~\bibnamefont {Blondel}}, \bibinfo {author} {\bibfnamefont
  {P.}~\bibnamefont {Prettenhofer}}, \bibinfo {author} {\bibfnamefont
  {R.}~\bibnamefont {Weiss}}, \bibinfo {author} {\bibfnamefont
  {V.}~\bibnamefont {Dubourg}}, \bibinfo {author} {\bibfnamefont
  {J.}~\bibnamefont {Vanderplas}}, \bibinfo {author} {\bibfnamefont
  {A.}~\bibnamefont {Passos}}, \bibinfo {author} {\bibfnamefont
  {D.}~\bibnamefont {Cournapeau}}, \bibinfo {author} {\bibfnamefont
  {M.}~\bibnamefont {Brucher}}, \bibinfo {author} {\bibfnamefont
  {M.}~\bibnamefont {Perrot}},\ and\ \bibinfo {author} {\bibfnamefont
  {E.}~\bibnamefont {Duchesnay}},\ }\bibfield  {title} {\bibinfo {title}
  {Scikit-learn: Machine learning in {P}ython},\ }\href
  {https://doi.org/10.48550/arXiv.1201.0490} {\bibfield  {journal} {\bibinfo
  {journal} {J. Mach. Learn. Res.}\ }\textbf {\bibinfo {volume} {12}},\
  \bibinfo {pages} {2825} (\bibinfo {year} {2011})}\BibitemShut {NoStop}%
\bibitem [{\citenamefont {Breiman}\ \emph {et~al.}(1984)\citenamefont
  {Breiman}, \citenamefont {Friedman}, \citenamefont {Olshen},\ and\
  \citenamefont {Stone}}]{breiman1984classification}%
  \BibitemOpen
  \bibfield  {author} {\bibinfo {author} {\bibfnamefont {L.}~\bibnamefont
  {Breiman}}, \bibinfo {author} {\bibfnamefont {J.}~\bibnamefont {Friedman}},
  \bibinfo {author} {\bibfnamefont {R.~A.}\ \bibnamefont {Olshen}},\ and\
  \bibinfo {author} {\bibfnamefont {C.~J.}\ \bibnamefont {Stone}},\ }\href
  {https://doi.org/10.1201/9781315139470} {\emph {\bibinfo {title}
  {Classification and Regression Trees}}}\ (\bibinfo  {publisher} {Taylor \&
  Francis},\ \bibinfo {address} {New York, NY, USA},\ \bibinfo {year}
  {1984})\BibitemShut {NoStop}%
\bibitem [{\citenamefont {Simpson}(1949)}]{simpson1949measurement}%
  \BibitemOpen
  \bibfield  {author} {\bibinfo {author} {\bibfnamefont {E.~H.}\ \bibnamefont
  {Simpson}},\ }\bibfield  {title} {\bibinfo {title} {Measurement of
  diversity},\ }\bibfield  {journal} {\bibinfo  {journal} {Nature}\ }\textbf
  {\bibinfo {volume} {163}},\ \href {https://doi.org/10.1038/163688a0}
  {10.1038/163688a0} (\bibinfo {year} {1949})\BibitemShut {NoStop}%
\bibitem [{\citenamefont {Clopper}\ and\ \citenamefont
  {Pearson}(1934)}]{clopper1934confidence}%
  \BibitemOpen
  \bibfield  {author} {\bibinfo {author} {\bibfnamefont {C.~J.}\ \bibnamefont
  {Clopper}}\ and\ \bibinfo {author} {\bibfnamefont {E.~S.}\ \bibnamefont
  {Pearson}},\ }\bibfield  {title} {\bibinfo {title} {On the use of confidence
  or fiducial limits illustrated in the case of the binomial},\ }\href
  {https://doi.org/10.1093/biomet/26.4.404} {\bibfield  {journal} {\bibinfo
  {journal} {Biometrika}\ }\textbf {\bibinfo {volume} {26}},\ \bibinfo {pages}
  {404} (\bibinfo {year} {1934})}\BibitemShut {NoStop}%
\bibitem [{\citenamefont {Tipping}(2001)}]{tipping2001sparse}%
  \BibitemOpen
  \bibfield  {author} {\bibinfo {author} {\bibfnamefont {M.~E.}\ \bibnamefont
  {Tipping}},\ }\bibfield  {title} {\bibinfo {title} {Sparse {Bayesian}
  learning and the relevance vector machine},\ }\href@noop {} {\bibfield
  {journal} {\bibinfo  {journal} {J. Mach. Learn. Res.}\ }\textbf {\bibinfo
  {volume} {1}},\ \bibinfo {pages} {211} (\bibinfo {year} {2001})}\BibitemShut
  {NoStop}%
\bibitem [{\citenamefont {Wipf}\ and\ \citenamefont
  {Rao}(2004)}]{wipf2004sparse}%
  \BibitemOpen
  \bibfield  {author} {\bibinfo {author} {\bibfnamefont {D.~P.}\ \bibnamefont
  {Wipf}}\ and\ \bibinfo {author} {\bibfnamefont {B.~D.}\ \bibnamefont {Rao}},\
  }\bibfield  {title} {\bibinfo {title} {Sparse {B}ayesian learning for basis
  selection},\ }\href@noop {} {\bibfield  {journal} {\bibinfo  {journal} {IEEE
  Trans. Signal Process.}\ }\textbf {\bibinfo {volume} {52}},\ \bibinfo {pages}
  {2153} (\bibinfo {year} {2004})}\BibitemShut {NoStop}%
\bibitem [{\citenamefont {Krogh}\ and\ \citenamefont
  {Hertz}(1991)}]{krogh1991simple}%
  \BibitemOpen
  \bibfield  {author} {\bibinfo {author} {\bibfnamefont {A.}~\bibnamefont
  {Krogh}}\ and\ \bibinfo {author} {\bibfnamefont {J.}~\bibnamefont {Hertz}},\
  }\bibfield  {title} {\bibinfo {title} {{A Simple Weight Decay Can Improve
  Generalization}},\ }in\ \href
  {https://proceedings.neurips.cc/paper_files/paper/1991/file/8eefcfdf5990e441f0fb6f3fad709e21-Paper.pdf}
  {\emph {\bibinfo {booktitle} {Advances in Neural Information Processing
  Systems}}},\ Vol.~\bibinfo {volume} {4},\ \bibinfo {editor} {edited by\
  \bibinfo {editor} {\bibfnamefont {J.}~\bibnamefont {Moody}}, \bibinfo
  {editor} {\bibfnamefont {S.}~\bibnamefont {Hanson}},\ and\ \bibinfo {editor}
  {\bibfnamefont {R.}~\bibnamefont {Lippmann}}}\ (\bibinfo  {publisher}
  {Morgan-Kaufmann},\ \bibinfo {address} {Burlington, MA},\ \bibinfo {year}
  {1991})\BibitemShut {NoStop}%
\bibitem [{\citenamefont {MacKay}(1992{\natexlab{a}})}]{mackay1992practical}%
  \BibitemOpen
  \bibfield  {author} {\bibinfo {author} {\bibfnamefont {D.~J.}\ \bibnamefont
  {MacKay}},\ }\bibfield  {title} {\bibinfo {title} {A practical {B}ayesian
  framework for backpropagation networks},\ }\href@noop {} {\bibfield
  {journal} {\bibinfo  {journal} {Neural Comput.}\ }\textbf {\bibinfo {volume}
  {4}},\ \bibinfo {pages} {448} (\bibinfo {year}
  {1992}{\natexlab{a}})}\BibitemShut {NoStop}%
\bibitem [{\citenamefont {MacKay}(1992{\natexlab{b}})}]{mackay1992bayesian}%
  \BibitemOpen
  \bibfield  {author} {\bibinfo {author} {\bibfnamefont {D.~J.}\ \bibnamefont
  {MacKay}},\ }\bibfield  {title} {\bibinfo {title} {Bayesian interpolation},\
  }\href@noop {} {\bibfield  {journal} {\bibinfo  {journal} {Neural Comput.}\
  }\textbf {\bibinfo {volume} {4}},\ \bibinfo {pages} {415} (\bibinfo {year}
  {1992}{\natexlab{b}})}\BibitemShut {NoStop}%
\bibitem [{\citenamefont {Borg}\ and\ \citenamefont
  {Groenen}(2005)}]{borg_modern_2005}%
  \BibitemOpen
  \bibfield  {author} {\bibinfo {author} {\bibfnamefont {I.}~\bibnamefont
  {Borg}}\ and\ \bibinfo {author} {\bibfnamefont {P.~J.~F.}\ \bibnamefont
  {Groenen}},\ }\href@noop {} {\emph {\bibinfo {title} {Modern multidimensional
  scaling: theory and applications}}},\ \bibinfo {edition} {2nd}\ ed.,\
  Springer {Series} in {Statistics}\ (\bibinfo  {publisher} {Springer},\
  \bibinfo {address} {New York},\ \bibinfo {year} {2005})\BibitemShut {NoStop}%
\bibitem [{\citenamefont {Lin}(2024)}]{lin_multidimensional_2024}%
  \BibitemOpen
  \bibfield  {author} {\bibinfo {author} {\bibfnamefont {D.}~\bibnamefont
  {Lin}},\ }\href {https://juliastats.org/MultivariateStats.jl/dev/mds/}
  {\bibinfo {title} {Multidimensional {Scaling}: {MultivariateStats}.jl}}
  (\bibinfo {year} {2024})\BibitemShut {NoStop}%
\bibitem [{\citenamefont {Bowman}\ and\ \citenamefont
  {Azzalini}(1997)}]{bowman1997applied}%
  \BibitemOpen
  \bibfield  {author} {\bibinfo {author} {\bibfnamefont {A.~W.}\ \bibnamefont
  {Bowman}}\ and\ \bibinfo {author} {\bibfnamefont {A.}~\bibnamefont
  {Azzalini}},\ }\href@noop {} {\emph {\bibinfo {title} {Applied smoothing
  techniques for data analysis: the kernel approach with S-Plus
  illustrations}}},\ Vol.~\bibinfo {volume} {18}\ (\bibinfo  {publisher} {OUP
  Oxford},\ \bibinfo {year} {1997})\BibitemShut {NoStop}%
\bibitem [{\citenamefont {Shiraishi}\ \emph {et~al.}(2018)\citenamefont
  {Shiraishi}, \citenamefont {Funo},\ and\ \citenamefont
  {Saito}}]{shiraishi2018speed}%
  \BibitemOpen
  \bibfield  {author} {\bibinfo {author} {\bibfnamefont {N.}~\bibnamefont
  {Shiraishi}}, \bibinfo {author} {\bibfnamefont {K.}~\bibnamefont {Funo}},\
  and\ \bibinfo {author} {\bibfnamefont {K.}~\bibnamefont {Saito}},\ }\bibfield
   {title} {\bibinfo {title} {Speed limit for classical stochastic processes},\
  }\href@noop {} {\bibfield  {journal} {\bibinfo  {journal} {Physical Review
  Letters}\ }\textbf {\bibinfo {volume} {121}},\ \bibinfo {pages} {070601}
  (\bibinfo {year} {2018})}\BibitemShut {NoStop}%
\bibitem [{\citenamefont {Hatano}\ and\ \citenamefont
  {Sasa}(2001)}]{hatano2001steady}%
  \BibitemOpen
  \bibfield  {author} {\bibinfo {author} {\bibfnamefont {T.}~\bibnamefont
  {Hatano}}\ and\ \bibinfo {author} {\bibfnamefont {S.-i.}\ \bibnamefont
  {Sasa}},\ }\bibfield  {title} {\bibinfo {title} {Steady-state thermodynamics
  of langevin systems},\ }\href@noop {} {\bibfield  {journal} {\bibinfo
  {journal} {Physical Review Letters}\ }\textbf {\bibinfo {volume} {86}},\
  \bibinfo {pages} {3463} (\bibinfo {year} {2001})}\BibitemShut {NoStop}%
\bibitem [{\citenamefont {Crooks}(2007)}]{crooks2007measuring}%
  \BibitemOpen
  \bibfield  {author} {\bibinfo {author} {\bibfnamefont {G.~E.}\ \bibnamefont
  {Crooks}},\ }\bibfield  {title} {\bibinfo {title} {Measuring thermodynamic
  length},\ }\href@noop {} {\bibfield  {journal} {\bibinfo  {journal} {Physical
  Review Letters}\ }\textbf {\bibinfo {volume} {99}},\ \bibinfo {pages}
  {100602} (\bibinfo {year} {2007})}\BibitemShut {NoStop}%
\bibitem [{\citenamefont {Kim}(2021)}]{kim2021information}%
  \BibitemOpen
  \bibfield  {author} {\bibinfo {author} {\bibfnamefont {E.-j.}\ \bibnamefont
  {Kim}},\ }\bibfield  {title} {\bibinfo {title} {Information geometry,
  fluctuations, non-equilibrium thermodynamics, and geodesics in complex
  systems},\ }\href@noop {} {\bibfield  {journal} {\bibinfo  {journal}
  {Entropy}\ }\textbf {\bibinfo {volume} {23}},\ \bibinfo {pages} {1393}
  (\bibinfo {year} {2021})}\BibitemShut {NoStop}%
\bibitem [{\citenamefont {Lee}\ \emph {et~al.}(2026)\citenamefont {Lee},
  \citenamefont {Koskelo}, \citenamefont {Gokhale}, \citenamefont {Li},
  \citenamefont {Fei}, \citenamefont {Liu}, \citenamefont {Lin}, \citenamefont
  {Dunkel}, \citenamefont {Skinner},\ and\ \citenamefont
  {Fakhri}}]{lee_2026_20492239}%
  \BibitemOpen
  \bibfield  {author} {\bibinfo {author} {\bibfnamefont {H.}~\bibnamefont
  {Lee}}, \bibinfo {author} {\bibfnamefont {E.}~\bibnamefont {Koskelo}},
  \bibinfo {author} {\bibfnamefont {S.}~\bibnamefont {Gokhale}}, \bibinfo
  {author} {\bibfnamefont {J.}~\bibnamefont {Li}}, \bibinfo {author}
  {\bibfnamefont {C.}~\bibnamefont {Fei}}, \bibinfo {author} {\bibfnamefont
  {C.-W.~J.}\ \bibnamefont {Liu}}, \bibinfo {author} {\bibfnamefont
  {L.}~\bibnamefont {Lin}}, \bibinfo {author} {\bibfnamefont {J.}~\bibnamefont
  {Dunkel}}, \bibinfo {author} {\bibfnamefont {D.}~\bibnamefont {Skinner}},\
  and\ \bibinfo {author} {\bibfnamefont {N.}~\bibnamefont {Fakhri}},\
  }\bibfield  {title} {\bibinfo {title} {{Dataset and codes for manuscript
  ``Topological flowscape reveals state transitions in nonreciprocal living
  matter'' }},\ }\href {https://doi.org/10.5281/zenodo.20492239}
  {10.5281/zenodo.20492239} (\bibinfo {year} {2026})\BibitemShut {NoStop}%
\end{thebibliography}%

\end{document}

\title{Supplementary Information: \\ Topological flowscape reveals state transitions in nonreciprocal living matter}% Force line breaks with \

%%%%%% Author list and affiliations %%%%%%%
%\normalsize
\author{Hyunseok~Lee$^{1,6}$}
\author{EliseAnne~Koskelo$^{1,2,6}$}
\author{Shreyas~Gokhale$^{1,6}$}
\author{Junang~Li$^{3,6}$}
\author{Chenyi~Fei$^4$}
\author{Chih-Wei Joshua~Liu$^1$}
\author{Lisa~Lin$^1$}
\author{J\"{o}rn~Dunkel$^4$}
\author{Dominic~J~Skinner$^5$}
\author{Nikta~Fakhri$^{1}$}
\email{Corresponding author: fakhri@mit.edu}

\affiliation{\vspace{0.25cm}\\ $^1$ Department of Physics, Massachusetts Institute of Technology, Cambridge, MA, USA \\ $^2$ Department of Physics, Harvard University, Cambridge, MA, USA \\ $^3$ Center for the Physics of Biological Function, Princeton University, Princeton, NJ, USA \\ $^4$ Department of Mathematics, Massachusetts Institute of Technology, Cambridge, MA, USA \\ $^5$ Center for Computational Biology, Flatiron Institute, New York, NY, USA \\ $^6$ These authors contributed equally and are joint first authors.}

\maketitle

\vspace{0cm}
\tableofcontents
\clearpage
%%%%%%%%%% Main part of supplementary information %%%%%%%%%%

\section{Experimental methods and image analysis} \label{sec1}

\subsection{Preparation of starfish embryos} \label{sec1:prep}
Starfish (\textit{Patiria miniata}) were obtained from Marinus Scientific and maintained in a saltwater aquarium at 15$^{\circ}$C. They were fed shrimp biweekly. Prior to fertilization, we extracted gonads (oocytes and sperm for female and male starfish, respectively) via small incisions on the underside of the adult starfish near the stomach. Spermatophores were stored in Eppendorf tubes at 4$^{\circ}$C before use. Ovaries were dissected with scissors to release oocytes, which were then washed once in calcium-free seawater to prevent premature maturation, followed by two rinses in filtered seawater (FSW). The cleaned oocytes were plated as monolayers into multiple sterilized 6-well culture plates (VWR, Catalog Number 10861-554) filled with FSW.
\\

For fertilization, 10mM 1-methyladenine (1-MA) was added to the oocyte cultures at a 1:1000 ratio with FSW. Most oocytes exhibited nuclear envelope breakdown within one hour, which was verified using a dissection microscope (Nikon SMZ745T). Within 2 hours of 1-MA addition, we introduced sperm to the oocyte cultures at an approximate ratio of 10 sperm per oocyte. Fertilized embryos were incubated at 15$^{\circ}$C for 24 hours before being transferred to room temperature (approx. 20$^{\circ}$C) for subsequent experiments.
\\

After 24 hours, successfully developed embryos began to swim and accumulated at the air-water interface. Only these embryos were collected and transferred into individual wells of a sterilized 24-well tissue culture plate (VWR, Catalog Number 10861-558; single well diameter: 15.7 mm). Each well contained about 2,000--3,000 embryos in 2~mL of FSW. 
\\

E1 and E2 embryos were both derived from the same starfish. However, E2 embryos were fertilized 24 hours before the E1 embryos and were kept at 15$^{\circ}$C in the interim.

\subsection{Binary pair experiment} \label{sec1:pair}

To identify pairwise nonreciprocity and the underlying connections to tilt precession dynamics, we performed image analysis of isolated E1-E2 pairs.

\subsubsection{Top-view experiment, image segmentation and tracking} \label{sec1:pair:top}
After the preparation of embryos, 30--50 embryos were transferred into each well of the 24-well plate. Once the microscope was focused on a field of view containing a pair of E1 and E2 embryos, time-lapse videos were captured at a frame rate of 10 frames per second and 1.25X magnification using the dissection scope (Nikon SMZ745T) (SI Video 1).
\\

We used \texttt{Cellpose} \cite{stringer2021cellpose, pachitariu2022cellpose}, a deep learning-based segmentation tool, to identify and segment the pair of starfish embryos~(E1 and E2) in each top-view image, applying the ``cyto3'' pre-trained model. To track the embryos over time, we utilized \texttt{Trackpy} \cite{allan2024trackpy, crocker1996methods}, adjusting the parameters to reconstruct embryo trajectories from many-embryo snapshots. Key parameters included the search range (maximum distance an object could move between frames) and the memory parameter (maximum number of consecutive missing frames allowed for linking trajectories).
\\

Gaps in the tracked embryo trajectories were filled using trajectory interpolation with the MATLAB function \texttt{interp1} and the \texttt{pchip} option, which applies shape-preserving cubic spline interpolation. If the number of missing frames at the endpoints was less than the memory parameter, extrapolation was performed to extend the trajectories.
\\

To estimate the tilt direction of embryos from top-view images, we identified the location of blastopores with the MATLAB function \texttt{circlefind}. When the AP axis is tilted, the top-view centroid position of the embryo deviates from the position of blastopore, and we use the direction from the blastopore to the centroid as a proxy for direction of the AP axis~(Fig. \ref{figSI:EmbryoNR}a). 

\subsubsection{Side-view experiment, image segmentation and tilt quantification}\label{sec1:pair:side}
To capture side-view images of starfish embryos, we pipetted the swimming embryos into a 25 mL tissue culture flask. The flask was positioned between the objective and the light source of a dissection microscope placed on its side. The images were taken at 10 frames per second and at 4X magnification (SI Video 2). 
\\

As in the top-view experiments, we used \texttt{Cellpose} with the ``cyto3'' pre-trained model. Segmentation fitted each embryo's side-view snapshot to an ellipsoid, and we used the direction of the major axis to estimate the tilt of the AP axis from an upright position. 

\subsubsection{Asymmetric precession underlies nonreciprocity between E1 and E2}\label{sec1:pair:nr}

We find that starfish embryos at different developmental stages, E1 and E2, exhibit nonreciprocal interactions through a run-and-chase dynamic, where E1-E2 pairs drift toward the E2 embryo. This is perhaps expected, as inter-embryo interactions are mediated by self-generated hydrodynamic flows that evolve with developmental stage~\cite{tan2022odd}. Nonetheless, identifying a clear mechanistic basis for E1-E2 nonreciprocity can provide insight into the underlying physics and inform the design of synthetic nonreciprocal systems. To this end, we propose a mechanism for nonreciprocity that emphasizes the inherently three-dimensional nature of the embryo system.
\\

In water, starfish embryos self-propel along their anterior-posterior (AP) axis due to their ciliary beating, while simultaneously rotating clockwise around this axis. At the air-water interface, their upward self-propelling force counteracts gravity, allowing the embryos to remain bounded at the surface. In a symmetric setup, the AP axis would be oriented upright, perpendicular to the interface. However, in E1--E2 pairs, this symmetry can be broken, leading to a biased tilt of the AP axis.
\\

\begin{figure}[!ht]
  \centering
  \small
\includegraphics[width=\linewidth]{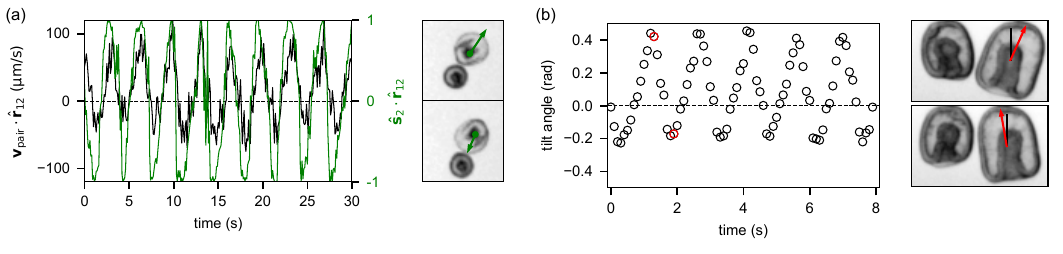}
    \caption{\textbf{Asymmetric tilt of E2's AP axis contributes to nonreciprocity between E1 and E2.} \panel{A} Left: Time series of the drift speed toward E2 (black) and E2's morphology polarization $\mathbf{\hat{s}}$'s alignment with $\mathbf{\hat{r}}_{12}\equiv \frac{\mathbf{r}_2-\mathbf{r}_1}{|\mathbf{r}_2-\mathbf{r}_1|}$ (green). Right: Estimation of $\mathbf{\hat{s}}$ (green arrow) from a displacement from blastopore (black dot) to centroid of E2 embryo images. The drift speed increases when E2 is tilted away from E1, indicating that E2's self-propulsion along its tilted AP axis significantly contributes to the nonreciprocal interaction. \panel{B} E2's tilt is asymmetric; it tilts more strongly when oriented away from E1 than toward it. Note: the embryo pair shown in \textbf{B} is different from that in \textbf{A}, and the oscillation period differs accordingly \cite{tan2022odd}. }
  \label{figSI:EmbryoNR}
\end{figure}

Our experiments show that the observed nonreciprocity between E1 and E2 is consistent with this biased tilt. Specifically, we find that the pairwise nonreciprocity oscillates over time, as indicated by a run-and-chase dynamic, and reaches a maximum when E2 is tilted away from E1 (Fig. \ref{figSI:EmbryoNR}a, SI Video 1), suggesting that E2's self-propulsion along its tilted AP axis influences the pair dynamics. If E2's tilt were symmetric over time, its net contribution to the dynamics would average to zero. However, side-view imaging reveals an asymmetric tilt: E2 tilts more strongly when directed away from E1 (Fig. \ref{figSI:EmbryoNR}b, SI Video 2). Taken together, these results indicate that the asymmetric tilt of E2's AP axis plays a key role in driving the observed nonreciprocal interaction between E1 and E2.

\ {
\subsubsection{{Asymmetric approach dynamics reveal nonreciprocal hydrodynamic interaction between E1 and E2}}\label{sec1:pair:approach}
As shown in Fig. 1b), starfish embryos at different developmental stages, E1 and E2, form a bound pair which drifts toward the E2 embryo. Here, we show additional evidence of an asymmetric interaction when the two embryos are far apart and approaching one other.\\\\
Specifically, Fig. S2 shows the longitudinal velocity of each embryo, $\mathbf{v}_1 \cdot \hat{\mathbf{r}}_{12}$ and  $\mathbf{v}_2 \cdot \hat{\mathbf{r}}_{12}$, when the pairwise distance  $r_{12}$ is between $400\mu m$ and $600\mu m$. This range of pairwise distances, which is 4-6 times the embryo radii, ensures that embryos are separated and that their drift dynamics are driven by the hydrodynamic attraction of the other embryo. If the effective attractive interaction between E1 and E2, mediated by self-generated hydrodynamic flows, is reciprocal, then the two embryos should approach each other with the same speed.
\begin{figure}[!ht]
  \centering
  \small
\includegraphics[width=\linewidth]{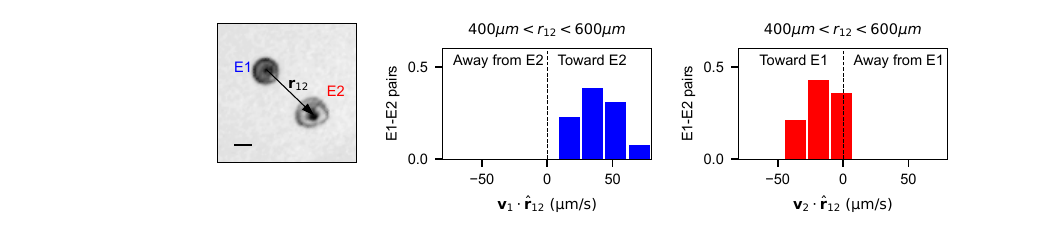}
    \caption{{\textbf{Asymmetric approach dynamics reveal nonreciprocal hydrodynamic interaction between E1 and E2.} Left: A snapshot of E1-E2 embryo pair when they are far apart. Scale bar: $\mathrm{100\ \mu m}$. 
    Center: The average drift speed of E1 along the longitudinal direction, $\mathbf{v}_1 \cdot \hat{\mathbf{r}}_{12}$, when $400 \mu m < r_{12} < 600 \mu m$, is $\mathrm{44 \pm 5\ \mu m/s}$ (SEM, n = 14). Its positive sign signifies that the E1 embryos approach E2 embryos in this regime.  
    Right: The average drift speed of E2 along the longitudinal direction, $\mathbf{v}_2 \cdot \hat{\mathbf{r}}_{12}$, when $400 \mu m < r_{12} < 600 \mu m$, is $\mathrm{-14 \pm 4\ \mu m/s}$ (SEM, n = 14). Its negative sign signifies that the E2 embryos approach E1 embryos in this regime. The magnitude of the speed is much smaller than that of E1, highlighting the nonreciprocity in their hydrodynamic attraction. }}
  \label{figSI:approachNR}
\end{figure}\\
Fig.~\ref{figSI:approachNR} shows that E1 approaches E2 much faster than E2 approaches E1 when they are far apart. This highlights that the effective attractive interaction between the two embryos is nonreciprocal. We note that since their interaction is mediated by self-generated hydrodynamic flows, nonreciprocity at the mesoscopic level is not a surprise. Along with our main systematic evaluation of nonreciprocal interactions based on Bayesian inference of interactions, this additional analysis demonstrates that it is most likely that the E1 and E2 embryos exchange nonreciprocal interactions. 
}

\subsection{\textcolor{black}{Emergence of a traveling velocity from pairwise hydrodynamic interactions of unequally-sized swimmers}}

\textcolor{black}{In this subsection, we consider a simple model of the hydrodynamic flows from different type embryos and its consequences for a bound pair state \cite{drescher2009dancing} at the air-water interface. Specifically, we consider 1) two Stokeslets with different strengths and show how they lead to nonreciprocal far-force fields, and 2) how under force balance, these nonreciprocal force fields give rise to a co-moving speed in the steady state. \\ }

\subsubsection{\textcolor{black}{Two Stokeslets with different strengths: nonreciprocal far-field forces}}

\textcolor{black}{Let us consider two different-type embryos at the air-water interface. The embryos can be approximated as two Stokeslets with different strengths $F_1$ and $F_2$. To stay near the air-water interface (which is essentially a stress-free boundary), the negatively buoyant embryos must push downward on the fluid \cite{tan2022odd}. Therefore, the force exerted on the fluid by each embryo is $-F_i \hat{z}$. We will assume that both Stokeslets originate from spheres with radii $R_1$ and $R_2$, with their tops touching the air-water interface. \\ }

\textcolor{black}{ With this setup, we calculate the flow created by one Stokeslet at the position of the other Stokeslet as a function of their horizontal distance $x$ to assess their interactions as in \cite{drescher2009dancing,squires_effective_2001}. With the stress-free boundary condition, we identify the flow as a sum of contributions from two isolated Stokeslets: a real one and an image on the opposite side of the boundary. The flow from an isolated Stokeslet is \cite{blake_fundamental_1974}: }

\textcolor{black}{
\begin{equation}
\begin{aligned}
u_j = \frac{f_k}{8 \pi \eta}\left( \frac{\delta_{jk}}{r} + \frac{r_j r_k}{r^3} \right),
\end{aligned}
\label{eq:Stokeslet}
\end{equation}}

\textcolor{black}{where $\eta$ is the fluid viscosity, $f_k$ is the strength of the Stokeslet, and $r$ is the distance to the Stokeslet. \\}

\textcolor{black}{ For instance, the flow generated by the real Stokeslet $-F_1\hat{z}$ at $(0,0,-R_1)$, examined at the position of the second embryo $(x, 0, -R_2)$, has the distance vector $\mathbf{r}_{\text{real}} = (x, 0, R_1-R_2)$. The flow is: \\}

\textcolor{black}{
\begin{equation}
\begin{aligned}
& u_x^{\text{real}} = -\frac{F_1}{8 \pi \eta} \frac{x(R_1-R_2)}{ (x^2+(R_1-R_2)^2)^{3/2} } \hat{x},\\
& u_y^{\text{real}} = 0,\\
& u_z^{\text{real}} = -\frac{F_1}{8 \pi \eta}\left( \frac{1}{(x^2+(R_1-R_2)^2)^{1/2}  } + \frac{(R_1-R_2)^2}{ (x^2+(R_1-R_2)^2)^{3/2}} \right)\hat{z}.
\end{aligned}
\label{eq:Stokeslet_real}
\end{equation}
}

\textcolor{black}{ Similarly, the flow generated by the image Stokeslet $+F_1\hat{z}$ at $(0,0,R_1)$, examined at the same position $(x, 0, -R_2)$, has the distance vector $\mathbf{r}_{\text{image}} = (x, 0, -R_1-R_2)$. The flow is: \\}

\textcolor{black}{
\begin{equation}
\begin{aligned}
& u_x^{\text{image}} = -\frac{F_1}{8 \pi \eta} \frac{x(R_1+R_2)}{ (x^2+(R_1+R_2)^2)^{3/2} } \hat{x},\\
& u_y^{\text{image}} = 0,\\
& u_z^{\text{image}} = \frac{F_1}{8 \pi \eta}\left( \frac{1}{(x^2+(R_1+R_2)^2)^{1/2}  } + \frac{(R_1+R_2)^2}{ (x^2+(R_1+R_2)^2)^{3/2}} \right)\hat{z}.
\end{aligned}
\label{eq:Stokeslet_image}
\end{equation}
}

\textcolor{black}{ Combined, the net flow that the second embryo feels at position $(x, 0, -R_2)$ is: \\}

\textcolor{black}{
\begin{equation}
\begin{aligned}
u_x &= -\frac{F_1}{8 \pi \eta} \left( \frac{x(R_1-R_2)}{ (x^2+(R_1-R_2)^2)^{3/2} } + \frac{x(R_1+R_2)}{ (x^2+(R_1+R_2)^2)^{3/2} } \right) \hat{x}.
\end{aligned}
\label{eq:Stokeslet_net_x}
\end{equation}
}

\textcolor{black}{ (We omit the full expression for $u_z$ here, as the vertical flow is largely canceled out by the surface tension of the interface \cite{tan2022odd}).\\ }

\textcolor{black}{ The horizontal flow is in most cases attractive ($u_x < 0$), and guaranteed to be so in the far-field regime, where $x \gg R_1, R_2$. By expanding Equation \ref{eq:Stokeslet_net_x} in this far-field limit, the leading-order term simplifies cleanly and the resulting horizontal flow velocity decays as $1/x^2$: }

 \textcolor{black}{
\begin{equation}
u_x \approx -\frac{F_1}{8 \pi \eta} \frac{2 R_1}{x^2} \hat{x}.
\label{eq:Stokeslet_farfield_x}
\end{equation}
}

\textcolor{black}{ As a result, the two embryos will approach each other to remain force-free in the fluid flow. Defining $U_1$ and $U_2$ as their respective horizontal velocities (with embryo 1 at the origin and embryo 2 at $+x$), they approach at different speeds: }

\textcolor{black}{
\begin{equation}
\begin{aligned}
& U_1 = \frac{F_2 }{8 \pi \eta}\frac{2 R_2}{x^2},\\
& U_2 = -\frac{F_1 }{8 \pi \eta}\frac{2 R_1}{x^2}.
\end{aligned}
\label{eq:speed}
\end{equation}
}

\textcolor{black}{ Because the speeds are not symmetric, the centroid of the system will translate. }

\subsubsection{\textcolor{black}{Force balance: co-moving speed in a steady state}}

\textcolor{black}{The embryo pair will eventually form a bound pair \cite{drescher2009dancing}, exchanging steric repulsion in addition to the hydrodynamic attraction. Since $F_1 \neq F_2$, the reciprocal steric repulsion is unable to satisfy the force-free condition for both embryos individually without a bulk translation. As a result, the bound pair will self-propel with a steady-state speed $U$. \\ }

\textcolor{black}{Assuming the pair moves in the $+x$ direction, the force each embryo feels in the horizontal direction becomes (using Stokes' law for the spherical embryos):}

\textcolor{black}{
\begin{equation}
\begin{aligned}
& f_1 = 6 \pi \eta R_1 \left( U_1 - U \right) - f_{\text{steric(horizontal)}},\\
& f_2 = 6 \pi \eta R_2 \left( U_2 - U \right) + f_{\text{steric(horizontal)}}.
\end{aligned}
\label{eq:force_balance}
\end{equation}}

\textcolor{black}{Under the force-free condition for the entire pair, $f_1 + f_2 = 0$, the internal steric forces cancel out, yielding:}

\textcolor{black}{
\begin{equation}
\begin{aligned}
U = \frac{R_1 U_1 + R_2 U_2}{R_1+R_2} &= \frac{1}{R_1+R_2} \left[ R_1 \left( \frac{F_2}{8\pi\eta} \frac{2 R_2}{x^2} \right) + R_2 \left( -\frac{F_1}{8\pi\eta} \frac{2 R_1}{x^2} \right) \right] \\
&= \frac{2 R_1 R_2 (F_2 - F_1)}{8 \pi \eta (R_1+R_2)}\frac{1}{x^2}.
\end{aligned}
\label{eq:pair_speed}
\end{equation}}

\textcolor{black}{ Note that when $R_1 = R_2$, the pair's self-propulsion speed is zero (if $F_1 = F_2$). Following \cite{short_flows_2006}, we make the assumption that the flagella of the swimming embryo exert a constant force per unit area at the embryo's spherical shell. In this case, the forces are proportional to the square of the embryo sizes, i.e., $F = k R^2$ \cite{drescher2009dancing}. Substituting $F_1 = k R_1^2$ and $F_2 = k R_2^2$ into Eq.~\ref{eq:pair_speed} gives: }

\textcolor{black}{ \begin{equation}
U = \frac{2k R_1 R_2 (R_2^2 - R_1^2)}{8 \pi \eta (R_1+R_2)} \frac{1}{x^2} = \frac{2k R_1 R_2 (R_2 - R_1)}{8 \pi \eta} \frac{1}{x^2}.
\end{equation} }

\textcolor{black}{ Under this assumption, the $(R_1+R_2)$ term in the denominator cancels. The pair speed $U$ scales linearly with the size difference $R_2-R_1$, weighted by the product of their sizes.\\ }

\textcolor{black}{ We note that here we keep the far-field flow for simplification, even though it is an approximation in the regime where the two embryos are in contact. In addition, we currently ignore that the steric repulsions act as external forces that can generate their own Stokeslet fields, whose effects could also be calculated using the method of images. }

\subsection{Binary mixture experiment} \label{sec1:many}

\subsubsection{Mixture experiment, image segmentation and tracking} \label{sec1:many:top}
After the preparation of embryos, 2000--3000 embryos were transferred to a single well of the 24-well plate. The ratio of E1 to E2 embryos was kept approximately equal, with 1,000--1,500 E1 embryos and 1,000--1,500 E2 embryos. The total volume was maintained at 2 mL.
\\

Time-lapse videos were captured at a frame rate of one frame every 5 seconds. Imaging was performed using a dissection microscope (Nikon SMZ745) equipped with a high-speed CMOS digital camera (Amscope MU500) mounted at the eyepiece (SI Video 3).
\\

We used \texttt{Cellpose} \cite{stringer2021cellpose, pachitariu2022cellpose}, a deep learning-based segmentation tool, to identify and segment individual starfish embryos in each top-view image. The ``cyto2'' pre-trained model was applied, with the embryo diameter set as 23~$\mathrm{\mu m}$ to guide the model. 
\\

Unlike the pair experiments where we could easily identify the tracks of E1 and tracks of E2, we needed a new way to classify the embryos based on their development stages. We describe the method in the next section.
\\

\subsubsection{Machine learning classification of E1 and E2 embryos} \label{sec1:many:ml}
As the mixed experiment contains a large number of embryos, we trained a machine learning model to classify embryos as either E1 (24 hours post fertilization) or E2 (48 hours post fertilization).
The model training set initially comprised crops around all embryo positions found by \texttt{Trackpy} \cite{allan2024trackpy, crocker1996methods} in snapshots of a living chiral crystal (LCC) \cite{tan2022odd}. This experiment (LCC) consists of embryos from a single developmental stage that remained in the field of view for a long enough time span including both 24 hours post fertilization and 48 hours post fertilization, which allowed us to label crops of the 24 hours post fertilization snapshot as class E1 and crops of the 48 hours post fertilization snapshot as class E2.
Model test sets initially comprised the sequences of crops around all embryo positions found by \texttt{Cellpose} \cite{stringer2021cellpose, pachitariu2022cellpose} from the mixed experiment.
Each crop sequence contained all observations of a single embryo.
Crops were preprocessed in \texttt{scikit-image} \cite{vanderwalt2014skimage} by rolling-ball background subtraction \cite{sternberg1983biomedical} using a ball radius of 8 pixels followed by CLAHE (contrast limited adaptive histogram equalization) \cite{pizer1987adaptive}.
All crops were $32\times 32$ pixels.
Crops around embryo positions not fully in the field of view (within 16 pixels of the boundary of LCC-snapshot or mixture experiment snapshot) were excluded from classification analyses.
Training datasets were balanced by randomly dropping crops of the majority class.
The final training dataset contained $N=2792$ observations with 1396 of each class.\\

A random forest \cite{breiman1996bagging} of 100 decision trees was then trained using the \texttt{RandomForestClassifier} class from the \texttt{scikit-learn} \cite{pedregosa2011sklearn} Python package.
All hyperparameters were set to defaults except the random state, which was set to 1234. Each decision tree was trained on a bootstrap sample of size $N$ from the training dataset.
Splits were chosen using the CART (classification and regression tree) algorithm \cite{breiman1984classification} by greedily decreasing the Gini impurity \cite{simpson1949measurement}.
During a growth step, 32 randomly chosen features (pixels) were considered at each leaf: the pixel threshold that most decreased the Gini impurity was used to split observations at the leaf into two new leaves.
Trees were grown until all leaves were pure (contained observations of only one class).
Pure leaves were assigned to the class of their observations.
During prediction, individual decision trees assigned each test-set crop to a leaf (hard class label) using the splits learned during training.
The proportions of the 100 trees assigning a crop to an E1 or E2 leaf were then assigned as the E1 or E2 class probabilities (soft labels) of the crop.
Class probabilities were averaged over all crops in a sequence, and the embryo imaged in the sequence was finally classified as belonging to the class (E1 or E2) with the highest average probability.\\

Evaluation using 970 manually labeled binary-mixture embryos with correct tracks of at least 30 observations revealed an accuracy of 0.959 (Clopper-Pearson 0.95 CI \cite{clopper1934confidence}: [0.944, 0.970]) (see SI Video 4).
%The model and code used to generate it are available at []\todo{need to update after we prepare the code repo}.

\subsubsection{Velocity polarization rotates clockwise during the traveling state}\label{sec1:many:pangle}
\begin{figure}[!ht]
  \centering
  \small
\includegraphics[width=\linewidth]{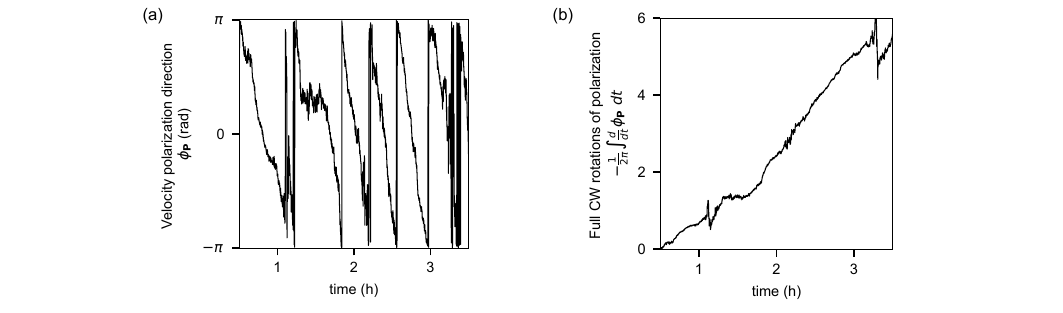}
    \caption{\textbf{Velocity polarization rotates clockwise during the traveling state.} \panel{A} Time series of $\phi_\mathbf{P} \equiv \arctan(\mathbf{P}_y/\mathbf{P}_x)$, the direction of velocity polarization $\mathbf{P}$. \panel{B} Cumulative number of full clockwise rotations of $\mathbf{P}$, starting from 0.5h in the experiment. }
  \label{figSI:Pangle}
\end{figure}

In the main text Fig.~1d, we show velocity polarization components $\mathbf{P}_x$ and $\mathbf{P}_y$. During the traveling state, the two components show oscillatory behavior with a time lag. Here, we show the direction of velocity polarization $\phi_\mathbf{P} \equiv \arctan(\mathbf{P}_y/\mathbf{P}_x)$. The time series of $\phi_\mathbf{P}$ shows a constant decline, indicating a clockwise rotation of the direction that embryos collectively flock toward (Fig. \ref{figSI:Pangle}). 

\subsubsection{E2 embryos lead E1 embryos in collective translation during the traveling state}\label{sec1:many:e1e2pangle}

In theory, the nonreciprocity-driven emergent polar order, such as traveling wave in mixed populations, should exhibit spatial asymmetry that reflects underlying run-and-chase dynamics. In our system, where E2 embryos ``run'' from E1, we expect E2 to consistently lead the traveling wave, with E1 following behind. \\

Although the direction of velocity polarization $\mathbf{P}$ changes over time, it does so in a predictable manner: the direction rotates clockwise. This allows us to identify the leading population at any moment by examining which embryo types are ahead in this rotation. Specifically, embryos with $\phi_\mathbf{v} - \phi_\mathbf{P} < 0$ (i.e., those rotated clockwise from $\mathbf{P}$) are leading in the direction of collective motion.\\

To further quantify which population leads, we define a global order parameter, the type-weighted position vector:
\begin{equation}
\boldsymbol{\psi}_q \equiv \sum_i q_i(\mathbf{r}_i-\frac{1}{N}\sum_j\mathbf{r}_j), \label{eqn:psiq}
\end{equation}
where $q_i=1$ for E2 and $q_i=-1$ for E1. For a single E1-E2 pair, this reduces to $\mathbf{r}_{12}\equiv \mathbf{r}_2-\mathbf{r}_1$. For greater than two embryos, $\boldsymbol\psi_q$ points toward regions enriched in E2 and depleted of E1. We compare the alignment between $\boldsymbol\psi_q$ and the average velocity $\langle \mathbf{v}_i \rangle_i$ to evaluate whether collective motion is biased toward E2.\\

\begin{figure}[!ht]
  \centering
  \small
\includegraphics[width=\linewidth]{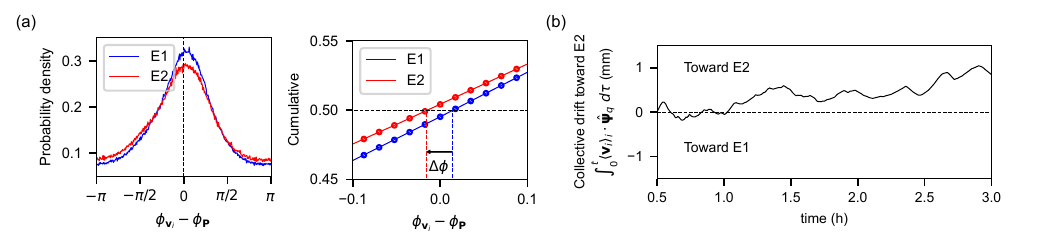}
\caption{\textbf{E2 embryos lead E1 embryos in collective translation during the traveling state.}  
\panel{A} Probability and cumulative distributions of $\phi_{\hat{\mathbf{v}}_i} - \phi_{\mathbf{P}}$, the angular deviation of each embryo's velocity direction from the global velocity polarization $\mathbf{P}$. E2 embryos are, on average, rotated slightly clockwise relative to E1s, with a median angular lead of $\Delta \phi = 0.03$ rad. Given that $\phi_\mathbf{P}$ rotates approximately 2.5 times per hour, this corresponds to a temporal lead of about 7 seconds for E2s in the flocking dynamics.  
\panel{B} Collective biased drift analogous to the pairwise biased drift shown in Fig. 1b. We define a type-weighted position vector $\boldsymbol{\psi}_q \equiv \sum_i q_i(\mathbf{r}_i - \frac{1}{N}\sum_j \mathbf{r}_j)$, with $q_i = \pm1$ for E2 and E1, respectively. Integrating the projection of average system velocity $\langle \mathbf{v}_i \rangle_i$ onto $\boldsymbol{\psi}_q$ over time confirms that the velocity polarization $\mathbf{P}$ is biased toward the E2-rich direction.}
  \label{figSI:E1E2Pangle}
\end{figure}

Together, these two independent analyses confirm that E2 embryos consistently lead E1 in the collective flocking during the traveling state (Fig. \ref{figSI:E1E2Pangle}). 

\subsubsection{Many-body behaviors beyond velocity polarization} \label{sec1:3:5}
In the main text, we focus on the velocity polarization $\mathbf{P}\equiv \langle\hat{\mathbf{v}}\rangle$, as a key signature of collective behavior in the embryo system. However, this is not the only emergent feature observed. In this section, we describe additional aspects of the many-body dynamics seen in the experiment.
\\

\begin{figure}[!ht]
  \centering
  \small
\includegraphics[width=\linewidth]{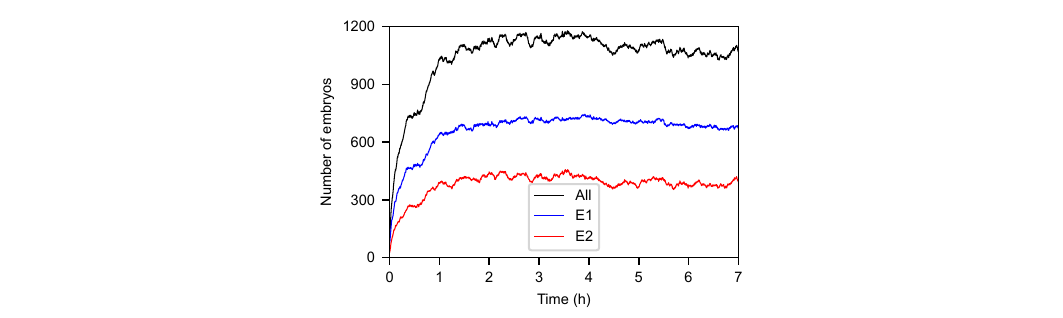}
    \caption{\textbf{Time series of the number of embryos at the air-water interface}  }
  \label{figSI:Ntime}
\end{figure}

At the start of the experiment (time 0), we mix thousands of E1 and E2 embryos. Some embryos quickly rise to the air-water interface, while others remain submerged. Over the first 30 minutes, the number of embryos visible at the interface increases rapidly, then plateaus around 1 hour~(Fig.~\ref{figSI:Ntime}). For consistency across measurements, we set the start time for analysis in the main text as 30 minutes after mixing (i.e., 0.5 h). \\

\begin{figure}[!ht]
  \centering
  \small
\includegraphics[width=\linewidth]{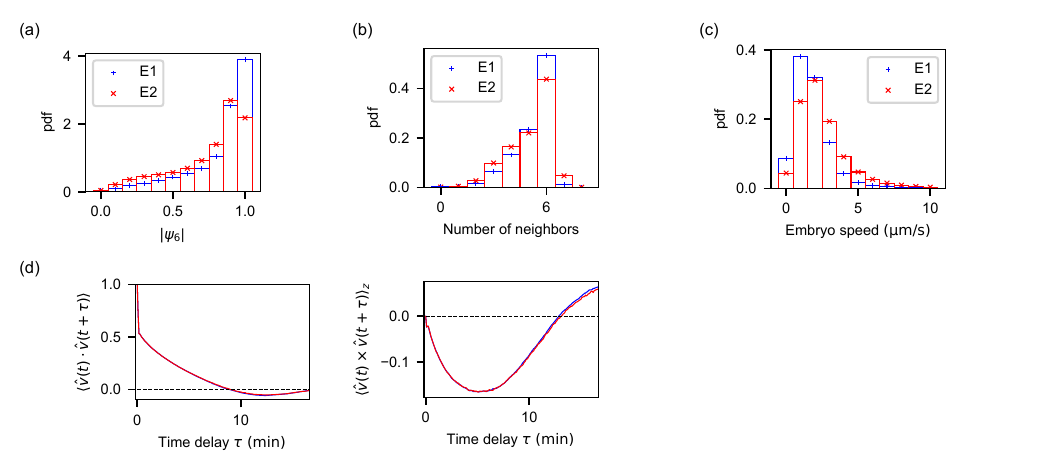}
    \caption{\textbf{Statistical differences between E1 and E2 embryos during the traveling state.} \panel{A} Distribution of the magnitude of the hexatic order parameter $|\psi_6|$. E1 embryos tend to exhibit higher structural order than E2.  
    \panel{B} Distribution of the number of neighbors. E1 embryos are more likely to have exactly six neighbors.  
    \panel{C} Distribution of embryo speeds. E1 embryos move more slowly on average than E2.  
    \panel{D} Velocity autocorrelation functions for E1 (blue) and E2 (red). Both types show similar temporal correlations in both longitudinal and transverse directions, persisting over several minutes.}
  \label{figSI:E1E2}
\end{figure}

For the first three hours of the experiment (the traveling state), the embryos exhibit persistent collective translation~(Fig.~1d). During this phase, they also form transient clusters that continuously break and merge (see SI Videos 3-4). This dynamic structural organization \ {can be} quantified using the hexatic order parameter, $\psi_6 = \frac{1}{n} \sum_{i=1}^n e^{6i\theta_i}$, where $\theta_i$ is the angle to the $i$th nearest neighbor. A value of $|\psi_6| = 1$ corresponds to perfect hexagonal packing, i.e., six neighbors separated by $\pi/3$. Interestingly, E1 embryos tend to have higher $|\psi_6|$ values than E2 (Fig. \ref{figSI:E1E2}a), and are also more likely to have exactly six neighbors (Fig. \ref{figSI:E1E2}b). In addition, E1 embryos generally move more slowly than E2 embryos (Fig. \ref{figSI:E1E2}c). Together, these findings suggest that E1s are more structurally ordered and less dynamic than E2s during the traveling state.
\\

Despite these differences, both E1 and E2 embryos exhibit similar velocity autocorrelation functions in the traveling state (Fig. \ref{figSI:E1E2}d), including a shared negative peak in the cross-correlation $\langle \hat{\mathbf{v}}(t) \times \hat{\mathbf{v}}(t+\tau) \rangle_z = \langle v_x(t)v_y(t+\tau) - v_y(t)v_x(t+\tau) \rangle$, which reflects the clockwise chirality of their trajectories. This indicates that, at long timescales, both types of embryos' long-time dynamics are governed by the same collective translation of the system. \\

\begin{figure}[!ht]
  \centering
  \small
\includegraphics[width=\linewidth]{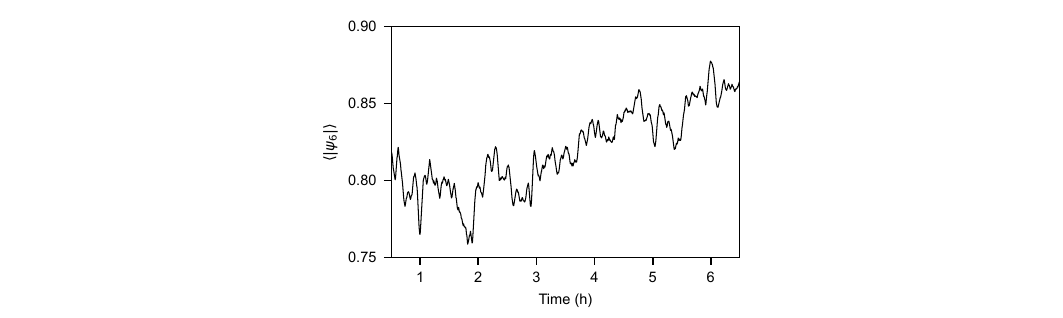}
    \caption{\ {\textbf{Time series of hexatic order parameter in the experiment.}  
Time series of the system-averaged hexatic order parameter $\langle |\psi_6| \rangle$.}}
  \label{fig:psi6}
\end{figure}

In the later hours of the experiment (the fluctuating state), the embryos cease their collective translation and instead self-organize into a more stable, ordered lattice. \ {We also find that the hexatic order parameter $\langle |\psi_6| \rangle$, which quantifies local sixfold symmetry, slowly rises in later hours (Fig.~\ref{fig:psi6}). Nevertheless, the significance of this overall change in  $\langle |\psi_6| \rangle$ (from 0.8 to 0.85) is weak, suggesting that we need a different structural order parameter to better characterize our system. }\\ % revisit this

\begin{figure}[!ht]
  \centering
  \small
\includegraphics[width=\linewidth]{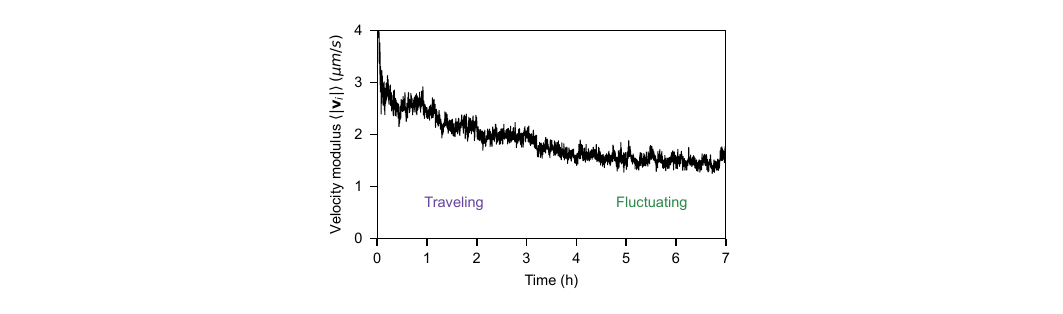}
    \caption{\textcolor{black}{\textbf{Time series of individual embryos' velocity modulus $\langle |\mathbf{v}_i|\rangle$.}  
The velocity modulus of individual embryos. }}
  \label{fig:vmod}
\end{figure}

\textcolor{black}{In late times, the average velocity modulus of individual embryos decreases only modestly (Fig.~\ref{fig:vmod}). In this fluctuating state, the positional displacements persist and their orientations become uncorrelated.} 

\subsection{Emergent collective states from additional experiments}\label{sec1:other}

Compared to a homogeneous embryo system---which typically forms a living chiral crystal (LCC)---the biologically tuned nonreciprocity between E1 and E2 embryos enables access to a much richer set of nonequilibrium behaviors, such as the traveling state discussed in the main text. 
\\

Here, we highlight two additional emergent collective states that further demonstrate the versatility of our experimental platform.
In both of the additional experiments, unless otherwise noted, we followed the same embryo preparation and data analysis pipeline as described in SI~Sec.~\ref{sec1:many:top}.

\subsubsection*{Demixed LCC}\label{sec1:other:demixed}
\begin{figure}[!ht]
  \centering
  \small
\includegraphics[width=\linewidth]{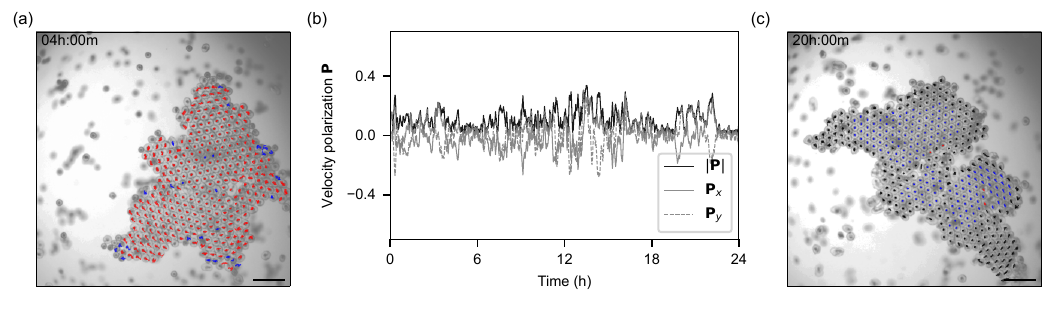}
    \caption{\textbf{Snapshots and velocity polarization of demixed LCC.}
\panel{A} Snapshot of the system shortly after E2 embryos have formed a nearly homogeneous crystal. Snapshot overlaid with 2-minute trajectories (blue: E1, red: E2).
\panel{B} Time series of velocity polarization, which remains low and fluctuates over time, indicating a lack of persistent collective motion.
\panel{C} Later snapshot showing an E1-dominated LCC after E2 embryos have disappeared. Embryos that appear after the first 12 hours of the experiment are labelled as black; for these embryos with age of 36h or beyond, our classifier based on 24h and 48h embryos may not be applicable. Given that most E2 embryos become unstable at the air-water interface at this point, these embryos that appear later are likely to be E1 embryos. }
  \label{figSI:demixed_1}
\end{figure}

In the main experiment, E1 and E2 embryos are mixed by stirring the well during their addition. However, if we skip this mixing step, the embryos remain initially demixed. Because E2 embryos are developmentally more advanced, they tend to reach the air-water interface earlier than E1 embryos.  This results in the early formation of a nearly homogeneous LCC composed mostly of E2 embryos. As time progresses, E1 embryos rise to the interface and attach to the periphery of the existing LCC, driving crystal growth. Eventually, E2 embryos lose stability at the interface and disappear, leading to an E1-dominated LCC. This results in a demixed regime that remains largely crystalline throughout (Fig. \ref{figSI:demixed_1} and SI Video 5).

\subsubsection*{Orbiting mixture}\label{sec1:other:orbit}
\begin{figure}[!ht]
  \centering
  \small
\includegraphics[width=\linewidth]{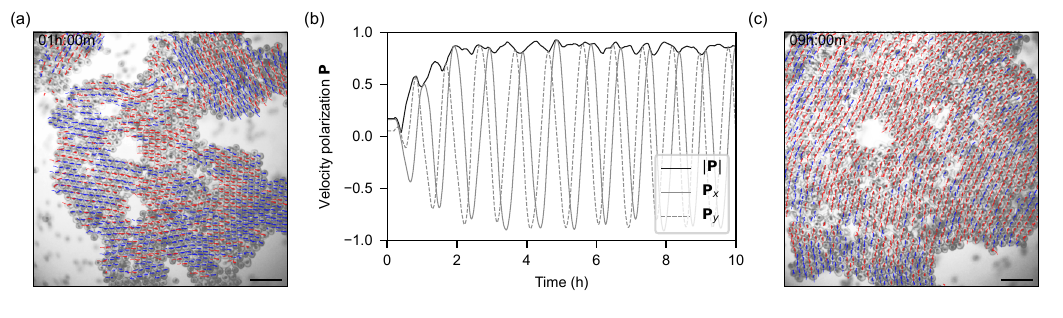}
    \caption{\textbf{Snapshots and velocity polarization of orbiting mixture.}
    \panel{A} Initial snapshot of a densely packed embryo system filling the well. Snapshot overlaid with 2-minute trajectories (blue: E1, red: E2).
    \panel{B} Time series of velocity polarization, which remains close to 1, indicating persistent, coherent translation along the boundary.
    \panel{C} Later snapshot showing continued orbiting motion with internal rearrangements.}
  \label{figSI:windshield_1}
\end{figure}

In the main experiment, the embryos primarily occupy the interior of the well and stay away from the circular boundary. However, if we introduce a significantly larger number of embryos, the system becomes densely packed, extending to the edge of the well. In this boundary-filled configuration, the embryos have little space to separate or break into distinct clusters, unlike in the traveling state. Instead, the entire system undergoes a persistent orbiting motion along the circular boundary, accompanied by continuous internal rearrangements. This high-density configuration gives rise to a distinct collective state: an orbiting mixture (Fig. \ref{figSI:windshield_1} and SI Video 6).
\\

\vfill
\clearpage
\section{Interaction inference}\label{sec2}
% HL: to check: r x z or z x r? gaussian noise explanation.
To describe the positional dynamics of starfish embryo mixtures, we employ the following equation:
\begin{equation}
\frac{d\mathbf{r}_i}{dt} = \sum_{j\neq i} f_L^{l_il_j}(r_{ij})\hat{\mathbf{r}}_{L,ij} + \sum_{j\neq i} f_T^{l_il_j}(r_{ij})\hat{\mathbf{r}}_{T, ij} + \boldsymbol{\xi}_i, \label{eqn:overdamped_dyn}
\end{equation}
where $\mathbf{r}_i$ denotes the in-plane coordinates of the $i^\mathrm{th}$ embryo, $l_i$ denotes the type of the $i^\mathrm{th}$ embryo ($l_i = 1$ for E1 embryo and $l_i=2$ for E2 embryo), $f_L$ and $f_T$ denote, respectively, the magnitudes of the longitudinal and transverse interactions, and $\boldsymbol{\xi}_i$ denotes a Gaussian white noise satisfying $\boldsymbol{\xi}_i(t)\boldsymbol{\xi}_j(t^\prime) = \Delta \mathbf{I} \delta_{ij}\delta(t-t^\prime)$, with $\Delta$ a scalar, and $\mathbf{I}$ the identity matrix. Here, the in-plane displacement vector between embryos $i$ and $j$ is given by $\mathbf{r}_{ij} \equiv \mathbf{r}_j - \mathbf{r}_i = r_{ij} \hat{\mathbf{r}}_{ij}$, where $r_{ij}$ is the distance between the embryos and $\hat{\mathbf{r}}_{ij}$ is the unit vector along this direction. Thus, the longitudinal unit vector is given by $\hat{\mathbf{r}}_{L,ij} = \hat{\mathbf{r}}_{ij}$, and the transverse unit vector is given by $\hat{\mathbf{r}}_{T,ij}=\hat{\mathbf{r}}_{ij}\times\hat{\mathbf{z}}$ where the unit vector $\hat{\mathbf{z}}$ is perpendicular to the plane in which the embryos reside. 
\\

To infer the force magnitudes $f_L$ and $f_T$ from time series of embryo positions, we parameterize $f$ using basis function expansions:
\begin{equation}
f(r_{ij}) = \sum_{n}w_n\theta_n(r_{ij}; R_0), \label{eqn:basis_function_expansion}
\end{equation}
where we choose $\theta_n(r_{ij}; R_0) = L_n(r_{ij}/R_0)\exp(-\frac{r_{ij}}{2R_0})$ indicating Laguerre polynomials $L_n$ with an exponential weighting factor, and we aim to learn the coefficients $w_n$ from data. The choice of the scale parameter $R_0$ will be discussed later. Introducing Eq.~(\ref{eqn:basis_function_expansion}) into Eq.~(\ref{eqn:overdamped_dyn}), we obtain:
\begin{align}
\frac{d\mathbf{r}_i}{dt} &= 
\sum_{j\neq i} \sum_n (w_n)_L^{l_il_j}\theta_n(r_{ij})\hat{\mathbf{r}}_{L,ij} + 
\sum_{j\neq i} \sum_n (w_n)_T^{l_il_j}\theta_n(r_{ij})\hat{\mathbf{r}}_{T,ij} + \boldsymbol{\xi}_i \notag \\ 
& = \sum_{\substack{O=L,T\\l=\mathrm{E1},\mathrm{E2}\\l^\prime=\mathrm{E1,E2}}}\sum_{n} (w_n)_O^{ll^\prime} \Bigg[\sum_{j\neq i}\delta_{l_il}\delta_{l_jl^\prime}\theta_n(r_{ij})\hat{\mathbf{r}}_{O,ij}\Bigg] + \boldsymbol{\xi}_i \label{eqn:linear_regression}
\end{align}
This essentially reduces the inference task to a linear regression problem.
We stack the eight sets of $(w_n)_O^{ll^\prime}~(O=L, T;~l=\mathrm{E1}, \mathrm{E2};~l^\prime=\mathrm{E1}, \mathrm{E2})$ into a coefficient vector $\mathbf{w}$. Similarly, Eqs. (\ref{eqn:linear_regression}) for all embryos at different time points can be stacked together into a matrix form:
\begin{equation}
\mathbf{v} = \boldsymbol{\Theta}\,\mathbf{w} + \boldsymbol{\xi}, \label{eqn:lr_matrix}
\end{equation}
where $\mathbf{v}$ and $\boldsymbol{\xi}$ are velocity and noise vectors in $\mathbb{R}^N$, $\boldsymbol{\Theta}$ is the library matrix in $\mathbb{R}^{N\times M}$, and $\mathbf{w}$ is the coefficient vector in $\mathbb{R}^M$. Here, each row corresponds to a spatial dimension of an embryo at a time point, and each column of $\boldsymbol{\Theta}$ corresponds to an interaction mode $(O,l,l^\prime)$ and computes the summation in the big square bracket in Eq.~(\ref{eqn:linear_regression}).  
\\

To perform Bayesian inference of $\mathbf{w}$, we minimize the negative log-posterior:
\begin{equation}
-\ln P(\mathbf{w} | \{\mathbf{r}_i(t)\} ) \sim -\ln P(\{\mathbf{r}_i(t)\} | \mathbf{w}) - \ln P(\mathbf{w}) \label{eqn:log_posterior}
\end{equation}
with respect to $\mathbf{w}$ given measurements $\{\mathbf{r}_i(t)\}$. To prevent overfitting, we follow previous work \cite{tipping2001sparse, wipf2004sparse} and impose a sparsity-promoting Gaussian prior over the coefficients:
\begin{equation}
P(\mathbf{w})
= \prod_{m} \mathcal{N}(w_m | 0, \gamma_m)
= \prod_{m} (2\pi \gamma_m)^{-1/2} \exp\Big( - \frac{w_m^2}{2\gamma_m}\Big), 
\end{equation}
where $\gamma_m$ are hyperparameters representing the variances of the Gaussian distributions, and we use a single index $m$ to represent the triplets $(O, l, l^\prime)$ in Eq.~(\ref{eqn:linear_regression}). Thus, the negative log-prior reads:
\begin{equation}
-\ln P(\mathbf{w}) = \sum_{m} \frac{w_m^2}{2\gamma_m} + \sum_{m} \frac{1}{2}\ln(2\pi\gamma_m), \label{eqn:log_prior}
\end{equation}
which is similar to an L2 regularization on the weights \cite{krogh1991simple, mackay1992practical}.
The negative log-likelihood function is given by:
\begin{equation}
-\ln P(\{\mathbf{r}_i(t)\}|\mathbf{w}) =  \frac{N}{2}\ln(2\pi) + \frac{1}{2}\ln |\boldsymbol{\Psi}| +  \frac{1}{2}(\mathbf{v}-\boldsymbol{\Theta}\mathbf{w})^T\boldsymbol{\Psi}^{-1}(\mathbf{v}-\boldsymbol{\Theta}\mathbf{w}), \label{eqn:log_likelihood}
\end{equation}
where $\boldsymbol{\Psi} = \Delta \mathbf{I}_N$ assuming Gaussian white noise. Introducing Eqs.~(\ref{eqn:log_prior}, \ref{eqn:log_likelihood}) into Eq.~(\ref{eqn:log_posterior}), we obtain the posterior that follows a Gaussian distribution $P(\mathbf{w} | \{\mathbf{r}_i(t)\}) = \mathcal{N}(\mathbf{w};\boldsymbol{\mu}, \boldsymbol{\Sigma})$, where the covariance matrix $\boldsymbol{\Sigma}$ and the mean $\boldsymbol{\mu}$ are given by:
\begin{gather}
\boldsymbol{\Sigma} = (\boldsymbol{\Theta}^T \boldsymbol{\Psi}^{-1}\boldsymbol{\Theta}^T + \boldsymbol{\Gamma}^{-1})^{-1} \label{eqn:Sigma}\\
\boldsymbol{\mu} = \boldsymbol{\Sigma}\boldsymbol{\Theta}^T\boldsymbol{\Psi}^{-1} \mathbf{v}, \label{eqn:mu}
\end{gather}
and $\boldsymbol{\Gamma}$ is a diagonal matrix with $\boldsymbol{\Gamma}_{mm}= \gamma_m$.\\

To determine the values of hyperparameters in Eqs.~(\ref{eqn:Sigma}, \ref{eqn:mu}), we employ a pragmatic procedure based on previous work \cite{mackay1992bayesian}, and choose $\gamma_m$ and $\Delta$ to maximize the marginal likelihood $P(\{\mathbf{r}_i(t)\} | \Delta, \{\gamma_m\}) = \int P(\{\mathbf{r}_i(t)\} | \mathbf{w}; \Delta) P(\mathbf{w} | \{\gamma_m\})  d\mathbf{w}$. We use the Expectation Maximization (EM) method to iteratively update the values of $\gamma_m$ and $\Delta$. Specifically, given $\gamma_m^{(n)}$ and $\Delta^{(n)}$ from the previous iteration, we compute the current estimate of $\boldsymbol{\mu}^{(n)}$ and $\boldsymbol{\Sigma}^{(n)}$ using Eqs.~(\ref{eqn:Sigma}, \ref{eqn:mu}). The EM approach gives the re-estimates:
\begin{gather}
\gamma_m^{(n+1)} = \mathrm{E}_{\mathbf{w}\sim\mathcal{N}(\boldsymbol{\mu}^{(n)}, \boldsymbol{\Sigma}^{(n)})}[\mathbf{w}_m^2]=(\boldsymbol{\mu}_m^{(n)})^2 + \boldsymbol{\Sigma}_{mm}^{(n)},~\mathrm{and} \label{eqn:update_mu}\\
\Delta^{(n+1)} = \mathrm{E}_{\mathbf{w}\sim\mathcal{N}(\boldsymbol{\mu}^{(n)}, \boldsymbol{\Sigma}^{(n)})}\Big[\frac{|\mathbf{v}-\boldsymbol{\Theta}\mathbf{w}|^2}{N}\Big] = \frac{1}{N}\Big[|\mathbf{v}-\boldsymbol{\Theta}\boldsymbol{\mu}^{(n)}|^2 + \Delta^{(n)}\sum_{m}(1-\boldsymbol{\Sigma}_{mm}^{(n)}/\gamma_m^{(n)}) \Big]. \label{eqn:update_Detla}
\end{gather}
We note that when the degrees of freedom of the data samples $N$ is much larger than the number of modes $M$, Eq.~(\ref{eqn:update_Detla}) can be approximated by $\Delta^{(n+1)}\approx |\mathbf{v}-\boldsymbol{\Theta}\boldsymbol{\mu}^{(n)}|^2 / N$. Finally, we note that the scale parameter $R_0$ is chosen to minimize the residual $|\mathbf{v}-\boldsymbol{\Theta}\boldsymbol{\mu}^{(n)}|^2 / N$ through an additional layer of optimization outside of EM.

\subsection{All inferred interactions beyond inter-type, longitudinal interactions }\label{sec2:1}

\begin{figure}[!ht]
  \centering
\includegraphics[width=\linewidth]{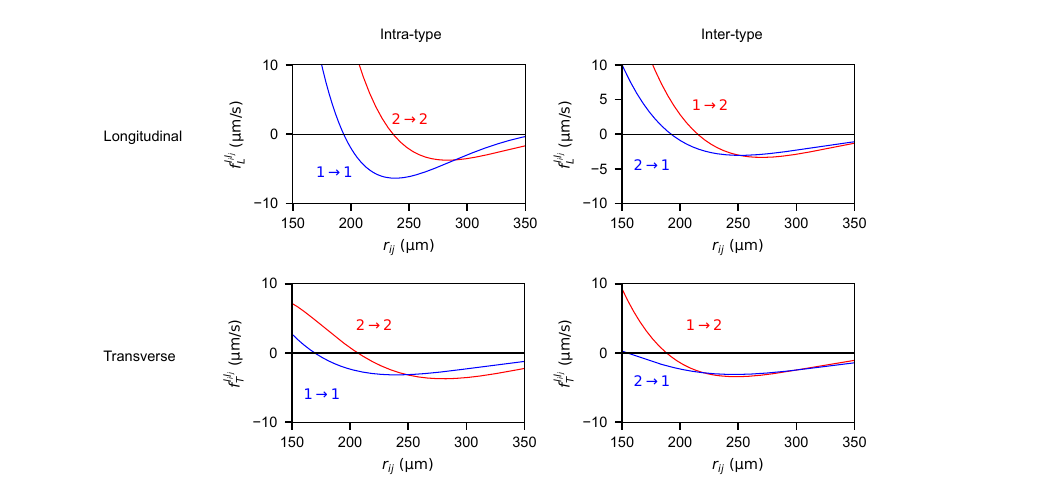}
    \caption{\textbf{All inferred interactions beyond inter-type, longitudinal interactions.}  }
  \label{figSI:Inference}
\end{figure}

We infer pairwise interactions between embryos in the binary mixture experiment by applying our inference method to the full trajectory dataset $\{\mathbf{r}_i(t), \dot{\mathbf{r}}_i(t)\}$ over all frames. 
%We rescale the spatial coordinates by 23 pixels ($203~\mu m$) and time by 80 experimental frames (400~s), such that both positions $\mathbf{r}_i$ and velocities $\dot{\mathbf{r}}_i$ are of order unity. After inference, all interactions are restored to physical units.\\
This procedure yields eight distinct interactions, corresponding to the four embryo-type combinations (E1--E1, E1--E2, E2--E1, E2--E2) and their longitudinal or transverse components, as shown in Fig.~\ref{figSI:Inference}.\\

The full set of interactions recapitulates several features of the embryo pairs. 
First, the longitudinal self-interactions (E1--E1 vs.\ E2--E2) differ in both magnitude and interaction range, reflecting the larger size of E2 embryos. 
Second, the transverse interactions are negative at the equilibrium pair distances ($195\ \mu m$ for E1--E1 and $237\ \mu m$ for E2--E2, based on longitudinal self-interactions), consistent with the observed clockwise rotation of embryo pairs. 
This transverse component reflects the chiral symmetry breaking arising from the ciliary beating of the embryos.

\begin{figure}[!ht]
  \centering
\includegraphics[width=\linewidth]{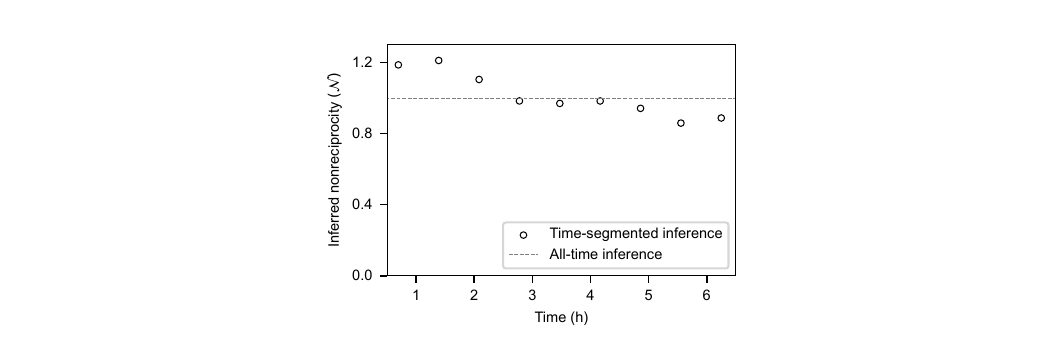}
    \caption{\textbf{Inferred nonreciprocity between E1-E2 pairs decreases over time} Over the course of experiment, inferred nonreciprocity decreases from $\mathcal{N}=1.2$ to $\mathcal{N}=0.9 $.  }
  \label{figSI:TimedNR}
\end{figure}

\subsection{Inferred nonreciprocity between E1-E2 pairs decreases over time}\label{sec2:2}

\ {In the main text, we present pairwise interactions inferred from all-time data and from early / late time windows}. Here, we \ {quantify} the nonreciprocity of longitudinal interactions between E1 and E2 embryos in time-segmented inference, where we use data from a segmented time-window (every 500 frames = 2500 seconds) to infer interactions. For each $f^{12,21}_L$ we obtain for a time-segmented data, we identify nonreciprocity $\mathcal{N}(t)$ by fitting $(f_L^{12}-f_L^{21})_{t} = \mathcal{N}(t)\times(f_L^{12}-f_L^{21})_{\mathrm{all\ time}}$.  Over the course of experiment, inferred nonreciprocity decreases from $\mathcal{N}=1.2$ to $\mathcal{N}=0.9 $ (Fig.~\ref{figSI:TimedNR}). 

%%%%%%%%%%%%%%%%%%%%%%%%%%%%%%%%%%%

\vfill
\clearpage
\section{Model analysis}\label{sec3}

\subsection{Inference-based model of pairwise interactions with tunable nonreciprocity $\mathcal{N}$}\label{sec3:1}

We construct a data-driven model based on pairwise interactions inferred from trajectories of starfish embryos. The dynamics of embryo $i$ are modeled as overdamped and governed by effective interactions with all other embryos:

\begin{equation} 
    \frac{d \mathbf{r}_i}{d t} = \sum_j  f_L^{l_i l_j}(r_{ij})\hat{\mathbf{r}}_{ij} +  f_T^{l_i l_j}(r_{ij})\hat{\mathbf{r}}_{ij} \times \hat{z} - f_r \mathbf{r}_i
\end{equation}

Here, $\hat{\mathbf{r}}_{ij}$ is the unit vector pointing from embryo $i$ to $j$, and $l_i \in \{1,2\}$ denotes the type (E1 or E2) of embryo $i$. The interaction functions $f_L^{l_i l_j}(r)$ and $f_T^{l_i l_j}(r)$ are the longitudinal and transverse components of the effective force, inferred from the main binary mixture experiment (SI~Sec.~\ref{sec2}). These forces are normalized by embryo drag coefficients, and we assume that inertia is negligible due to strong overdamping. We also neglect noise, based on the assumption that self-generated flows captured by pairwise interactions dominate the embryo dynamics.\\

To explore how nonreciprocity affects collective behavior, we introduce a one-parameter family of interaction models that systematically vary the degree of asymmetry between E1-E2 interactions. Specifically, we only modify the longitudinal interaction terms $f_L^{12}(r)$ and $f_L^{21}(r)$, while keeping all other interaction functions fixed at their experimentally inferred values.\\

To enable this, we decompose the inferred E1-E2 longitudinal interactions into symmetric and antisymmetric parts:
\begin{equation}
    f_L^S(r) \equiv \frac{f_L^{12}(r) + f_L^{21}(r)}{2}, \quad
    f_L^A(r) \equiv \frac{f_L^{12}(r) - f_L^{21}(r)}{2}
\end{equation}

We then construct a tunable model with nonreciprocity parameter $\mathcal{N}$ as:
\begin{equation} 
    f_L^{12}(r;\mathcal{N}) = f_L^S(r) + \mathcal{N} f_L^A(r), \quad 
    f_L^{21}(r;\mathcal{N}) = f_L^S(r) - \mathcal{N} f_L^A(r)
\end{equation}

\ {A value of $\mathcal{N} = 0$ indicates reciprocal forces, while $\mathcal{N} = 1$ indicates forces that match the inferred all-time interactions. To study the effects of nonreciprocity on emergent structure and dynamics, we modeled a range of values including weaker ($\mathcal{N} < 1$) and stronger ($\mathcal{N} > 1$) nonreciprocity.}
\\

To mimic the curved air-water interface of the experimental system, as reported in \cite{tan2022odd}, we introduce a weak central potential term, modeled by $f_r = 2 \times 10^{-4}$. For embryos within 10 interaction units of the center, the magnitude of this confinement force is approximately three orders of magnitude weaker than a typical pairwise interaction. This central potential prevents embryos from drifting arbitrarily far, capturing the effect of the confining geometry of the experimental well. Additionally, we reduce the strength of transverse interactions $f_T$ to 10\% of the inferred values, reflecting the size-dependent slowdown of cluster rotations reported in \cite{tan2022odd}.

\subsection{Implementation of numerical simulations}\label{sec3:2}

We implemented the model in Julia and performed simulations using a forward Euler method with a time step of $\Delta t = 0.1$ model time unit. Based on the experimental distribution of embryo displacement per frame (relative to distance between embryos), we record simulation snapshots every 7.5 model time units (75 time steps). Simulations were run in an open domain without periodic or reflective boundaries. Although the domain was unbounded, the weak central potential was sufficient to confine the dynamics within a finite area, as observed experimentally.\\

We explored nine levels of nonreciprocity, $\mathcal{N} \in \{0, 0.25, 0.5, 0.75, 1, 1.25, 1.5, 1.75, 2\}$, and ran 20 independent simulations for each value. Each simulation was initialized with 200 E1 and 200 E2 embryos placed randomly within a square domain of size $[-20, 20]\times [-20, 20]$, where one unit length corresponds to $267~\mu m$.\\

Each simulation was run for 3000 frames. To remove transient effects, we analyzed only the second half of each trajectory (frames 1501-3000). For these data, we computed time-averaged metrics including velocity polarization $\mathbf{P}$, the structural order parameter $\langle d_{\mathrm{hex}} \rangle$, and the frequency of topological motifs, which together characterize collective motion and spatial organization.\\

All simulations were performed on the MIT Supercloud cluster using Julia v1.9.1. We verified that results are compatible with Julia v1.11, and that all package dependencies are reproducible and documented. SI Video 7 shows nine representative simulations for the nine nonreciprocity values. \\

\subsection{Neighbor exchanges become frequent at $\mathcal{N}=1$}\label{sec3:3}
\begin{figure}[!ht]
  \centering
  %\small
\includegraphics[width=\linewidth]{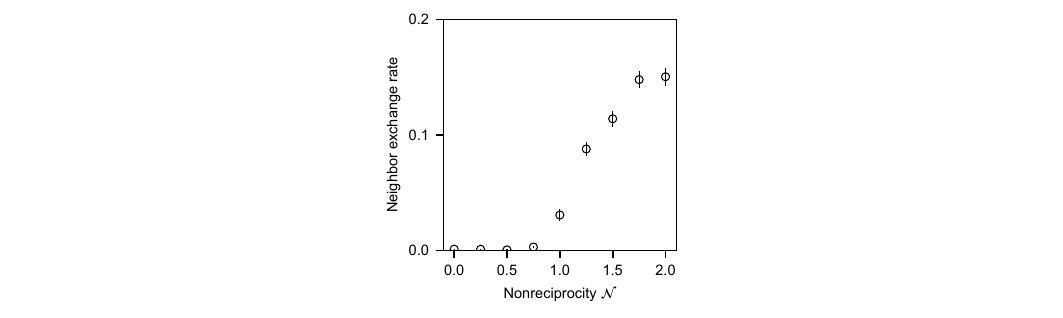}
    \caption{\textbf{Neighbor exchanges rates as a function of nonreciprocity $\mathcal{N}$.} The rate remains near zero for $\mathcal{N}<1$ and rises linearly until saturating near $\mathcal{N}\approx1.75$.  }
  \label{figSI:nnexchange}
\end{figure}

Our inference-based simulations reveal that sufficiently strong nonreciprocity disrupts crystalline order, producing repeated cycles of cluster merging and breaking. To quantify this structural instability, we measure the \emph{neighbor exchange rate}, defined as the probability that a given particle changes at least one of its nearest neighbors between consecutive snapshots (corresponding to 5 seconds in experimental time)(Fig.~\ref{figSI:nnexchange}).
\\

For weak nonreciprocity ($\mathcal{N}<1$), the system remains crystalline with minimal neighbor exchanges. Then, the neighbor exchange rate rises approximately linearly with $\mathcal{N}$, reflecting increasingly dynamic rearrangements. The rate eventually saturates around $\mathcal{N}\approx 1.75$, coinciding with the regime of highly transient fragments and continuous structural reorganization.

\ {
\subsection{\ {Number of connected components and nonreciprocity}}\label{sec3:4}
Yet another way to quantify the breaking of cluster is to count the number of connected components in the system, which will be larger than one if the system deviates from crystalline or self-propelled crystalline states (Fig.~\ref{figSI:ncluster}). \\
For weak nonreciprocity ($\mathcal{N}<1$), the system remains crystalline with a single connected component. Then, the number of connected components rises with $\mathcal{N}$, reflecting highly fragmented configuration. 
\begin{figure}[!ht]
  \centering
  %\small
\includegraphics[width=\linewidth]{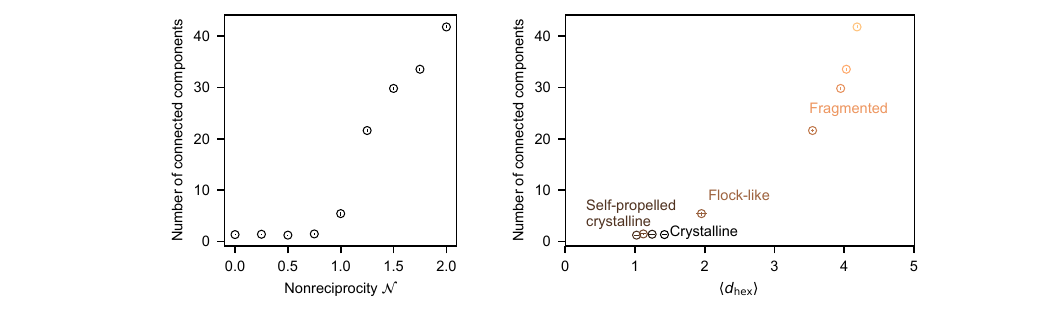}
    \caption{\ {\textbf{Number of connected components increases with $\mathcal{N}$.} \textit{Left:} Number of connected components as a function of nonreciprocity $\mathcal{N}$. Error bars are SEM for time-averaged number of connected components across simulations (n=20). The system remains fully connected as a single cluster for $\mathcal{N}<1$. For $\mathcal{N}>1$, clusters start to break into smaller fragments. \textit{Right:} State diagram of $\langle d_{\mathrm{hex}} \rangle$ and number of connected components. Unlike velocity polarization $\mathbf{P}$, the number of connected components increases along with  $\langle d_{\mathrm{hex}} \rangle$. As a result, the $\mathcal{N}=0$ and $0<\mathcal{N}<1$ states are not separated in the diagram. }}
  \label{figSI:ncluster}
\end{figure}
}

%%%%%%%%%%%%%%%%%%%%%%%%%%%%%%%%%%%

\vfill
\clearpage
\section{Topological metric and structural order parameter}\label{sec4}

\subsection{What is a motif?}\label{sec4:1}
Self-organized structures are often characterized by leveraging symmetries: for example, the structure factor $S(q)$ or the pair distribution function $g(r)$ which detect translational symmetry, and the hexatic order parameter $\psi_6$ which measures sixfold rotational symmetry. However, many self-organized structures---as observed in the dynamic clusters of starfish embryo mixtures in their traveling state and the nonreciprocal mixture simulations with a sufficiently strong nonreciprocity $\mathcal{N}\ge1$---can be far away from any obvious symmetries, limiting the applicability and interpretation of conventional, symmetry-based order parameters. 
\\

\begin{figure}[!ht]
    \centering
    \includegraphics[width=\linewidth]{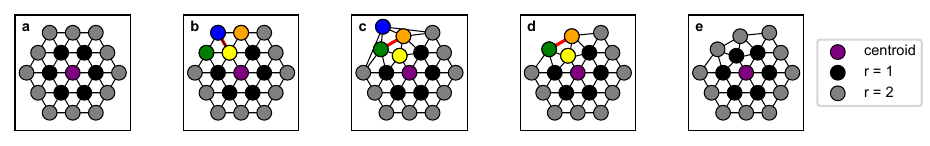}
    \caption{\textbf{Illustration of the T1 transition from the $\textrm{M}_1$ motif to the $\textrm{M}_2$ motif.} \panel{A} The hexagonal motif ($\textrm{M}_1$) has a central node (centroid) surrounded by 6 nearest neighbors ($r = 1$) and 12 next-nearest-neighbors ($r = 2$) (19 embryos total). \panel{B}-\panel{C} In a T1 transition, a bond flip (shown in red) occurs. During the transition, the bond between embryos which are originally nearest neighbors flips, e.g. so that the blue and yellow embryos become next-nearest-neighbors from nearest neighbors, and the green and orange embryos become nearest neighbors from next-nearest-neighbors. \panel{D} Since a motif is comprised of embryos which are within two bonds from the centroid ($r = 2$), the blue embryo is eliminated from the motif. \panel{E} The resulting motif is a hexagon with a 5-defect ($\textrm{M}_2$) that has 18 embryos. }
    \label{figSI:motif}
\end{figure}

To overcome this limitation, we characterize the self-organized structures via the network topology of local neighborhoods. Specifically, each particle's local neighborhood (here we choose next-to-nearest neighbors, i.e. $r=2$) corresponds to a \emph{topological motif}, a network in which particles are nodes and nearest-neighbor connections are edges. For example, a hexatic crystal corresponds to a topological motif with the central node (representing the focal particle) with edges to its six neighbors, whose edges connect to their subsequent neighbors~(Fig.~\ref{figSI:motif}a). 
\\

Even for an $r=2$ neighborhood, there are many possible motifs; we consider $\mathcal{O}$(15,000) motifs, which are only a subset of the observed possible motifs. 
Although motifs do not possess intrinsic coordinates, they can be naturally quantified at the pair level: the distance between any two motifs is defined as the number of topological (T1) transitions required to transform one motif into the other~(Fig.~\ref{figSI:motif}). This distance measure highlights the topological nature of the formalism and offers a unique connection to structural dynamics in living systems, such as tissue dynamics and morphogenesis~\cite{skinner2021topological}. 

\subsection{Identification of topological motifs from data}\label{sec4:2}
\begin{figure}[!ht]
    \centering
    \includegraphics[width=\linewidth]{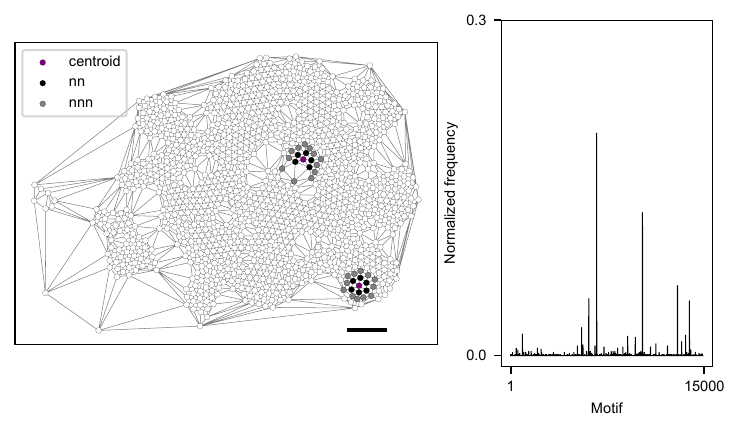}
    \caption{\textbf{Example topological motif analysis of a configurational snapshot.} Shown here: the network of embryos at 3 hours in the mixed experiment from the main text (scale bar: $10\ \mathrm{\mu m}$). Embryo positions are triangulated using Delaunay triangulation, and motifs are extracted for the $r = 2$ (next-nearest-neighborhood) around each embryo. Two example motifs are highlighted (center: a non-hexagonal motif, bottom-right: a hexagonal motif). The centroid of each motif is depicted in purple, its nearest neighbors in black, and next-nearest-neighbors in gray. The probability distribution of all motifs from the snapshot is shown on the right.  }
    \label{fig:motif_examples}
\end{figure}

We used the Julia topological packing statistics tool \texttt{TopologicalAnalysis} developed in \cite{skinner2021topological} to extract the $r = 2$ (next-nearest-neighborhood) topological motifs and their empirical probability distributions based on embryo positions for all experiments and model simulations. Topological motifs are determined from the Delaunay triangulation of embryo positions at a single time point. Likewise, the probability of a motif corresponds to the normalized frequency of that motif at a single time point. While we do not filter triangulation edge lengths, we choose not to keep track of the motifs centered on embryos included in the convex-hull (outer envelope) of the Delaunay triangulation. Therefore, the extracted probability distributions comprise motifs centered only on embryos within the convex hull, but can include edges that connect embryos across clusters (Fig. \ref{fig:motif_examples}). %Figure XXXX shows an example Delaunay triangulation and r = 2 motif for time XXXX in the mixed experiment and its corresponding probability distribution. %Add in a callout here for video of relevant motifs in experiment
\\

From our inference-based model of nonreciprocal mixtures, we extract the data-driven topological flip graph \cite{skinner2021topological} from the steady-states of nine representative simulations with varying nonreciprocity $\mathcal{N}$ (from $\mathcal{N} = 0$ to 2.0 in steps of 0.25). %This metric space captures the pairwise distances, defined by the number of topological T1 transitions, between motifs. 
Probability distributions are projected onto this flip graph which comprises $\mathcal{O}(15,000)$ topological motifs. 
\\

Only motifs which occurred more than ten times over the course of all simulations were incorporated into the data-driven flip graph. For all but the highest nonreciprocity values ($\mathcal{N} < 1.5$), this cutoff captured well over 80\% of the motifs observed in the nonequilibrium steady-state, Table \ref{table:cutoff_TEM_main}. For the higher nonreciprocity values this cutoff reflected a lower total probability, around 65\% for $\mathcal{N} = 1.5$ and half of the probability of the more diverse $\mathcal{N} = 2.0$  state. This is due to the fact that higher nonreciprocity leads to greater dynamic instability. For example, for the $\mathcal{N} = 2.0$ state, motifs observed only once in the simulation account for around 30\% of its total probability. However, we find that these probability cutoffs adequately capture the diversity of the higher nonreciprocity states relative to the flocking and low reciprocity states, as evidenced by high values of $\langle d_{hex} \rangle$ and the shallower (more diverse) topological landscapes shown in the main text (Figs. 2 and 3). \\ %Furthermore, an analysis of the PC1 distance (a proxy for $\langle d_{hex} \rangle$, see Figure \ref{fig:PCA_9ptfg_restrict2}) between $\mathcal{N}$ = 1.0 and $\mathcal{N}$ = 1.5, $\Delta \textrm{PC}1$, as a function of cutoff shows that incorporating these less frequent motifs leads to a change of 10\% or less in $\Delta \textrm{PC}1$. 

\begin{table}[tb]
\caption{Total number of topological motifs observed for each of the nine representative simulations i) in the absence of a cutoff, and ii) projected onto the flip graph with a cutoff that each motif must occur greater than 10 times. For all but $\mathcal{N} \geq 1.5$, a cutoff of greater than 10 occurrences per motif captures 80\% or more of the probability.}
\label{table:cutoff_TEM_main}
\begin{tabular}{c|l|l|l}
Nonreciprocity & Number of motifs & Number of motifs& Total probability \\
 & (without cutoff) & (with cutoff) & reflected with cutoff   (\%) \\
 \hline 
 \hline 
0              & 71                                & 70                             & 99.7                                           \\
0.25           & 159                               & 136                            & 99.9                                           \\
0.5            & 37                                & 36                             & 99.7                                           \\
0.75           & 386                               & 260                            & 99.8                                           \\
1              & $\sim$27,000                      & $\sim$2,400                    & 90                                             \\
1.25           & $\sim$74,000                      & $\sim$3,800                    & 80                                             \\
1.5            & $\sim$130,000                     & $\sim$4,300                    & 65                                             \\
1.75           & $\sim$154,000                     & $\sim$4,200                    & 60                                             \\
2              & $\sim$187,000                     & $\sim$4,200                    & 50                                            
\end{tabular}
\end{table}

Further, as shown in Section \ref{sec5:3} (Fig. \ref{figSI:MotifAtlas}) for $\mathcal{N} = 1$, probability drops off quickly with motif rank for the top 14 most probable motifs, indicating that a relatively small subset of the motifs are needed to capture the complexity of the $\mathcal{N} = 1$ state.

\subsection{Calculation of structural order parameter}
\label{sec4:3}
Leveraging the motif probability distributions, $p$, we define the structural order parameter $\langle d_{\mathrm{hex}} \rangle$ which describes the system-average topological distance to a hexagonal crystal:
\begin{equation}
    \langle d_{\mathrm{hex}} \rangle = \sum_{j} p(j) \hspace{1pt } d_{\mathrm{hex}}(j).   
    \label{eq:def_dhex}
\end{equation}
Here, $j$ is taken to be the motif index in the flip graph, $p(j)$ is its probability, and $d_{\mathrm{hex}}(j)$ is the topological distance between the motif and the hexagonal motif. As we will show below, this order parameter has another meaningful physical interpretation- it corresponds to the topological earth mover's distance \cite{skinner2021topological} to the perfect hexagonal crystal, and thus describes the minimum cost (in units of T1 transitions) to transform the system configuration into the hexagonal crystal (SI~Sec.~\ref{sec4:1}). By plotting the $d_{\mathrm{hex}}$ value associated with each motif for each embryo (motif centroid), we find that $d_{hex}$ is spatially distributed throughout the crystal, with low values ($d_{hex} \leq 2$) concentrated at the center of the crystal, and higher $d_{hex}$ values ($2 \leq d_{hex} \leq 6$) located at crystal boundaries (see SI Video 8 for the model simulations and SI Video 10 for the mixed experiment). 
\\

%\subsection{Structural order parameter is equivalent to Topological Earth-Mover distance to crystal}\label{sec4:1}
Computationally, we calculate the structural order parameter $\langle d_{\mathrm{hex}} \rangle$ by leveraging the recently developed \texttt{TopologicalAnalysis} package in Julia~\cite{skinner2021topological}. The topological earth mover (TEM) distance in this package is a metric to quantify the difference between two probability distributions of motifs projected onto the flip graph. It can be described as the minimum cost (in units of T1 transitions) to transform one probability distribution, $p_A$, into another, $p_B$, and is given by:
\begin{equation}
    \mathrm{TEM}(p_A, p_B) = \min_{\gamma} \sum_{i,j} \gamma_{ij} \hspace{1pt } d(i,j),
    \label{eq_TEM_dist}
\end{equation}
where $\gamma$ corresponds to a transport map between the distributions that satisfies $\sum_j \gamma_{ij} = p_A(i)$ and $\sum_i \gamma_{ij} = p_B(j)$ with $\gamma_{ij} \geq 0$. Here, $d(i,j)$ denotes the topological distance (number of connecting edges) between two motifs on the flip graph.\footnote{Note that the TEM distance can take on any value greater than or equal to zero, while the topological distances $d(i,j)$ are constrained to nonnegative integers.}\\

In the case where one of the probability distributions contains a single motif, e.g.:
\begin{equation}
 p_A(i) =  p_{\mathrm{hex}}(i) = 
    \begin{cases}
        1 & \text{if  $i =$ index for hexagonal motif} \\
        0 &  \text{else},
    \end{cases}
\end{equation}
then the TEM distance given in Eq.~(\ref{eq_TEM_dist}) reduces to $\langle d_{hex} \rangle$, a weighted average of the distances $d_{hex}(j) = d(\text{hex},j)$ of the motifs present in distribution $p_B$ from that single motif (the hexagon). That is,
\begin{equation}
    \langle d_{\mathrm{hex}} \rangle = \sum_{j} p_B(j) \hspace{1pt } d_{\mathrm{hex}}(j)  =  \mathrm{TEM}(p_{\mathrm{hex}},p_B) = \min_{\gamma} \sum_{i,j} \gamma_{ij} \hspace{1pt } d(i,j).
\end{equation}
This follows from the fact that there is only one way to minimally transform the single motif distribution, $p_{hex}$, into any other distribution $p_B$:  $\gamma_{hex,j} = p_B(j)$.\footnote{Consider the simple test case $p_A = [1,0,0]$ and $ p_B= [0.7,0.1,0.2]$ where we find that $\gamma = ( (0.7,0.1,0.2),(0,0,0),(0,0,0) ) = (\vec{p_B},\vec{0},\vec{0})$.}
\\

\begin{figure}[!ht]
    \centering
    \includegraphics[width = \linewidth]{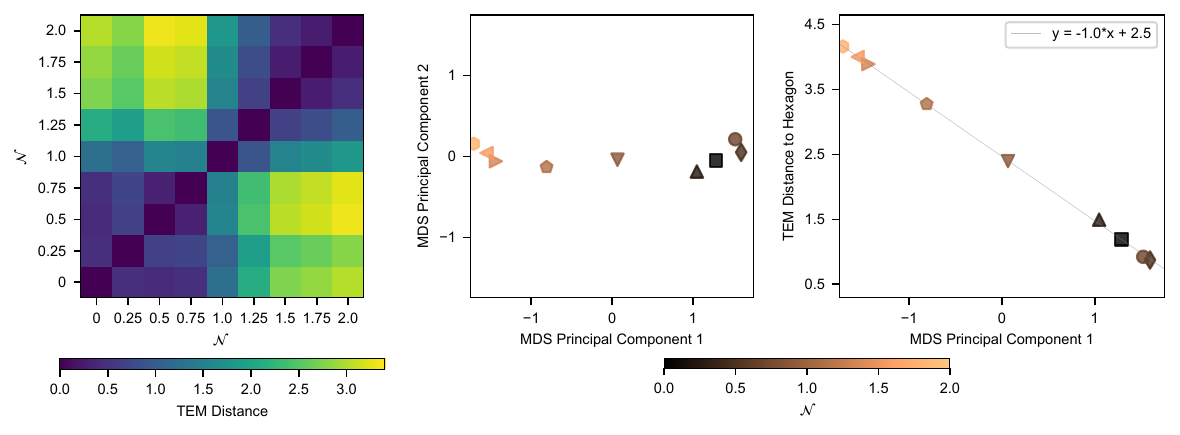}
    \caption{\textbf{TEM distance matrix between simulations clusters based on nonreciprocity.} Left: TEM distance between the nine representative simulations used to determine the data-driven flip graph. There are distinct clusters that distinguish low nonreciprocity values ($\mathcal{N} < 1$) from higher nonreciprocity values ($\mathcal{N} > 1$), with $\mathcal{N} = 1$ acting as a boundary between these clusters. Right: A two-dimensional embedding of the TEM distance matrix using classical multidimensional scaling. The first principal component corresponds to the TEM distance to the hexagon (line of best fit shown in gray).}
    \label{fig:TEM_dist_9pt_restrict10}
\end{figure}

As further evidence that $\langle d_{hex} \rangle$ is the relevant structural order parameter, we perform a clustering analysis on the TEM distance matrix between the nine representative simulations for the nine different nonreciprocity values~(Fig.~\ref{fig:TEM_dist_9pt_restrict10}). The matrix shows two distinct clusters for low nonreciprocity values ($\mathcal{N} < 1$) and high nonreciprocity values ($\mathcal{N} > 1$), separated by the $\mathcal{N} = 1$ state. A two-dimensional embedding of the TEM distance matrix performed using multidimensional scaling (MDS) further delineates these clusters~(Fig.~\ref{fig:TEM_dist_9pt_restrict10}). In particular, the first principal component of the MDS embedding scales linearly with the TEM distance to hexagon ($\langle d_{hex} \rangle$). %Performing a principal components analysis \cite{PCA_ref}, we find that 95\% of the variance was described by the first principal component, which is associated with the metric $\langle d_{hex} \rangle$, Figure %\ref{fig:PCA_9ptfg_restrict2}.

%\begin{figure}[htbp]
%    \centering
%    \includegraphics[]{TEM_distance_bw_simulations_9x20.pdf}
%    \caption{Left: TEM Distance between the nine representative simulations used to determine the data-driven flip graph (shown here: distances based on motifs that occur more than two times in the steady state). There are distinct clusters that distinguish low nonreciprocity values ($\mathcal{N} < 1$) from higher nonreciprocity values ($\mathcal{N} > 1$) with $\mathcal{N} = 1$ acting as a boundary between these clusters. Right: These clusters are consistent across initial conditions (shown here: distances based on motifs which occur more than ten times across the steady state).}
%    \label{fig:TEM_dist_9ptfg_restrict2}
%\end{figure}

%\begin{figure}[htbp]
%    \centering
%    \includegraphics[]{MDS_on_9pt_TEM_distances_restrict2.pdf}
 %   \caption{Left: The first two principal components of the representative simulations determined from a principal components analysis of the TEM distance matrix in Figure \ref{fig:TEM_dist_9ptfg_restrict2}. Right: The order parameter, $\langle d_{hex} \rangle =$ TEM Distance to Hexagon, is identified from its correspondence with PC1. As in the TEM distance matrix, there are two clear clusters of low nonreciprocity values and high nonreciprocity values, separated by the $\mathcal{N} = 1$ state.}
 %   \label{fig:PCA_9ptfg_restrict2}
%\end{figure}

\subsection{Nonreciprocal \ {error correction} }\label{sec3:4}
Our interaction inference reveals that E1 and E2 embryos experience distinct self-interactions~(SI Sec.~\ref{sec2:1}). Consequently, even in the reciprocal case ($\mathcal{N}=0$), model simulations produce a bi-disperse crystal containing a finite fraction of defects.\\

\begin{figure}[!ht]
  \centering
  \small
\includegraphics[width=\linewidth]{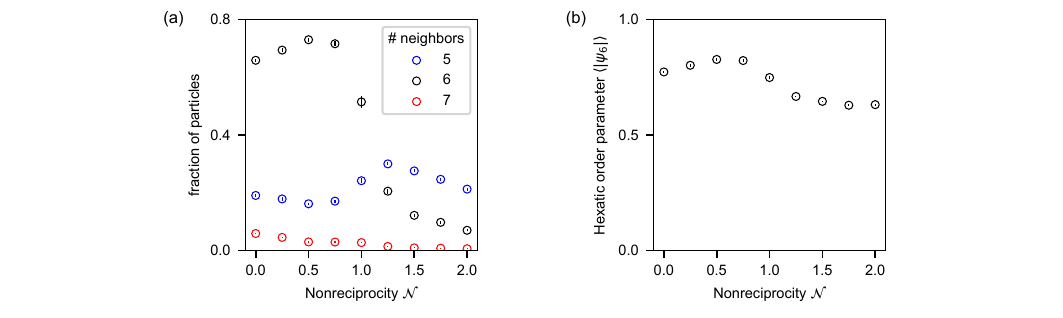}
    \caption{\textbf{Number of defect supports nonreciprocal \ {error correction}. } 
  \panel{A}~Weak nonreciprocity ($\mathcal{N}<1$) reduces the number of 5- and 7-fold defects, driving the system toward a more ordered crystal. 
  \panel{B}~The same regime enhances the hexatic order $\langle |\psi_6|\rangle$, indicating improved crystalline order.}
  \label{figSI:nrselfhealing}
\end{figure}

Surprisingly, introducing a weak nonreciprocity enhances crystalline order. This effect, captured in the main text via our topological metric, is corroborated here by both the reduction of 5- and 7-fold defects and the enhancement of the hexatic order $\langle |\psi_6| \rangle$~(Fig.~\ref{figSI:nrselfhealing}). This \emph{nonreciprocal \ {error correction}} is counterintuitive, given that nonreciprocity is known to promote instabilities and dynamics. \\

We interpret this effect as analogous to annealing. Nonreciprocity introduces local stresses that destabilize low-symmetry or weakly ordered regions, while highly symmetric crystalline neighborhoods remain robust. As these defective regions reorganize, the system approaches a near-perfect crystal. Stronger nonreciprocity, in contrast, can destabilize even high-symmetry structures and promotes cluster breakage.
\\

This selective stabilization is reminiscent of order-by-disorder phenomena and dissipative engineering, where low-symmetry configurations may be preferentially removed to enhance overall order. A quantitative understanding of this mechanism, particularly how collective effects from particles outside the local neighborhood contribute to \ {error correction}, remains an interesting direction for future work.

%%%%%%%%%%%%%%%%%%%%%%%%%%%%%%%%%%%

\vfill
\clearpage
\section{Topological landscape}\label{sec5}

We take advantage of the topological metric space of the flip graph (SI~Sec.~\ref{sec4}) to elucidate underlying differences in the spread across motif compositions of the probability distributions, by constructing topological landscapes.

\subsection{Defining a low-dimensional manifold from topological flipgraph via multidimensional scaling}\label{sec5:1}

 We compute a two-dimensional embedding of the metric space using multidimensional scaling (MDS) \cite{borg_modern_2005} that forms the foundation ($xy$ plane) for all our landscapes. Here, each motif $j$ corresponds to a single point ($x_j$,$y_j$) in the foundation. The height $z_j$ in the landscape is given by the observed motif probability $p(j)$.  We use classical MDS on the pairwise distance matrix $D$ (with $D_{ij} = d(i,j)$) between topological motifs, determined from their graph distances. This two-dimensional embedding method is chosen to best preserve pairwise distances \cite{borg_modern_2005} and serves our purpose of data visualization. \\

\ {Multidimensional scaling (MDS) aims to find a lower $m$-dimensional representation of a higher $k$-dimensional space ($m < k$). It takes as input the pairwise distances between coordinates in the higher dimension, $d_{ij}$, for pair $(i,j)$, in matrix $D = [d_{ij}]$ of size $k \times k$. The procedure then outputs a coordinate matrix $X = [\mathbf{x_i}]$ of size $k \times m$ ($i \in [1,k]$) that lists the coordinates of the original $k$ points in their lower $m$-dimensional representation. Depending on the goal, $X$ is determined by a particular optimization of the resulting pairwise distances in the lower dimensional representation $\tilde{d}(X)$. For example, $X$ can be found such that $\tilde{d}(X)$ reflects the ordinal order of pairwise distances in the higher dimensional space (ordinal MDS) or the metric nature of the pairwise distance (classical MDS and other metric MDS methods) \cite{borg_modern_2005}. We used classical multidimensional scaling, also known as principal coordinates analysis, to estimate a low-dimensional representation of the motif coordinates based on their flipgraph distances in order to best preserve pairwise distances. \\ }

\subsubsection{\ {Classical Multidimensional Scaling Procedure}}
\label{sec:classical_MDS}

\ {
Classical MDS finds the coordinate matrix $X$ that optimizes the following strain loss function in a least-squares sense:
\begin{equation}
    Strain(\mathbf{x}_1,\mathbf{x}_2,...,\mathbf{x}_n) = \sqrt{\frac{\sum_{i,j} \left(g_{ij}^2 - \mathbf{x}^T_i\mathbf{x}_j\right)^2}{\sum_{i,j}g_{ij}^2}}. 
    \label{eq:strain}
\end{equation}
Here, $g_{ij}$ is the $(i,j)$ entry of $G$ (related to the pairwise distance $d_{ij}$ between motifs $i$ and $j$ on the $k$-dimensional flipgraph, Equation \ref{eq:g_ij}) and $\mathbf{x}_j$ and $\mathbf{x}_i$ are their $m$-dimensional representations (i.e. the $j^{th}$ and $i^{th}$ column of $X$, respectively). \\}

\ {The coordinates matrix $X$ in classical MDS can be derived as follows. First, assume that there exists a coordinate matrix $X$ that is size $k \times N$, where $N$ is a sufficiently high dimension that embeds all $k$ coordinates in an $\mathbb{R}^N$ Euclidean vector space.\footnote{Note that the rank of $X$ must be less than or equal to $\textrm{min}(N,k)$} The Gram matrix, $G  = X X^T$ can be constructed from the coordinate matrix $X$ and is related to the $k \times k$ pairwise distance matrix $D$ via:
\begin{equation}
    D^{(2)} = \mathbf{c} \mathbf{1}^T + \mathbf{1} \mathbf{c}^T - 2 G,
    \label{eq:highdim_inner_product}
\end{equation}
where $D^{(2)} = [d_{ij}^2]$, $\mathbf{1}$ is a $k \times 1$ vector of ones, and $\mathbf{c}$ is the vector of the diagonal elements of $G$ \cite{lin_multidimensional_2024,borg_modern_2005}. This is a high-dimensional generalization of the identity $||\mathbf{u}-\mathbf{v}||^2 = ||\mathbf{u}||^2 + ||\mathbf{v}||^2 - 2 \mathbf{u} \cdot \mathbf{v}$ for two vectors $\mathbf{u}$ and $\mathbf{v}$ in a Euclidean vector space $\mathbb{R}^N$ \cite{borg_modern_2005,lin_multidimensional_2024}. Now to solve for $X$, we first need to find $G$ from $D$. It can be shown from Equation \ref{eq:highdim_inner_product} that $G$ is given by \cite{borg_modern_2005}:
\begin{equation}
    G = -\frac{1}{2} C D^{(2)}C,
\end{equation} 
where $C$ is the centering matrix $C = I_k - \frac{1}{k}J_k$ where $J_k$ is the $k \times k$ matrix of all ones and $I_k$ is the $k \times k$ identity matrix. For each $(i,j)$ pair, 
\begin{equation}
    g_{ij} = -\frac{1}{2}\left( d_{ij}^2 - \frac{1}{k}\sum_l d_{lj}^2 - \frac{1}{k}\sum_l d_{il}^2 + \frac{1}{k^2} \sum_{l,m} d_{lm}^2 \right).
\label{eq:g_ij}
\end{equation} }

\ {Using the fact that $G$ is real, symmetric and positive semi-definite, it can be written in the eigendecomposition:
\begin{equation}
    G = U \Lambda U^*,
\end{equation}
where $U$ is the unitary matrix of its eigenvectors, $U^*$ is its conjugate transpose, and $\Lambda$ is the diagonal matrix of its eigenvalues. \\}

\ {Let $X = U \Lambda^{1/2}$ with $\Lambda^{1/2} = [\lambda_i^{1/2}]$, the diagonal matrix composed of the square root of the eigenvalues $\lambda_i$ of $G$. It can be shown that:
\begin{equation}
    G = U \Lambda U^* = U \Lambda^{1/2} \Lambda^{1/2} U^*= (U \Lambda^{1/2})((\Lambda^{1/2})^*U^*)= (U \Lambda^{1/2})(U \Lambda^{1/2})^* = X X^T.
\end{equation}} \ {Thus, $X = U \Lambda^{1/2}$ a $k \times k$ coordinate matrix that derives from the $k$-dimensional pairwise distance $D$. Thus far, all expressions have been equalities. Low-dimensionalization is introduced by selecting the first $m$ eigenvalues and eigenvectors of $G$ to construct $X$, so that $X_m = U_m \Lambda^{1/2}_m \approx X$. Since $G$ is positive semi-definite, its eigenvalues are also its singular values. Thus, $X_m = U_m \Lambda^{1/2}_m$ is the rank-$m$ representation of the higher dimensional coordinates matrix $X$. It corresponds to the least-squares solution that minimizes the strain (Equation \ref{eq:strain}) for $m$ of $k$ dimensions. \\}

\begin{figure}[htbp]
    \centering
    \includegraphics[]{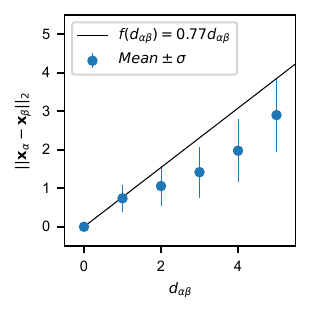}
   \caption{\textbf{Larger flip graph distances correlate with larger distances in the MDS embedding}. Shepard plot comparing the pairwise distances between motifs in the MDS embedded space, $||\mathbf{x}_\alpha - \mathbf{x}_\beta||_2$, to the true distances, $d_{\alpha \beta}$, on the flip graph for the 2-dimensional embedding. On average, larger distances in the MDS space correspond to larger distances on the flip graph, and this trend is monotonic. Errorbars correspond to one standard deviation from the mean. }
    \label{fig:Sheppard_plot_MDS_main}
\end{figure}

\subsubsection{\ {Classical MDS Embedding of the Flipgraph}}

\ {To construct the probability landscapes, we perform classical MDS in Julia using the \texttt{MultivariateStats} package \cite{lin_multidimensional_2024} to find a two-dimensional representation ($m = 2$) of the motif coordinates based on their flipgraph distances.} On average, increasing distances in the flipgraph correspond to increasing distances in the embedded space and follow a monotonic trend (Fig. \ref{fig:Sheppard_plot_MDS_main}). %In practice, graph distances are rarely fully Euclidean. We chose classical MDS over other forms of metric MDS because our distance measure itself is metric (number of T1 transitions) and physically meaningful.

\begin{figure}[htbp]
    \centering
    \includegraphics[]{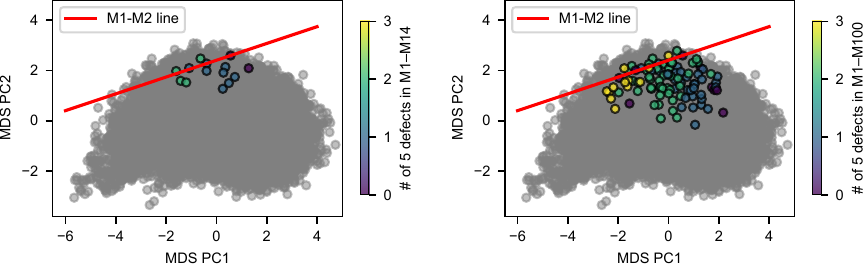}
    %\hspace{10pt}
    %\includegraphics[width=0.4\linewidth]{FigS20_Y_MDS_colored_by_5_defects_M1toM100_modified.png}
    \caption{\ {\textbf{The M$_1$-M$_2$ slices defines an emergent order parameter- the number of 5-defects in a motif.} Left: The M$_1$-M$_2$ line (shown in red) sorts the top 14 topological motifs by the number of 5 defects contained in that motif. Right: This sorting persists for the top 100 motifs, as well. Gray indicates less probable motifs (those that have rank higher than 100). }}
    \label{fig:SX_emergent_order}
\end{figure}

\subsubsection{\ {Emergent order parameter in the low-dimensional MDS space}}

\ {In this subsection we demonstrate the emergence of a physical order parameter $M$ along the M$_1$-M$_2$ line in MDS space. As Figure \ref{fig:SX_emergent_order} shows, this 1-dimensional line effectively sorts the motifs by the number of 5 defects contained in that motif.} \\

\subsection{Mapping frequency of motifs to probabilities on the manifold via kernel density estimation}\label{sec5:2}

To transform the discrete probability landscape into a smooth landscape, we use kernel density estimation \cite{bowman1997applied}. At each motif coordinate, we place a Gaussian kernel, weighted by its observed probability. The estimator for the density is given by:
\begin{equation}
    \hat{f}_{b}(x,y) = \frac{1}{N} \sum_{j = 1}^N K_{b}(x - x_j,y - y_j) = \frac{1}{N} \sum_{j = 1}^N p(j) \exp{\left(-\frac{r(j)^2}{b^2} \right)},
\end{equation}
where $p(j)$ is the probability of motif $j$ ($j \in [1,N]$), $r(j) = \sqrt{(x-x_j)^2 + (y-y_j)^2}$, and $b$ is the smoothing bandwidth. Adapting the rule of thumb by Bowman and Azzalini \cite{bowman1997applied}, we use the following bandwidth for the Gaussian kernel:
\begin{equation}
    b = \left(\frac{4}{N(d+2)}\right)^{1/(d+4)}\frac{\tilde{\sigma}}{0.6745}
    \label{eqn:kernel}
\end{equation} The term $\tilde{\sigma}$ is the mean absolute deviation estimator, which here we approximate by $\tilde{\sigma} = \sqrt{\left(\overline{|x_j - \overline{x_j}|}\right)^2 + \left(\overline{|y_j - \overline{y_j}|}\right)^2}$ where $\overline{(\cdot)}$ represents the mean with respect to the MDS data (the motifs). \\

Kernel density estimates were computed on grids with a resolution much lower than the average distance between points (128$\times$128 grid with $x$ and $y$ each ranging from [-4,4] for a bin size of 0.0625$\times$0.0625). For 1-dimensional landscapes along $\mathrm{M}_1$-$\mathrm{M}_2$ axis, all probability distributions are discretized with a bin size of $0.0101$, corresponding to $1\%$ of the $\mathrm{M_1}$--$\mathrm{M_2}$ distance.

\subsection{Atlas of topological motifs in simulation and experiment}\label{sec5:3}

\begin{figure}[!ht]
  \centering
\includegraphics[width=\linewidth]{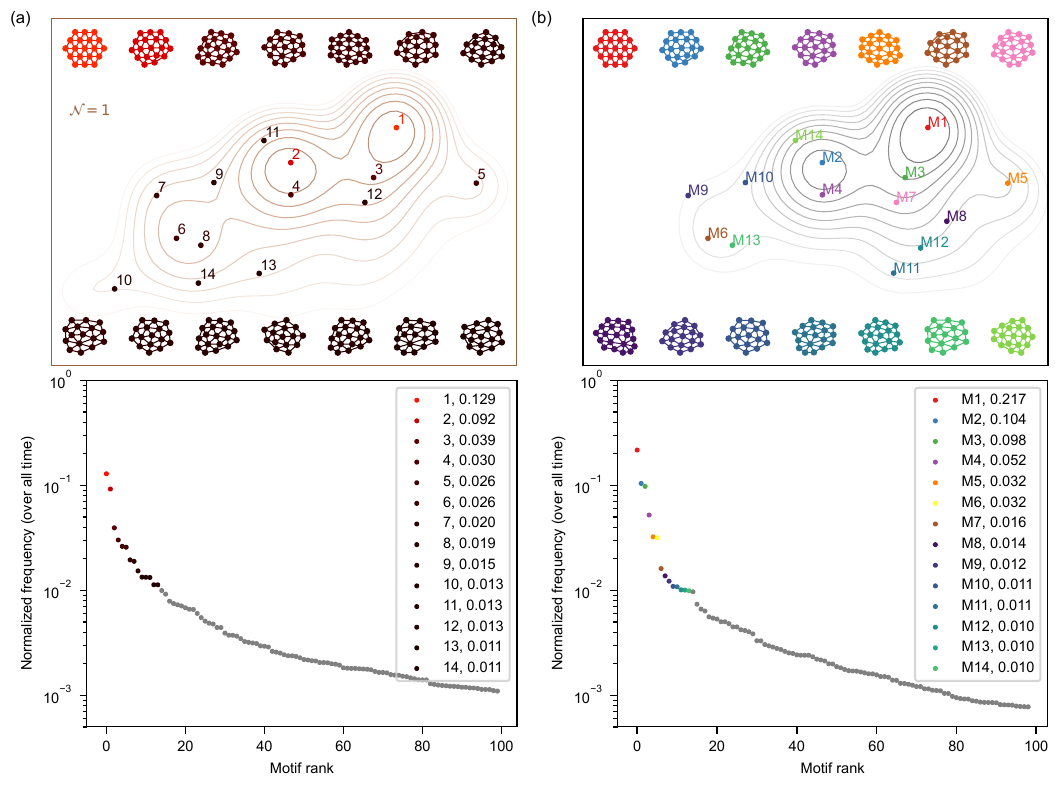}
    \caption{\textbf{Atlas of topological motifs from simulation and experiment } \panel{A} \emph{Top:} Atlas of motifs from simulation with inferred nonreciprocity $\mathcal{N}=1$. This contains identical information with Fig. 3c, but is enlarged with labels up to the 14th most observed motif. \emph{Bottom:} Normalized motif frequencies over all time in model with $\mathcal{N} = 1$.
    \panel{B} \emph{Top:} Atlas of motifs from experiment. Motifs up to $\textrm{M}_6$ (6th most observed motif) are identical to \textbf{A}.
    \emph{Bottom:} Normalized motif frequencies over all time in experiment.
    }
  \label{figSI:MotifAtlas}
\end{figure}

In the main text, we show the 14 motifs that have frequency over 1\% in simulations with $\mathcal{N}=1$. Here we show the same motif atlas with full labels on the low-dimensional manifold~(Fig.~\ref{figSI:MotifAtlas}a). 
\\

In addition, we show the 14 most-observed motifs in the experiment~(Fig.~\ref{figSI:MotifAtlas}b). We label these top 14 motifs as $\mathrm{M}_1$ to $\mathrm{M}_{14}$ in the order of decreasing frequencies. Notably, the top 6 motifs are identical to the top 6 motifs observed in the simulation with $\mathcal{N}=1$; thus $\mathrm{M}_1$ and $\mathrm{M}_2$ found in $\mathcal{N}=1$ are consistent with the top two motifs defined based on the frequency in the experiment. 

\subsection{Spatiotemporal distributions of topological motifs $\mathrm{M_1}$ and $\mathrm{M_2}$}\label{sec5:4}

$\mathrm{M_1}$ and $\mathrm{M_2}$ are the two most dominant topological motifs in both experiment and simulation with experimentally inferred $\mathcal{N}=1$. 
Characterizing their spatial distributions may provide insight into how local structure relates to global organization in nonreciprocal mixtures. 
\\

\begin{figure}[!ht]
  \centering
  \small
\includegraphics[width=\linewidth]{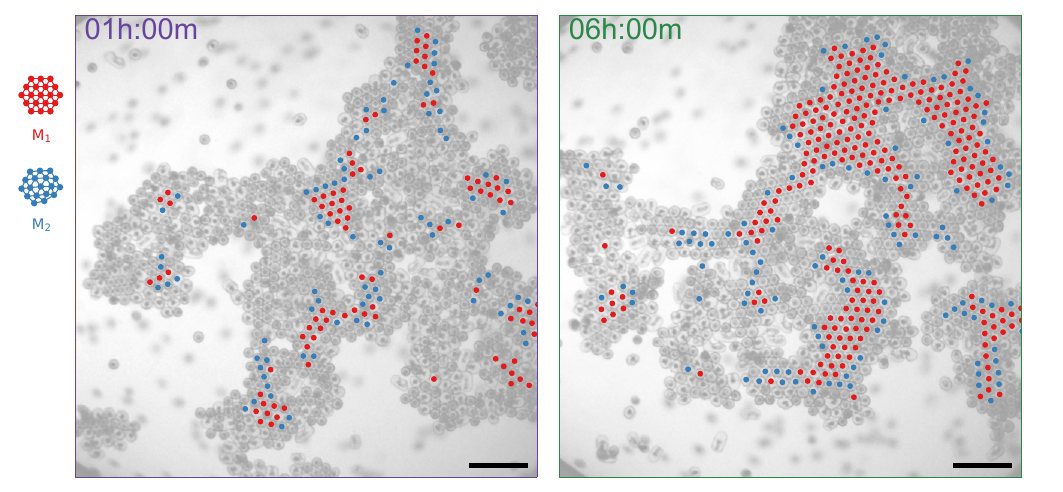}
  \caption{\textbf{Spatial distributions of topological motifs M$_1$ and M$_2$.} $\mathrm{M_1}$, a perfect crystal, is labeled as red, while $\mathrm{M_2}$, an almost-perfect crystal with a 5-fold defect, is labeled as blue. Compared to early time, the late-time snapshot from the experiment shows many more $\mathrm{M_1}$ motifs. }
  \label{figSI:motiflocation}
\end{figure}

As shown in Fig.~\ref{figSI:motiflocation} and SI Videos 9 and 11, $\mathrm{M_1}$ motifs are primarily located in the crystalline cores of clusters, whereas $\mathrm{M_2}$ motifs appear preferentially at their peripheries. 
Consequently, the relative abundance of these motifs reflects the system's global organization: 
large, well-ordered crystals are dominated by $\mathrm{M_1}$, while fragmented or loosely packed configurations exhibit an increased fraction of $\mathrm{M_2}$. 
\\

\begin{figure}[!ht]
    \centering
    \includegraphics[]{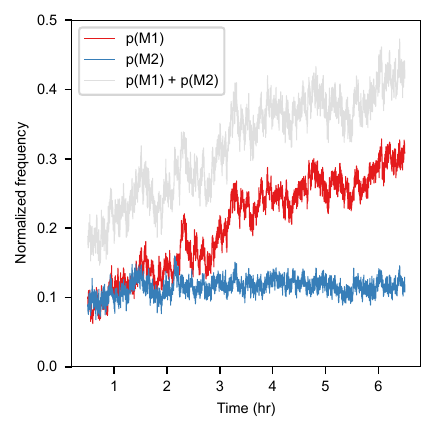}
    \caption{\textbf{Probability of M$_1$ and M$_2$ motifs over time in main experiment.} The transition at $\sim 3$ hr coincides with an increase in the probability of $\textrm{M}_1$.}
    \label{fig:M1M2_prob_v_time_exp}
\end{figure}

In addition to the spatial distribution, we also elucidate here how the frequencies of $\mathrm{M}_1$ and $\mathrm{M}_2$ motifs change over time. Over time, $\mathrm{M}_1$ increases in its proportion in the experiment while the frequency of $\mathrm{M}_2$ stays constant (Fig. \ref{fig:M1M2_prob_v_time_exp}).

\subsection{Transition between $\mathrm{M}_1$ and $\mathrm{M}_2$ in experiment is indirect}\label{sec5:5}

In the main text, we find that the structural transition near nonreciprocity $\mathcal{N}=1$ is a first-order-like transition that involves a coexistence state between $\mathrm{M}_1$ and $\mathrm{M}_2$. Since many features of phase transitions only become meaningful in thermodynamic limit, this analogy between the nonreciprocal transition in structure and first-order phase transitions does not imply that the nonreciprocal transition \emph{is} a first-order phase transition. Instead, by connecting the complex biological system to a well-established theoretical framework, the analogy opens new avenues of testable predictions. Here, we will demonstrate one example of these predictions and its validation.
\\

\begin{figure}[!ht]
  \centering
  \small
\includegraphics[width=\linewidth]{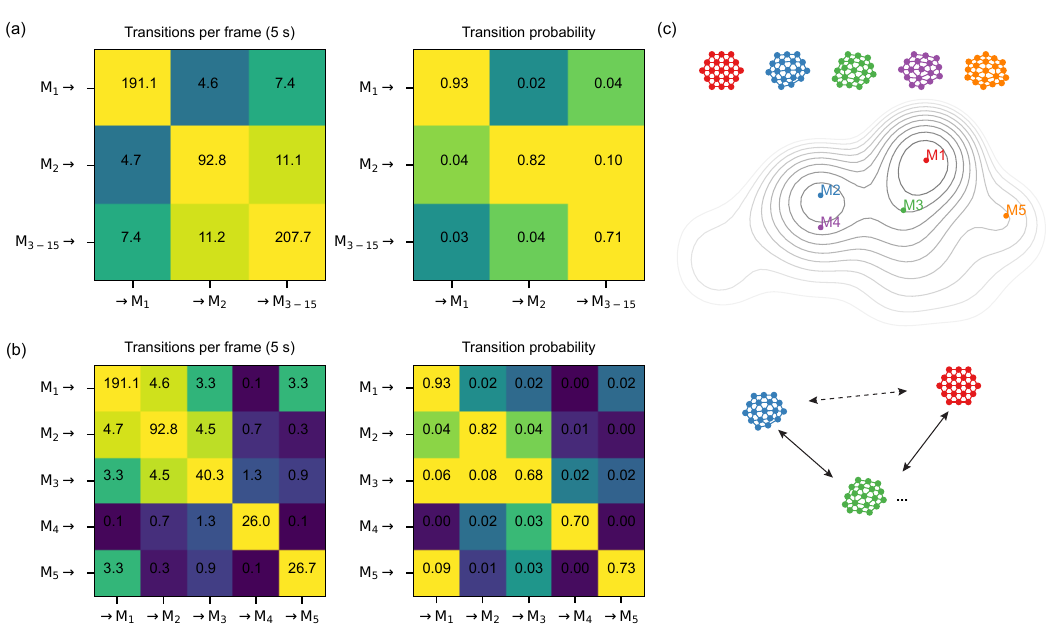}
    \caption{\textbf{Transition between $\textrm{M}_1$ and $\textrm{M}_2$ in experiment is indirect. } \panel{A} Transition between $\mathrm{M_1}$ and $\mathrm{M_2}$ is often mediated by other motifs, even though the direct transition is allowed. \panel{B} $\mathrm{M_3}$ and $\mathrm{M_5}$ are major intermediate motifs that mediate the transition between $\mathrm{M_1}$ and $\mathrm{M_2}$. \panel{C} Top: Shapes of motifs $\mathrm{M_1}$-$\mathrm{M_5}$ and their locations in low-dimensional manifold. Bottom: A cartoon of transition between motifs. }
  \label{figSI:M1M2flux}
\end{figure}

The coexistence state of $\mathrm{M}_1$ and $\mathrm{M}_2$ imply that the two configurations are separated by a set of configurations that are less likely than either. While the two motifs are in fact apart by a single T1 transition, we find that the transition between two motifs is more often indirect (mediated by other motifs) than direct~(Fig.~\ref{figSI:M1M2flux}a). In particular, despite its lower frequency, $\mathrm{M}_3$ exchanges significant probability flux with both $\mathrm{M}_1$ and $\mathrm{M}_2$ that is comparable to the direct flux between $\mathrm{M}_1$ and $\mathrm{M}_2$. 

\ {
\subsection{\ {Nonreciprocal hysteresis}}\label{sec5:6}
As shown in the previous subsection SI.~\ref{sec5:5}, the coexistence state of $\mathrm{M}_1$ and $\mathrm{M}_2$ in nonreciprocity-driven structural transitions, revealed by topological landscapes in the main text, provides many testable hypotheses that help us better understand our system. Here we verify yet another testable hypothesis stemming from this analogy to first-order phase transitions, which we call nonreciprocal hysteresis. \\\\
As shown in Fig.~2c, a strong nonreciprocity leads to a steep increase in the topological order parameter $\langle d_{\textrm{hex}} \rangle$, which quantifies how far the local network topology is from a perfect crystal with hexatic symmetry. The analogy to first-order phase transition raises a testable hypothesis: does the $\langle d_{\textrm{hex}} \rangle$ exhibit a hysteresis loop if we slowly change nonreciprocity $\mathcal{N}$? In other words, do self-organized structures `heated' from a weak $\mathcal{N}$ break at nonreciprocity stronger than $\mathcal{N}=1$, and similarly structures `cooled' from a strong $\mathcal{N}$ stabilize at nonreciprocity weaker than $\mathcal{N}=1$? \\\\
To test this idea, we performed numerical simulations where nonreciprocity $\mathcal{N}$ is slowly increased ($\mathcal{N}=0.75$ to $\mathcal{N}=1.25$) or slowly decreased ($\mathcal{N}=1.25$ to $\mathcal{N}=0.75$), Fig.~3e and Supplementary Video 12. In either direction, $\mathcal{N}$ is changed by $0.05$ every 3000 frames (the full timescale of original simulations with fixed $\mathcal{N}$) to ensure that $\mathcal{N}$ varies slowly.\\} 

\ {That is, in addition to the trend of $\langle d_{\textrm{hex}}\rangle$ from simulations initialized at each $\mathcal{N}$ (black), which we also show in Fig.~2c, Fig.~3e shows two new curves from simulations under time-varying $\mathcal{N}$ that either slowly increases from 0.75 (red) or slowly decreases from 1.25 (blue). Every 3000 frames, $\mathcal{N}$ changes by $0.05$. As in the main text simulations in Fig.~2c, we analyze the local network topologies in the steady state (the last 1500 frames at each $\mathcal{N}$) to remove the effect of transients. Error bars correspond to the SEM ($n = 20$). The deviation of red and blue curves from the black curve demonstrates nonreciprocal hysteresis, in which the self-organized structure depends on the history of the nonreciprocity.}
%\begin{figure}[!ht]
%  \centering
%  \small
%\includegraphics[width=\linewidth]{Sx_hysteresis.pdf}
%    \caption{\ {\textbf{Nonreciprocal hysteresis.} In addition to the trend of $\langle d_{\textrm{hex}}\rangle$ from simulations initialized at each $\mathcal{N}$ (black), which we also show in the main text, here we show two new curves from simulations under time-varying $\mathcal{N}$ that either slowly increases from 0.75 (red) or slowly decreases from 1.25 (blue). Every 3000 frames, $\mathcal{N}$ changes by $0.05$. As in the main text simulations, we analyze the local network topologies in the steady state (the last 1500 frames at each $\mathcal{N}$) to remove the effect of transients. Error bars correspond to the SEM ($n = 20$). The deviation of red and blue curves from the black curve demonstrates nonreciprocal hysteresis, in which the self-organized structure depends on the history of the nonreciprocity. }}
%  \label{figSI:hysteresis}
%\end{figure}

\vfill
\clearpage
\section{Topological flowscape}\label{sec6}

Before presenting the technical details of our flowscape construction~(SI~Sec.~6.2), we first outline the general concept of a \emph{flowscape}, which is broadly applicable to any series of probability landscapes.
\\

We introduce the flowscape as a framework for visualizing the temporal evolution of high-dimensional structural configurations in a reduced and interpretable space. Specifically, we consider a time series of probability distributions $P(X; t)$ defined over a discrete set of states $x_i$, with corresponding probabilities $p_i(t) \equiv P(x_i; t)$, satisfying $\sum_i p_i(t) = 1$. In the main text (Fig.~4), $x_i$ corresponds to coordinates in a coarse-grained, low-dimensional manifold constructed from topological motif distributions (topological landscapes), and $t$ denotes experimental time. The flowscape embeds these evolving distributions into a reference-based information space, enabling intuitive visualization of time-dependent landscapes.\\

To define the flowscape, we select one or more reference distributions $\{Q^a(X)\}$, each represented as a discrete probability vector $\{q^a_i\}$. These references may correspond to canonical structural states (e.g., $\mathrm{M_1}$-like, $\mathrm{M_2}$-like, initial state, or final state), serving as anchors against which the current system state is compared. \\

For each time point $t$, we compute the Kullback-Leibler (KL) divergence from the instantaneous distribution $P(t)$ to each reference $Q^a$:

\begin{equation}
    D_{\mathrm{KL}}(P(t)\|Q^a) = \sum_i p_i(t) \log\left( \frac{p_i(t)}{q^a_i} \right)
    \label{eqn:kld}
\end{equation}

We then use the KL divergence from each reference to create an embedding of each distribution $P(t)$ as a point in $\mathbb{R}^n$, where $n$ is the number of reference states. For instance, in Fig.~4c of the main text, we embed the evolving state as a trajectory in a 2D space defined by divergences to two reference distributions. Each coordinate quantifies how structurally dissimilar the current state is from a corresponding reference. In this paper, since $P$---the landscape of topological motifs---represents a state of self-organized structure, each flowscape point reflects the instantaneous structural dissimilarity from designated reference structural state.

\subsection{Conceptual basis of the flowscape}\label{sec6:1}

Before addressing the specific construction of the topological flowscape used in our system, we discuss general features of the framework that apply to any evolving probability distribution. \\

\paragraph{Choice of divergence metric.}
We primarily use Kullback-Leibler (KL) divergence due to its interpretability and strong foundation in information theory. However, the conceptual framework of the flowscape does not depend on this specific choice. Alternative measures, including the symmetric Jensen-Shannon divergence and Wasserstein (earth mover's) distance, can also be used, each with different sensitivities to distributional changes. For instance, in SI Sec.~\ref{sec7:TEM}, we construct a variant of the flowscape using the topological earth mover (TEM) distance. \\ 

\paragraph{KL asymmetry and interpretation.}
KL divergence is not symmetric, i.e., $D_{\mathrm{KL}}(P\|Q) \ne D_{\mathrm{KL}}(Q\|P)$. In our case, the references correspond to well-defined structural states, and the KL divergence measures how much information is ``lost'' when approximating the current state with that reference. \\ 

%\paragraph{Flowscape vs. Motif Frequency Embedding.}
%It's important to distinguish between raw motif frequency embeddings (e.g., PCA over frequency vectors) and flowscape embeddings. While both provide dimensionality reduction, flowscape coordinates take into account of relative distances between motifs, through their spatial distributions in a low-dimensional embedding. 

%\paragraph{Flowscape and information bottleneck.}
%Although technically distinct, the flowscape construction shares conceptual similarity with the information bottleneck framework, in which a representation is optimized to preserve relevant information with respect to a target variable. In flowscapes, each coordinate projects the system by encoding it based on a particular reference state, being agnostic to deviation that is orthogonal to the reference.

\paragraph{Flowscape and information geometry.}
Flowscapes do not explicitly encode information flow or entropy production, as they are constructed solely from snapshots of probability distributions without requiring transition rates. Nevertheless, KL divergences can bound or approximate informational quantities, drawing conceptual parallels to results in information geometry and stochastic thermodynamics. These connections suggest that flowscape trajectories may carry latent thermodynamic structure, even when derived from static distributions. In SI Sec.~\ref{sec7:infolength}, we illustrate this link by analyzing information rate transitions, demonstrating how flowscapes can reflect underlying dynamical constraints.

%\paragraph{Illustrative examples.}
%To build intuition, consider two flowscape trajectories starting from the same state but undergoing different transitions, as shown in Fig.~\ref{figSI:FlowscapeCartoon}. If mass in the probability distribution shifts ``horizontally,'' this may generate a different flowscape trajectory than a ``vertical'' shift, depending on the structure of the reference distributions.

\subsection{Details of topological flowscape introduced in the main text}\label{sec6:2}

In Fig.~4c of the main text, we construct a flowscape using two reference distributions chosen to highlight the transition between the self-organized states. Specifically, the references are defined as one-dimensional Gaussian distributions along the $\mathrm{M_1}$--$\mathrm{M_2}$ axis:
\[
Q^x = \text{Normal}(\mathrm{M_2}, \sigma), \quad
Q^y = \text{Normal}(\mathrm{M_1}, \sigma),
\]
where the 1-dimensional time-dependent distributions $P(t)$ are taken from Fig.~4b. The width $\sigma$ of each reference is set to the time-averaged kernel width of the experimental landscape, obtained via Eq.~(\ref{eqn:kernel}).\\

This construction reflects a stepwise simplification of the complex structural landscape toward an interpretable representation. While a flowscape could, in principle, be constructed using two-dimensional distributions over the low-dimensional manifold---or even higher---dimensional distributions retaining additional manifold directions, our 1D construction specifically emphasizes transitions along the $\mathrm{M_1}$--$\mathrm{M_2}$ axis, primarily visualizing transitions between these two structural states with less emphasis on dynamics far from this axis.\\

For the KL divergence calculation, all probability distributions are discretized with a bin size of $0.0101$ in the low-dimensional coordinate, corresponding to $1\%$ of the $\mathrm{M_1}$--$\mathrm{M_2}$ distance. The reference width $\sigma = 0.0606$ matches the kernel width of the experimental landscape. All distributions are normalized to satisfy $\sum_i p_i = 1$, and the flowscape coordinates (KL divergences) are then evaluated using Eq.~(\ref{eqn:kld}). %Changing the bin size primarily rescales the flowscape coordinates, as the apparent distinguishability of discrete distributions depends on binning; however, qualitative trends---such as the rate of transitions and the directional turn at the traveling--fluctuating transition---remain robust.

\subsection{Flowscape of inference-based model over the change in nonreciprocity}\label{sec6:3}

Our flowscape in the main text uses topological landscapes from experiment over time. Here, we show flowscape of topological landscapes from model simulations over nonreciprocity $\mathcal{N}$. 

\begin{figure}[!ht]
  \centering
  \small
\includegraphics[width=\linewidth]{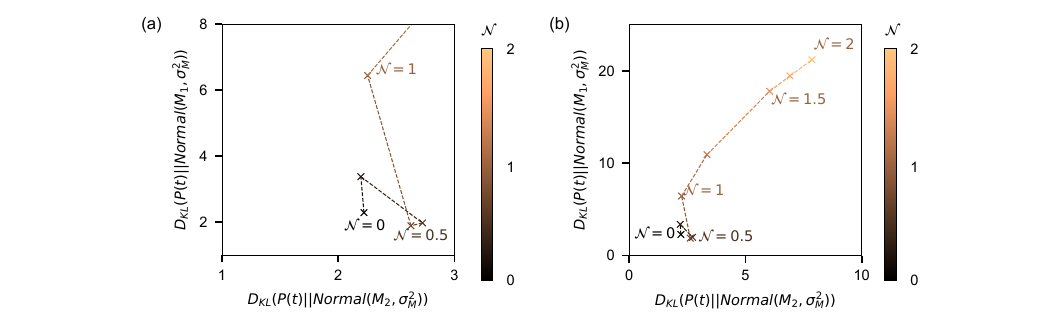}
    \caption{\textbf{Flowscape of inference-based model over the change in nonreciprocity. } \panel{A} Flowscape of inference-based model, with the same x- and y- axes ranges as in the main text Fig. 4c. The trajectory passes through the lower-right corner unlike the experimental flowscape, as the intermediate $\mathcal{N}=\{0.5,\ 0.75\}$ bring the system closest to a perfect crystal. \panel{B} The same flowscape with larger ranges of x- and y- axes. As nonreciprocity exceeds $\mathcal{N}=1$, the system moves away from both $\mathrm{M}_1$ and $\mathrm{M}_2$. }
  \label{figSI:NRflowscape}
\end{figure}

The flowscape of the inference-based model recapitulates how nonreciprocity leads to different self-organized structures~(Fig.~\ref{figSI:NRflowscape}). First, at weak nonreciprocity $\mathcal{N}<1$, the trajectory traverses into the lower-right corner of the flowscape where it is closest to $\mathrm{M}_1$, as nonreciprocity anneals the crystal and removes defects~(Fig.~\ref{figSI:NRflowscape}a). Second, at strong nonreciprocity $\mathcal{N}>1$, the trajectory moves away from both axes, indicating a strong deviation from a crystalline structure as the system becomes more fragmented~(Fig.~\ref{figSI:NRflowscape}b).

%As in the experimental flowscape, this model flowscape also reveals detailed features of the transition in self-organized structures via the quantification. 
% N=1 is imbalanced, N=1.5, 1.75, 2.0 are still different.

\subsection{Characterization of the transition state}\label{sec6:4}

Unlike the flowscape of model simulations where $\mathcal{N}$ is varied, the experimental flowscape reveals a transition trajectory from $\mathcal{N}=1$ to $\mathcal{N}=0$ that passes through states located in the lower-left corner of the flowscape, near both $\mathrm{M_1}$ and $\mathrm{M_2}$. These ``transition states'' do not correspond to a steady state in model simulations under fixed $\mathcal{N}$.\\

\begin{figure}[!ht]
  \centering
  \includegraphics[width=\linewidth]{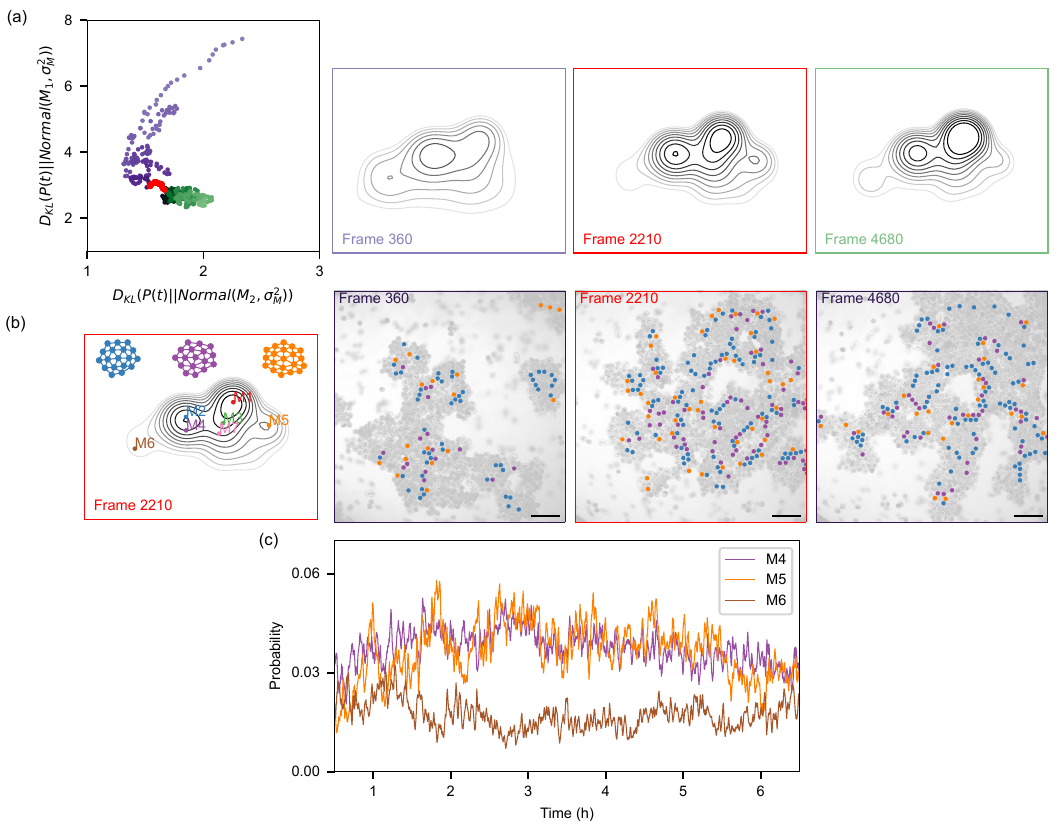}
  \caption{\textbf{Traveling--fluctuating transition proceeds through hollow crystals.} 
  \panel{A} Flowscape trajectory indicates that the transition between $\mathcal{N}=1$-like and $\mathcal{N}=0$-like regimes passes through a distinct intermediate state around Frame 2210 ($\sim 3$\,h), marked by the appearance of a third peak.
  \panel{B} This peak corresponds to motif $\mathrm{M_5}$, which contains a 7-fold defect and whose frequency maximizes at the transition. 
  Motifs $\mathrm{M_2}$ (5-defect), $\mathrm{M_4}$ (5--7 defect pair), and $\mathrm{M_5}$ (7-defect) organize into defect lines surrounding voids, producing a hollow crystal not observed in simulations.
  \panel{C} The frequencies of $\mathrm{M_4}$ and $\mathrm{M_5}$ peak at intermediate times, whereas other motifs, such as $\mathrm{M_6}$, do not follow this trend.
  }
  \label{figSI:CornerState}
\end{figure}

Inspection of the topological landscape at the transition (e.g., Frame 2210) shows the emergence of a third peak at $\mathrm{M_5}$, corresponding to a motif containing a 7-fold defect~(Fig.~\ref{figSI:CornerState}). The spatial distribution of motifs further reveals that $\mathrm{M_2}$ (5-defect), $\mathrm{M_4}$ (5-7 defect pair), and $\mathrm{M_5}$ (7-defect) form extended lines of defects as the embryo mixture transitions from the traveling to the fluctuating state. Consistently, the frequencies of $\mathrm{M_4}$ and $\mathrm{M_5}$ peak at intermediate times, whereas other motifs, such as $\mathrm{M_6}$, do not follow this trend.\\

Based on these observations, we interpret the transition state as a \emph{hollow crystal}. During the transition, multiple clusters merge gradually rather than instantaneously. When two clusters meet, they initially leave an empty void between them, which fills only slowly over time. This contrasts with the model simulations, where steady states either form a nearly perfect crystal or exhibit zipper-like fractures and continuous line mergers, but do not produce hollow clusters.

\subsection{Motif cycles in cluster merging and fragmentations}\label{sec6:5}

In the main text Fig.~4d, we show that flowscape admits a versatile choice of reference states. In particular, by taking experimental snapshots from different timepoints as references, we can use flowscape as a temporal microscope that can zoom into small-scale, rapid changes in structures. \\

\begin{figure}[!ht]
  \centering
  \small
\includegraphics[width=\linewidth]{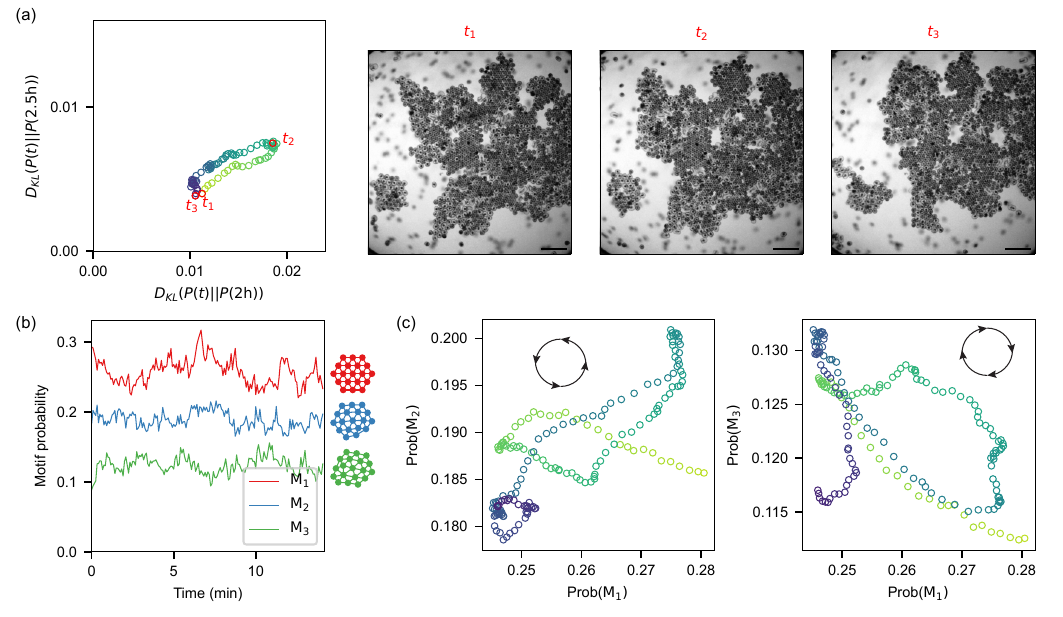}
    \caption{\textbf{Motif cycles in cluster merging and fragmentation.} \panel{A} \emph{Left}: a portion of the flowscape shown in main Fig.~4d, focusing on a single cycle. \emph{Right}: Snapshot images of experiment at early, middle, and late points of the cycle. \panel{B} Time series of probabilities of three dominant motifs $\mathrm{M}_1$, $\mathrm{M}_2$, and $\mathrm{M}_3$. Shapes of the motifs are shown at the right side of the time series. \panel{C} Trajectory of the probabilities shown in b, averaged over 3-minute time windows. }
  \label{figSI:MotifCycle}
\end{figure}

In our experiment, this fine signature turns out to represent repeated cluster merging and fragmentation. Here, we elaborate on this observation by focusing on a particular cycle in the flowscape with a short (30 minutes) time gap between references~(Fig.~\ref{figSI:MotifCycle}). From experimental snapshots, we find that the cycle corresponds to a merger-and-breaking of multiple clusters, which accompanies the closing-and-opening of gaps between them~(Fig.~\ref{figSI:MotifCycle}a). The time series of frequencies of individual motifs offers a detailed view of this phenomenon, marked by the peak of the crystalline $\mathrm{M}_1$ motif probability at the middle of the cycle~(Fig.~\ref{figSI:MotifCycle}b). \\

This time series of motifs reveals interesting interactions between structures~(Fig.~\ref{figSI:MotifCycle}c). Specifically, the chirality of trajectories in $\mathrm{M}_1$-$\mathrm{M}_2$ space and $\mathrm{M}_1$-$\mathrm{M}_3$ space suggests that an increase in $\mathrm{M}_2$ follows an increase in $\mathrm{M}_1$, while an increase in $\mathrm{M}_3$ precedes an increase in $\mathrm{M}_1$. This cyclic relation also connects to an emergent nonreciprocity: high probability of $\mathrm{M}_1$ leads to an increase in $\mathrm{M}_2$, while high probability of $\mathrm{M}_2$ accompanies a decrease in $\mathrm{M}_1$. Understanding these intricate nonequilbrium dynamics between the topological motifs could be an interesting future direction to follow.

%\subsection{Diagonal displacement recapitulates full-dimensional trajectory of topological landscapes}\label{sec6:4}

\subsection{Topological flowscape of additional experiments}\label{sec6:6}

Here, we apply our conceptual toolbox---topological landscape and topological flowscape---to additional experiments described in SI~Sec.~\ref{sec1:other}, which exhibited two distinct states: a demixed LCC and an orbiting mixture.

\subsubsection{Demixed LCC}\label{sec6:6:1}

\begin{figure}[!ht]
  \centering
\includegraphics[width=\linewidth]{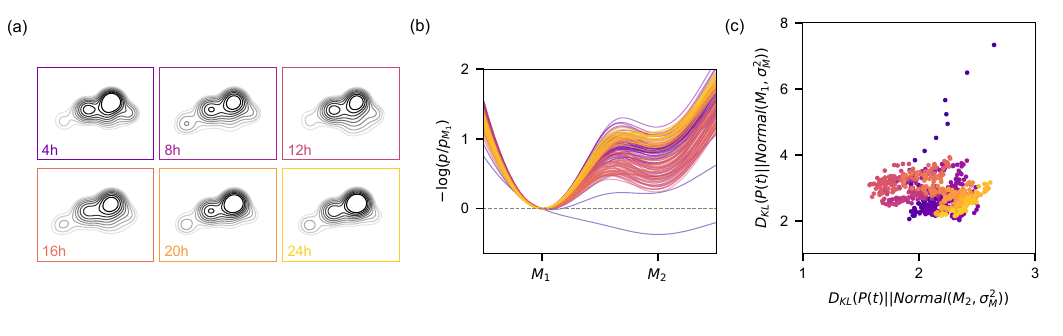}
    \caption{\textbf{Topological landscape and flowscape of a demixed E1-E2 mixture experiment.} 
    \panel{A} Topological landscapes of the demixed experiment over time
    \panel{B} $\textrm{M}_1$-$\textrm{M}_2$ slice of topological landscapes over time. 
    \panel{C} Topological flowscape. The system exits the crystalline state at 16h, followed by a reentrance.}
  \label{figSI:Expm_demixed_2}
\end{figure}

In the demixed LCC experiment where E2 embryos swim to the air-water interface before E1 embryos, the self-organized structure remained almost crystalline. This is recapitulated in Fig.~\ref{figSI:Expm_demixed_2}a: the peak at the crystalline motif $\mathrm{M}_1$ remains in the topological landscapes over time. Nevertheless, Fig.~\ref{figSI:Expm_demixed_2}b suggests that the structure is still dynamic over time. The landscape slices along $\mathrm{M}_1$--$\mathrm{M}_2$ axis reveal that, while $\mathrm{M}_1$ remains the most dominant peak, the $\mathrm{M}_2$ peak grows and shrinks over time. 
\\

The flowscape in Fig.~\ref{figSI:Expm_demixed_2}c resolves the structural dynamics over time. At first, the system quickly develops a crystalline structure, approaching $\mathrm{M}_1$. At around 16 hours, the system drifts away from $\mathrm{M}_1$ and towards $\mathrm{M}_2$, followed by a re-entrance towards $\mathrm{M}_1$ and away from $\mathrm{M}_2$. As shown in SI Video 5, we find that a large-scale crystal rearrangement occurs at this time (16 hours), in which the system almost divides into two crystals and then merges back together. This highlights that this demixed LCC regime still exhibits distinct dynamics from a homogeneous LCC that was previously reported~\cite{tan2022odd}.

\subsubsection{Orbiting mixture}\label{sec6:6:2}

\begin{figure}[!ht]
  \centering
\includegraphics[width=\linewidth]{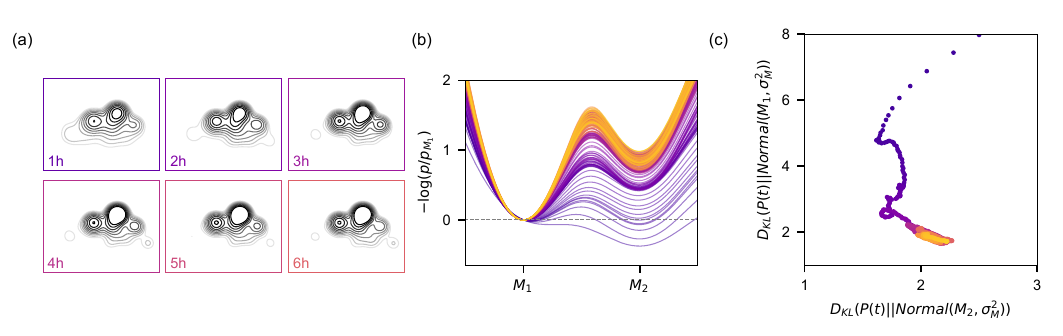}
    \caption{\textbf{Topological landscape and flowscape of an orbiting E1-E2 mixture experiment} \panel{A} Topological landscapes of the experiment over time. There is a pronounced third peak at $\textrm{M}_5$, which contains a single seven-fold defect. 
    \panel{B} $\textrm{M}_1$-$\textrm{M}_2$ slice of topological landscapes over time. 
    \panel{C} Topological flowscape of the experiment over time.  }
  \label{figSI:Expm_windshield_2}
\end{figure}

In the orbiting mixture experiment where the embryos form a boundary-filling crystal, the structure remains mostly crystalline but with significant internal rearrangements. This is recapitulated in Fig.~\ref{figSI:Expm_windshield_2}a: while the peak occurs at the crystalline motif $\mathrm{M}_1$ in the topological landscapes, a third peak other than $\mathrm{M}_2$ emerges at the lower-right regime of the $\mathrm{M}_1$ peak. This third peak corresponds to a topological motif $\mathrm{M}_5$ that is extensively discussed in SI~Sec.~\ref{sec6:4}. This $\mathrm{M}_5$-peak emerges from the presence of holes in the crystalline structure, which are surrounded by $\mathrm{M}_5$ and other defect-containing motifs (see also SI Video 11 for $\textrm{M}_1$-$\textrm{M}_5$ in the main experiment and SI Video 9 for $\textrm{M}_1$-$\textrm{M}_5$ in the model simulations). 
\\

In Fig.~\ref{figSI:Expm_windshield_2}b, landscape slices along the $\mathrm{M}_1$--$\mathrm{M}_2$ axis exhibit a transition that is also first-order-like through $\mathrm{M}_1$--$\mathrm{M}_2$ coexistence. However, in contrast to the main experiment, the $\mathrm{M}_2$-dominance is short-lived as $\mathrm{M}_1$ starts to dominate over $\mathrm{M}_2$ within the first 30 minutes in the experiment.
\\

These structural dynamics are represented on a flowscape~(Fig.~\ref{figSI:Expm_windshield_2}c). We find a rapid initial descent and turn, like in our main experiment. Later, the system shows an interesting oscillation over a same short line segment on the flowscape, over which $\textrm{M}_1$ increases while $\textrm{M}_2$ decreases (and vice versa). 

\vfill
\clearpage
\section{Additional analysis}\label{sec7}

\subsection{Estimation of entropy production rate}\label{sec7:epr}

We observe that the flowscape of the main experiment exhibits a rate transition coinciding with the macroscopic transition in velocity polarization. 
To further quantify this synchrony, we sought an information-theoretic rate estimate that does not rely on the sequential dimensional reduction of the topological landscape and flowscape framework. 
We achieve this by applying a well-known speed limit for stochastic processes. \\

Shiraishi et al.~\cite{shiraishi2018speed} derived a speed limit for classical stochastic Markov processes using the Hatano--Sasa entropy production~\cite{hatano2001steady}, which remains valid for nonequilibrium dynamics with finite stationary currents:

\begin{equation}
    \tau \ge \frac{c^* L(\mathbf{p}(0), \mathbf{p}(\tau))^2 }{2 \Sigma_{HS}\langle A \rangle_\tau}. 
    \label{eqn:speedlimit}
\end{equation}

The timescale $\tau$ sets a speed limit for a transition between two probability distributions $\mathbf{p}(0)$ and $\mathbf{p}(\tau)$ from different times $t=0$ and $t=\tau$, with the net Hatano-Sasa entropy production $\Sigma_{HS}$ and the time-averaged activity $\langle A \rangle \equiv \frac{1}{\tau}\int_0^\tau dt \sum_{i \neq j}W_{ij}(t)p_j(t)$. Here, $c^*=0.896...$ is a numerical constant and $L(\mathbf{p}(0), \mathbf{p}(\tau))\equiv \sum_i |p_i(0)-p_i(\tau)|$ is the statistical distance with the $L^1$ norm. \\

By rearranging this relation, we obtain a lower bound for the Hatano--Sasa entropy production over the time interval $\tau$: 
\begin{equation}
    \Sigma_{HS} \ge \frac{c^* L(\mathbf{p}(0), \mathbf{p}(\tau))^2 }{2 \langle A \rangle_\tau \tau}. 
    \label{eqn:epr}
\end{equation}
Because the Hatano--Sasa entropy production bounds the total entropy production from below, this provides a conservative estimate of the entropy production rate. \\

We used the full motif probability vector $\mathbf{p}$ (length $\sim 15{,}000$) with a time gap $\tau = 20\ \mathrm{s}$ (4 frames) to compute the statistical distance $L(\mathbf{p}(t),\mathbf{p}(t+\tau))$ and activity $\langle A(t) \rangle_\tau $ over time, and used Eq.~(\ref{eqn:epr}) to estimate the Hatano--Sasa entropy production bound $\dot{\Sigma}_{HS}$. 
Importantly, this procedure relies solely on motif frequencies to calculate $\mathbf{p}(t)$ and the activity $\langle A(t) \rangle_\tau $, derived from the transition rates $W_{ij}(t)$ between motifs; it does not require constructing the topological landscape or flowscape. \\

\begin{figure}[!ht]
  \centering
  \small
\includegraphics[]{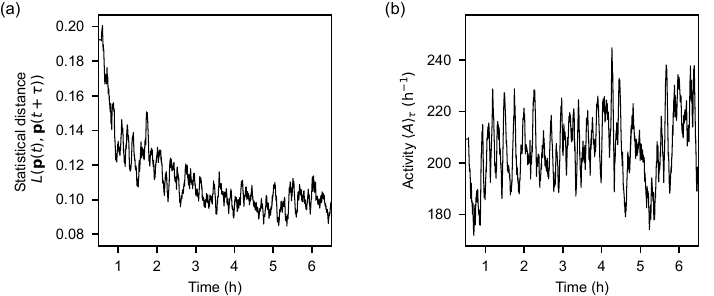}
    \caption{\textbf{Statistical distance and activity over time.} 
    \panel{A} Statistical distance $L(\mathbf{p}(t),\mathbf{p}(t+\tau))$ with $\tau=20\ \mathrm{s}$, computed using the $L^1$ norm. 
    \panel{B} Time-averaged activity $\langle A(t) \rangle_\tau = \frac{1}{\tau}\int_0^\tau d\tau' \sum_{i \neq j}W_{ij}(t+d\tau')p_j(t+d\tau')$. 
    The combination of the activity and statistical distance yields a lower bound on the Hatano--Sasa entropy production rate.
    }
  \label{figSI:EPR}
\end{figure}

Fig.~\ref{figSI:EPR} shows the two components required to estimate the entropy production rate (EPR): statistical distance and activity. 
The EPR transition, shown in Fig. 4f of the main text, is primarily driven by the statistical distance, while the activity remains relatively constant over time.

\subsection{Information rate over time exhibits a rate transition}\label{sec7:infolength}

In addition to KL divergence-based distinguishability and an independent entropy production rate estimate, here we present yet another quantification of an information rate which is based on information geometry. \\

%Specifically, we treat the set of topological landscapes from experiment as a one-parameter family of distributions, where this single parameter is time. Then we construct its Fisher-Rao geometry---a statistical manifold where the metric is governed by infinitesimal... In the end, the information rate quantifies ...

Specifically, for the main experiment, we calculate the information rate~\cite{crooks2007measuring, kim2021information}:
\begin{equation}
    \Gamma(t) \equiv \sqrt{\frac{2}{dt}J(p(x,t+dt)|p(x,t)) } .
    \label{eqn:inforate}
\end{equation}
with timestep $dt = 5\ \mathrm{s}$ and Jensen-Shannon Divergence $J(P|Q)\equiv \frac{1}{2}(D_{KL}(P|Q)+D_{KL}(Q|P))$. 
\\

\begin{figure}[!ht]
  \centering
  \small
\includegraphics[width=\linewidth]{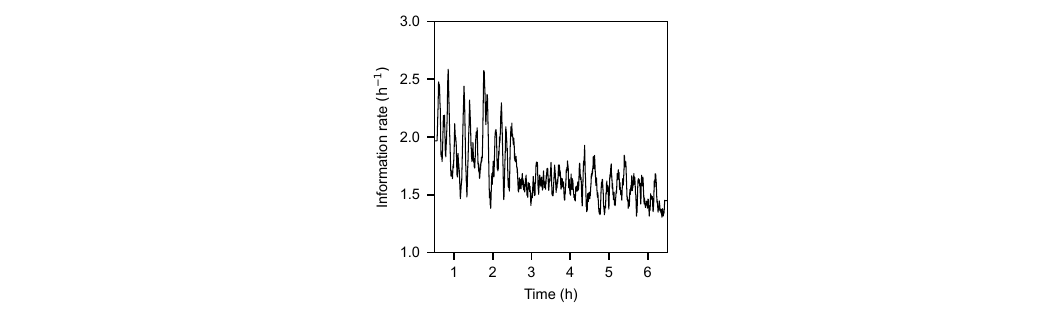}
    \caption{\textbf{Information rate $\mathbf{\Gamma}(t)$ (based on the Jensen-Shannon Divergence) over time exhibits a rate transition. } }
  \label{figSI:infolength}
\end{figure}

In Fig.~\ref{figSI:infolength}, we find that the information rate both slows down and fluctuates less at around 3 hours, the time of traveling-fluctuating transition. This reinforces our observation that an information rate shift accompanies the state transition in the main experiment~(Fig.~4f, Fig.~\ref{figSI:EPR}). Moreover, the suppression of fluctuation offers an additional informational signature that could be investigated in future work.

\begin{figure}
    \centering
    \includegraphics[]{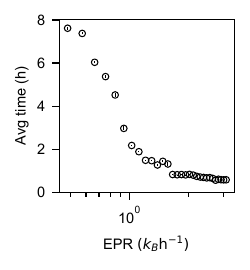}
    \caption{\ {Average time (h) of each EPR bin (error bars = standard deviation).}}
    \label{fig:epr_bin_v_time}
\end{figure}

\subsection{\ {Dissipation correlates with enhanced information flow and motif transition rates}}

\ {In this section, we investigate the correlation of directed flows between configurations with dissipation. Since we cannot obtain the full entropy production rate (EPR) of our experiment, we base our analysis on the estimated lower bound for the EPR from thermodynamic speed limits (Supplementary Section \ref{sec7:epr}). This estimator depends on both the activity $A(t)$, which describes the overall level of transitions, and change in the motif distribution via $L(\mathbf{p}(0),\mathbf{p}(\tau))$. \footnote{\ {Note that this estimate is not accurate in the special case of a  nonequilibrium steady state. Under this condition, $L(\mathbf{p}(0),\mathbf{p}(\tau))$ = 0, and the EPR estimate is zero, despite true dissipation being large. However, in the experiment, we generally observe at least a small nonzero bias, so that constant bias and increasing activity will always lead to an increasing EPR estimate, as expected. }} \\}

\ {On the flowscape, the rate at which the trajectory approaches a given reference distribution corresponds the speed of information flow towards that reference, $-\frac{d}{dt}D_{KL}\left(P(t)||P_\mathrm{ref}\right)$. As shown in the main text Fig. 4f, we find that increasing EPR estimates correlate with an increased approach speed towards the M$_1$-dominated reference, $-\frac{d}{dt}D_{KL}\left (P(t)||\mathcal{G}(M_1,\sigma_M^2) \right)$. Here, EPR values are binned into 30 bins ranging evenly from 0.48 to 3.19 $k_B h^{-1}$. Data points (error bars) show the average (standard deviation of) information speed corresponding to that EPR bin. Points are colored by the average time of that EPR bin. Figure \ref{fig:epr_bin_v_time} shows the average time and its standard deviation for each EPR bin. The time derivative of the information flow, $-\frac{d}{dt}D_{KL}\left(P(t)||P_\mathrm{ref}\right)$, is computed using a finite difference method with $\Delta t = 5$ frames (25 sec) and then smoothed with a moving average with a window size of 45 frames (3 min 45 sec). \\}

\ {Transition biases, $B^i_\tau(t)$, towards motif $i$ over a time frame $\tau$, are computed as: 
\begin{equation}
    B^i_\tau(t) = \frac{\sum^{t+\tau}_{t' = t} \sum_{j\neq i} J_{ij}(t') - J_{ji}(t') }{\sum^{t+\tau}_{t' = t} \sum_{k\neq j}J_{kj}(t') },
\end{equation}
where $J_{ij}(t) \geq 0$ is the flux (in counts) from motif $j$ to motif $i$ from frame $t$ to $t+1$. Transition biases $B^i_\tau(t)$ for a given motif $i$ are found over windows of $\tau = 5$ frames (25 sec) and then smoothed with a moving average with a window size of 45 frames (3 min 45 sec).\\
}

\begin{figure}
    \centering
    \includegraphics[]{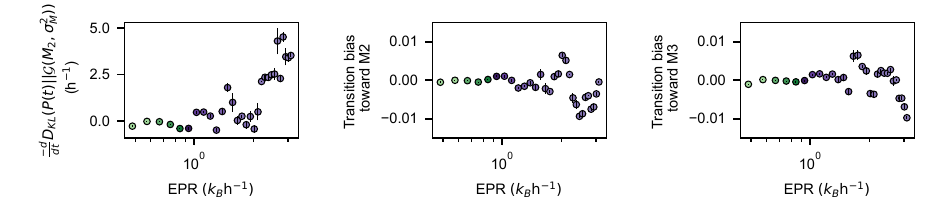}
    \caption{\ {\emph{Left:} Large dissipation correlates with fast information flow towards M$_2$ at early times in the traveling state and slow information flow away from M$_2$ at the transition and in the subsequent traveling state. \emph{Center:} Transition bias towards the M$_2$ motif binned by the EPR. \emph{Right:} Transition bias towards the M$_3$ motif binned by the EPR. }} 
    \label{fig:M2_info_flow}
\end{figure}

\ {Figure \ref{fig:M2_info_flow} shows the information speed of approach towards M$_2$ binned over EPR estimates as well as transition bias towards M$_2$. The speed of information flow towards M$_2$ is initially fast in the early time of the traveling state. At the transition, the speed slows and information flows away (on average) from M$_2$. This figure also shows that high dissipation correlates with biased transitions both towards and away from the M$_2$ motif and M$_3$ motif.\\ }

\subsection{Topological earth mover distance in the main experiment}\label{sec7:TEM}

The topological earth mover (TEM) distance \cite{skinner2021topological}, discussed in SI~Sec.~\ref{sec4:3}, is the minimum number of topological (T1) transitions to transform one distribution of topological motifs to another distribution. While we mainly use the TEM distance to compute the structural order parameter $\langle d_{\mathrm{hex}} \rangle$, this TEM distance can quantify topological distance between any pair of structures, beyond fixing one of them to be a perfect hexagonal crystal. This implies that, in addition to the KL divergence between topological landscapes, the TEM distance can also be used as a measure of dissimilarity between self-organized structures.
\\

%We note that the two measures of dissimilarity, KL Divergence and TEM distance, are essentially complementary given their crucial conceptual distinctions. First, TEM distance does not take account of a low-dimensional manifold of topological motifs. Second, TEM distance is a distance, satisfying symmetry and triangle inequality. Given this, it is useful to see how TEM distance characterizes structural changes in experiment over time. 
% KLD: the genearality, and information theoretic interpretation. 
% For topological systems, this can be a good measure. But it's less generalizable. 

\begin{figure}[!ht]
  \centering
\includegraphics[width=\linewidth]{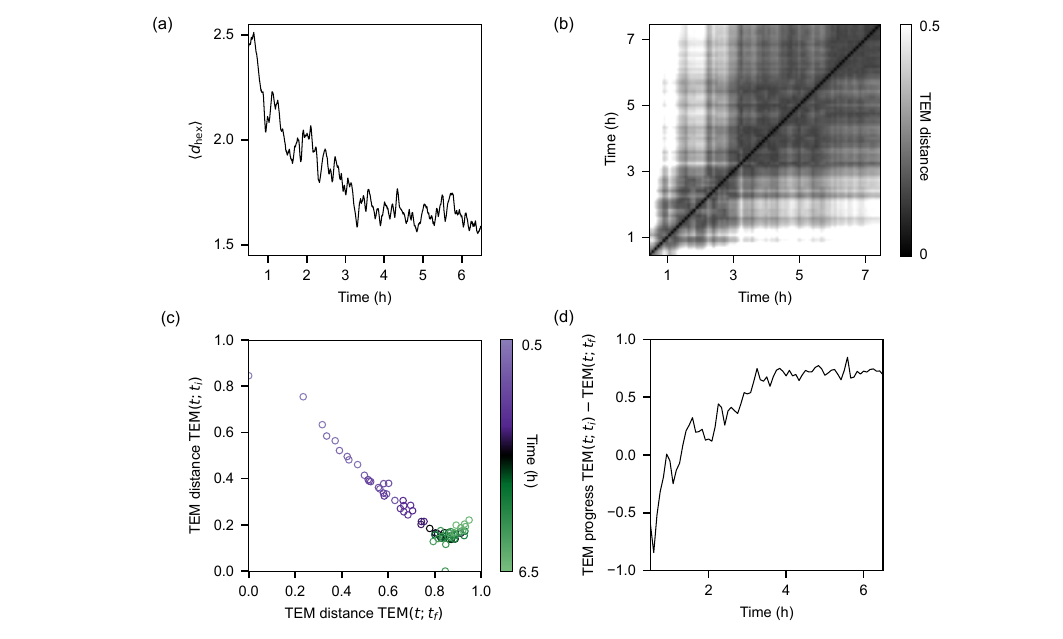}
    \caption{\textbf{TEM distance provides an additional measure for dissimilarity between self-organized structures.} \panel{A} Topological order parameter $\langle d_{\mathrm{hex}}\rangle$ decreases over time. 
    \panel{B} Pairwise topological earth mover (TEM) distances between self-organized structures in the experiment at different times. The structure exhibits a bipartite block pattern between early time (before 3h) and late time (after 3h), which correspond to traveling and fluctuating states.
    \panel{C} Topological flowscape using TEM distances from two reference states (the initial and final structures). 
    \panel{D} Diagonal displacement of TEM distances. 
    }
  \label{figSI:TEMtrajectory}
\end{figure}

We first note that $\langle d_{\mathrm{hex}}\rangle$--- equivalent to the TEM distance from each experimental snapshot and a perfect crystal---exhibits a transition at around 3.5 hours when the traveling state to fluctuating state transition occurs~(Fig.~\ref{figSI:TEMtrajectory}a). Similarly, the pairwise distance map between experimental snapshots exhibits a bipartite pattern that separates structures before and after 3 hours~(Fig.~\ref{figSI:TEMtrajectory}b). 
\\

Moreover, like in the topological flowscape, we can visualize the structural transition on coordinates of TEM distances from the initial and final structures~(Fig.~\ref{figSI:TEMtrajectory}c).  Interestingly, the sum of the two distances stay almost constant over the experiment. In addition, the self-organization progress, quantified as a diagonal displacement from initial to final states, exhibits a rate shift at the state transition as in other rate measures~(Fig.~\ref{figSI:TEMtrajectory}d) (Secs. \ref{sec7:epr}-\ref{sec7:infolength}).  

\ {
\subsection{\ {Analyzing E$_1$-E$_2$ pattern complexity in the main experiment}}
The landscape/flowscape framework is generalizable to any physical property with a metric that describes distances between configurations. For example, in topological landscapes, that physical property is the local network topology (motif) of each embryo, and the metric is the number of T1 transitions between motifs. In this section, we present pattern-based landscapes-- where configurations now correspond to the pattern of E1-E2 embryos, and the metric that distinguishes patterns is the number of ``spin flips'' to transform one pattern into another. We study patterns that occur on M$_1$ (hexagonal) topological motifs. \\\\
Examples of eight observed E1-E2 patterns are shown in Fig. \ref{fig:type_type_analysis_def}a) (left). Blue (red) embryos indicate E1 (E2) embryos present in observed $\mathrm{M}_1$ motifs, i.e. neighborhoods with hexatic symmetry, whereas gray embryos are not part of an $\mathrm{M}_1$ motif. Edges are only plotted for nearest neighbor embryos that are part of the examples. The right hand side of the figure shows the three all-time most probable patterns. Overall, the probability distributions over patterns are much more diverse than those for topological motifs (Fig. \ref{fig:type_type_analysis_def}b), with $\sim 2000$ patterns exhibiting comparable probability to the top 3 patterns (P$_1$,P$_2$,P$_3$, from top to bottom). Here, patterns are extracted over 5 min time bins. However, as shown in Fig. \ref{fig:type_type_analysis_def}c), the probability distributions over patterns can be grouped using the Jensen-Shannon distance into early time ($<3$ hr) and late time ($> 3$ hr) blocks, coincident with the traveling and fluctuating states. Compared to early times, the later time block of distributions has a lower diversity, indicated by a decrease in the Shannon entropy (Fig. \ref{fig:type_type_analysis_def}d). Thus, even at the level of probability distributions alone, this pattern-based analysis uncovers statistical differences in the spatial patterning of young and old embryos in the early time traveling state, and late time fluctuating state in the main experiment.\\\\
Next, we incorporate the metric that distinguishes individual patterns to construct a pattern-based landscape. Here, we start from the spin flip distance between two patterns, which describes the minimum number of flips\footnote{\ {We define one flip as the tranformation between an E1 embryo into an E2 embryo and vice versa at the same node location.}} to transform one pattern into another, allowing for reflections and rotations. For example, for the top 3 most probable patterns shown in Fig. \ref{fig:type_type_analysis_def}a) (right), the spin-flip distances are $d_{12} = 3$, $d_{23} = 2$, $d_{13} = 1$.\\\\
Based on the spin-flip distances, we map the E1-E2 patterns onto a low-dimensional manifold for type-pattern landscape. Specifically, we use classical MDS (Supplementary Section \ref{sec:classical_MDS}) of the spin-flip distance space to map each pattern to a location ($x_i$,$y_i$) in 2-dimensional embedding. The height $z_i$ of the landscape at each point ($x_i$,$y_i$) is equal to the probability $p_i$ of the $i^{th}$ pattern at a given time. Like for topological landscapes, we used kernel density estimation to find a smoothed probability landscape (Supplementary Section \ref{sec5:2}).\\\\
Fig. \ref{fig:type_type_results}a) depicts the atlas of the top-10 most probable patterns (over all time) in an example landscape at 3hr:05m. Clearly, many more than just these top-10 patterns contribute to the shallow landscape. The two axes $x$ (MDS PC1) and $y$ (MDS PC2) correspond to emergent order parameters. $x$ decreases with an increasing number of E2 embryos in a pattern, while increasing $y$ corresponds to the an increase in the number of E2 embryos in the boundary (next nearest neighbors) minus the number of E2 in the bulk (nearest neighbors and centroid). Equivalently, Fig. \ref{fig:type_type_results}b) shows that there is an orthonormal basis about the $y = x$ and $y = -x$ lines that indicates whether a pattern is E1 or E2 rich in the boundary and in the bulk. By examining the time-evolving probability landscapes, Fig. \ref{fig:type_type_results}c), it is evident that most observed patterns are E1 dominant in the bulk and in the boundary. At early times, in the traveling state ($t<3$ hr), the landscape stretches into the regions where E2 can dominate the bulk or boundary, but that this diversity becomes much more suppressed at later times in the fluctuating state ($t > 3 $ hr). Finally, we see a systematically lower proportion of E2 embryos present in the hexagonal motif over time compared to the system-wide proportion. This may suggest that hexagonal motifs form more easily when the local proportion of E2 embryos is lower. \\\\
Overall, this analysis presents the first steps towards analyzing the role of higher-order interactions in a nonreciprocal system and how spatial patterning can interplay with structure and dynamics to give rise to distinct topological motifs, like the hexagon. Future work is needed to better understand these mechanisms.
\begin{figure}[h]
    \centering
    \includegraphics[]{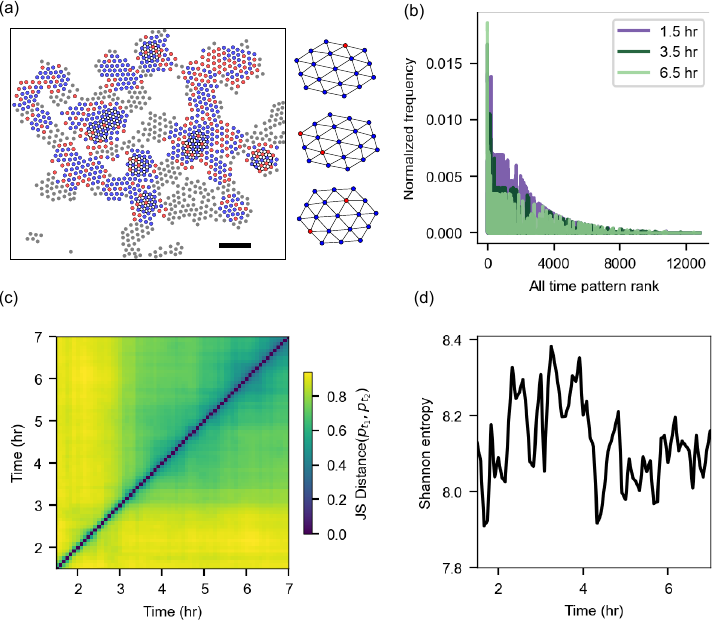}
    \caption{\ {\textbf{Distributions of patterns of E1 and E2 embryos on hexagonal topological motifs are distinguishable between the traveling state and the fluctuating state in experiment.} (a) For each instance of a hexagonal topological motif (M$_1$), there is a corresponding pattern of E1 and E2 embryos. The example snapshot shown here has 8 such instances highlighted (of $\sim 150$ total). The three patterns shown on the right are the top three most common patterns over all time in the experiment (E1 = blue, E2 = red). (b) The probability distributions over patterns, e.g. at 1.5 hr, 3.5 hr, and 6.5 hr, are more diverse than those for topological motifs. From early times to late times, this diversity decreases as lower-ranked patterns grow in probability. (c) The Jensen-Shannon distances between probability distributions of patterns at time $t_1$ and time $t_2$ shows a block structure that is divided at the transition time, $\sim 3.5$ hr, in the main experiment. (d) The Shannon entropy of the probability distribution of patterns is larger in the traveling state ($< 3.5$ hr) and decreases in the fluctuating state ($> 3.5$ hr). }}
    \label{fig:type_type_analysis_def}
\end{figure}
\begin{figure}[h]
    \centering
    \includegraphics[width=\linewidth]{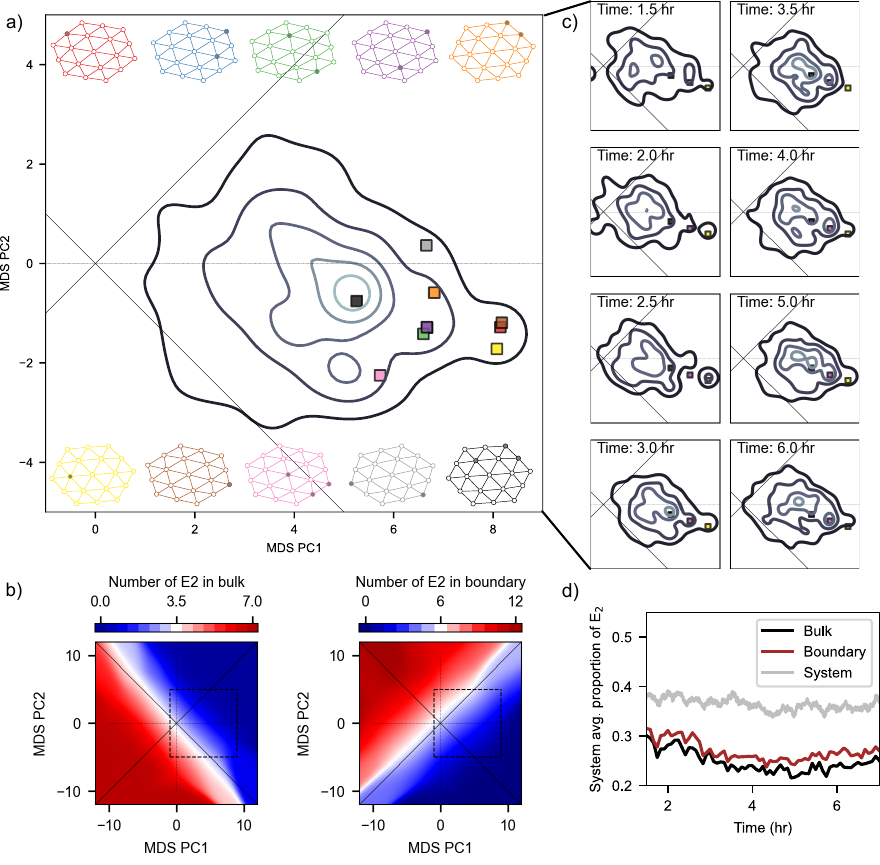}
    \caption{\ {\textbf{Pattern-based landscapes exhibit non-monotonic diversity from the traveling to fluctuating state.} (a) Atlas of top ten most probable patterns (over all time) shown on the probability landscape from 3hr:05m (E1 embryos = white nodes, E2 embryos = gray nodes). The high entropy distribution contains significant contributions from many more than these top ten patterns, leading to a shallow landscape. (b) Weighted kernel density estimates for the net number of E2 embryos in the bulk or boundary. White indicates an equal ratio of E1:E2, while blue indicates E1 rich and red indicates E2 rich. The dashed square region indicates the zoomed in region shown in a) and c). These demonstrate that the MDS coordinates correspond to emergent order parameters. Principal component (PC) 1 describes the number of E2 embryos in a pattern, while PC2 describes the spatial ``moment'' of these E2 in a pattern, i.e. the number of E2 in the boundary ($r=2$) minus those in the bulk ($r\leq 1$). Equivalently, there is an orthonormal basis rotated 45 degrees from PC1 and PC2 about the origin whose two directions indicate i) whether the bulk is E1 or E2 rich and ii) whether the boundary is E1 or E2 rich. (c) Time evolving pattern-based landscapes. Over all times, most probability is contained in the quadrant where both the boundary and bulk are E1 rich. However, at early times ($t< 3.5$ hr), the distributions also show patterns that are E2 rich in the bulk or boundary. At the transition time $t \geq 3.5$ hr, these contributions decrease, before growing slightly again after 4 hr. (d) Average proportion of E2 embryos in bulk of M$_1$, boundary of M$_1$, and system-wide, across time. At the transition $t \sim 3.5$ hr, both the relative proportion of E2 in the bulk and boundary decay, before increasing at later times in the fluctuating state. These proportions are consistently lower than the system-wide proportion at all times. }}
    \label{fig:type_type_results}
\end{figure}
}

%they can still be binned according to the same time blocks we saw in the topological evolution in the experiment

%Key point: Doing flowscape in the spin-flip graph tells us something physical- over time, the motifs tend towards one lower complexity pattern, and get there by first passing through a regime where pattern complexity increases and is more diverse (uniform distribution). This transition point coincides with the polar and structural transition in the main experiment.

%Key point: The approach towards the reference pattern tracks with the approach towards a hexagonal motif in the experiment. (Compared to the approach to a uniform distribution). 

\clearpage

\subsection{\ { Mean field theory} }
\subsubsection{\ {Continuum theory for monodisperse 2D systems of interacting particles} }
\ {We first develop a highly simplified mean field description of an overdamped monodisperse 2D system of particles that experience long-ranged attraction, short-ranged repulsion, as well as non-conservative transverse forces. While we assume the attractive and repulsive forces are central forces, we do not require them to be derivable from a potential. They can be nonequilibrium in origin, such as the Stokeslet mediated attraction between starfish embryos. Under these assumptions, the equation of motion for the position $\mathbf{r}_i$ of particle $i$ can be written as 
\begin{equation} \label{eq:1}
    \frac{d\mathbf{r}_i(t)}{dt} = \sum_{j \neq i} \left[ v_{\parallel}f_{\parallel}\left(\frac{|\mathbf{r}_i(t)-\mathbf{r}_j(t)|}{R}\right)\hat{\mathbf{r}_{ij}} + v_{\perp}f_{\perp}\left(\frac{|\mathbf{r}_i(t)-\mathbf{r}_j(t)|}{R}\right)(\hat{\mathbf{e}_z}\times\hat{\mathbf{r}_{ij}})\right] 
\end{equation}}
\ { where the sum runs over all particles $j$ that interact with particle $i$, $v_{\parallel}$ and $v_{\perp}$ are longitudinal and transverse speed scales that reflect the strength of longitudinal (central) and transverse forces, and $R$ is a microscopic length scale such as the embryo radius for the starfish system. Further, the unit vector $\hat{\mathbf{r}_{ij}}$ points from the position of particle $j$ to the position of particle $i$, and $\hat{\mathbf{e}_z}$ is a unit normal to the plane containing the particles. We will further assume that the functions $f_{\parallel}$ and $f_{\perp}$ are continuously differentiable over all physically realizable inter-particle distances. In a system with long-ranged attractions and short-ranged repulsions, $f_{\parallel}$ is typically of the form:
\begin{equation}\label{eq:2}
    f_{\parallel}(r) = \frac{C}{r^{a}} - \frac{D}{r^b}
\end{equation}
where $a,b,C,D \in \mathbb R_{> 0}$, i.e. they are positive real numbers, and $a>b$. The function therefore generically crosses zero at some equilibrium separation $r_0$, which can be a convenient point around which to perform a Taylor expansion. \\}

\ {Our goal is to coarse grain Eqn.\ref{eq:1} such that it describes a velocity field $\mathbf{v}(\mathbf{r})$ in terms of the particle number density field $\rho(\mathbf{r})$ and the density gradient $\nabla\rho(\mathbf{r})$. As the first step, following standard practice \cite{marchetti2013hydrodynamics} we define coarse grained position vectors, as well as density and velocity fields by averaging over an area $\partial\Omega$ as follows: 
\begin{equation}\label{eq:3}
\rho(\mathbf{r},t) = \sum_{i\in \partial\Omega} \delta(\mathbf{r}(t)-\mathbf{r}_i(t));\hspace{0.8cm}
    \mathbf{r}(t) = \frac{1}{\rho(\mathbf{r},t)}\sum_{i\in \partial\Omega} \mathbf{r}_i(t)\delta(\mathbf{r}(t)-\mathbf{r}_i(t));\hspace{0.8cm}
    \mathbf{v}(\mathbf{r},t) = \frac{1}{\rho(\mathbf{r},t)}\sum_{i\in \partial\Omega} \frac{d\mathbf{r}_i(t)}{dt}\delta(\mathbf{r}(t)-\mathbf{r}_i(t))
\end{equation} }
\ {Multiplying both sides of Eqn.\ref{eq:1} by $\delta(\mathbf{r}(t)-\mathbf{r}_i(t))$, summing over $\partial\Omega$ and using Eqn.\ref{eq:3}, we get
\begin{equation} \label{eq:4}
   \rho(\mathbf{r})\mathbf{v}(\mathbf{r}) = \sum_{i\in\partial\Omega}\sum_{j \neq i} \left[ v_{\parallel}f_{\parallel}\left(\frac{|\mathbf{r}_i-\mathbf{r}_j|}{R}\right)\hat{\mathbf{r}_{ij}} + v_{\perp}f_{\perp}\left(\frac{|\mathbf{r}_i-\mathbf{r}_j|}{R}\right)(\hat{\mathbf{e}_z}\times\hat{\mathbf{r}_{ij}}) \right]\delta(\mathbf{r}(t)-\mathbf{r}_i(t))
\end{equation} }
\ {where we have chosen to not show the time dependence of $\mathbf{v}$, $\rho$, and $\mathbf{r}$ explicitly for convenience of notation. To obtain a simplified expression for Eqn.\ref{eq:4}, we replace $f_{\parallel/\perp}(|\mathbf{r}_i-\mathbf{r}_j|)$ by their respective values at a distance corresponding to the local mean separation between particles, $r_m$. This is equivalent to assuming that the dominant contribution to $f_{\parallel/\perp}$ comes from particles that are nearest to each other. With this simplification, using $r_m = |\mathbf{r}_i-\mathbf{r}_j|$, we get
\begin{equation} \label{eq:5}
   \rho(\mathbf{r})\mathbf{v}(\mathbf{r}) = \frac{v_{\parallel}}{r_m}f_{\parallel}\left(\frac{r_m}{R}\right)\sum_{i\in\partial\Omega}\sum_{j \neq i} (\mathbf{r}_i-\mathbf{r}_j)\delta(\mathbf{r}-\mathbf{r}_i) + 
   \frac{v_{\perp}}{r_m}f_{\perp}\left(\frac{r_m}{R}\right)\sum_{i\in\partial\Omega}\sum_{j \neq i}(\hat{\mathbf{e}_z}\times(\mathbf{r}_i-\mathbf{r}_j)) \delta(\mathbf{r}-\mathbf{r}_i)
\end{equation}}
\ {Using the definitions of $\rho(\mathbf{r})$ and $\mathbf{r}$ from Eqn.\ref{eq:3} in Eqn.\ref{eq:5}, we can evaluate the sum over $i$ explicitly to get
\begin{equation}\label{eq:6}
    \mathbf{v}(\mathbf{r}) = \frac{v_{\parallel}}{r_m}f_{\parallel}\left(\frac{r_m}{R}\right)\sum_{j \neq i}(\mathbf{r}-\mathbf{r}_j) + \frac{v_{\perp}}{r_m}f_{\perp}\left(\frac{r_m}{R}\right)\sum_{j \neq i}(\hat{\mathbf{e}_z}\times(\mathbf{r}-\mathbf{r}_j))
\end{equation} }
\ { we can convert the sums in Eqn.\ref{eq:6} to integrals over a circular neighborhood centered on $\mathbf{r}$ with radius equal to the interaction range $R_c$. Denoting $r' = |\mathbf{r}-\mathbf{r}'|$, we can write 
\begin{equation} \label{eq:7}
    \mathbf{v}(\mathbf{r}) = \frac{v_{\parallel}}{r_m}f_{\parallel}\left(\frac{r_m}{R}\right)\int_{2R}^{R_c}\int_0^{2\pi}\rho(\mathbf{r'})(\mathbf{r}-\mathbf{r'})r'dr'd\theta + \frac{v_{\perp}}{r_m}f_{\perp}\left(\frac{r_m}{R}\right)\int_{2R}^{R_c}\int_0^{2\pi}\rho(\mathbf{r'})\left[\hat{\mathbf{e}_z}\times(\mathbf{r}-\mathbf{r'})\right]r'dr'd\theta
\end{equation} }
\ {The integral over $dr'$ has a lower limit of $2R$ rather than $0$, as the sum in Eqn.\ref{eq:1} is over $j\neq i$. We first focus on the integral over $\theta$ in the first term on the RHS of Eqn.\ref{eq:3}. By summing contributions from vectors that point in opposite directions, we have 
\begin{equation}\label{eq:8}
    \int_0^{2\pi}\rho(\mathbf{r'})(\mathbf{r}-\mathbf{r'})d\theta = \int_0^{\pi}\left[\rho(\mathbf{r+(r'-r)})-\rho(\mathbf{r-(r'-r)})\right](\mathbf{r}-\mathbf{r'})d\theta
\end{equation}}
\ {Using the Taylor expansion of the density field around $\mathbf{r}$ to linear order, we get
\begin{equation}\label{eq:9}
    \int_0^{2\pi}\rho(\mathbf{r'})(\mathbf{r}-\mathbf{r'})d\theta = \int_0^{\pi}2(\mathbf{r'}-\mathbf{r})^{\text{T}}\nabla\rho(\mathbf{r})(\mathbf{r}-\mathbf{r'})d\theta
\end{equation}}
\ {where $()^{T}$ denotes the matrix transpose operation. We can rewrite the RHS Eqn.\ref{eq:5} in terms of vector components $x-x' = r'\text{cos}\theta$ and $y-y' = r'\text{sin}\theta$ as follows:
\begin{equation}\label{eq:10}
   - \int_0^{\pi}\left[r'^2\text{cos}^2\theta\frac{\partial\rho(\mathbf{r})} {\partial x} + r'\text{sin}\theta\text{cos}\theta\frac{\partial\rho(\mathbf{r})}{\partial y}\right]\hat{\mathbf{e}_x} + \left[r'^2\text{sin}^2\theta\frac{\partial\rho(\mathbf{r})} {\partial y} + r'\text{sin}\theta\text{cos}\theta\frac{\partial\rho(\mathbf{r})}{\partial x}\right]\hat{\mathbf{e}_y}
\end{equation}}
\ {Using $\int_0^{\pi}\text{cos}^2\theta d\theta = \int_0^{\pi}\text{sin}^2\theta d\theta = \pi/2$, and $\int_0^{\pi}\text{sin}\theta\text{cos}\theta d\theta = 0$, Eqn.\ref{eq:5} simplifies to
\begin{equation}\label{eq:11}
    \int_0^{2\pi}\rho(\mathbf{r'})(\mathbf{r}-\mathbf{r'})d\theta = -\pi r'^2\nabla\rho(\mathbf{r})
\end{equation}}
\ {Eqn.\ref{eq:11} also implies that the integration over $r'$ in Eqn.\ref{eq:7} can be easily evaluated as 
\begin{equation}\label{eq:12}
    \int_{2R}^{R_c}r'^3dr' = \frac{R_c^4 - 16R^4}{4}
\end{equation}}

\ { Using the results from Eqns.\ref{eq:11} and \ref{eq:12}, Eqn.\ref{eq:3} can be written as
\begin{equation}\label{eq:13}
    \mathbf{v}(\mathbf{r}) = \frac{-\pi(R_c^4-16R^4)}{4}\left[ \frac{v_{\parallel}}{r_m}f_{\parallel}\left(\frac{r_m}{R}\right)\nabla\rho(\mathbf{r}) + \frac{v_{\perp}}{r_m}f_{\perp}\left(\frac{r_m}{R}\right)(\hat{\mathbf{e}_z}\times\nabla\rho(\mathbf{r}))\right]
\end{equation} }

\ {The final step is to use the fact that the local mean separation between particles is related to the local density $\rho(\mathbf{r})$, through the relation $r_m = 1/\sqrt{\rho(\mathbf{r})}$. This enables us to write Eqn.\ref{eq:13} in a convenient matrix form as}

\ {
\begin{equation}\label{eq:14}
    \begin{bmatrix}
    v_x\\
    v_y
    \end{bmatrix}
    =
     \begin{bmatrix}
    g_{\parallel}(\rho) & g_{\perp}(\rho)\\
    -g_{\perp}(\rho) & g_{\parallel}(\rho)
    \end{bmatrix}
    \begin{bmatrix}
        \partial_x \rho\\
        \partial_y \rho
    \end{bmatrix}
\end{equation} }

\ {where we have defined the functions 
\begin{equation}\label{eq:15}
    g_{\parallel/\perp}(\rho(\mathbf{r})) = \frac{-\pi(R_c^4-16R^4)}{4}v_{\parallel/\perp}\sqrt{\rho(\mathbf{r})}f_{\parallel/\perp}\left(\frac{1}{\sqrt{\rho(\mathbf{r})R^2}}\right)
\end{equation}}

\ {Eqn.\ref{eq:14} describes a general linear response relationship where density gradient is the driving force and velocity is the response. The two are related by a mobility matrix that in general has a complicated dependence on density. If the total number of particles in the system is conserved, the full dynamical evolution of the system can be described by combining Eqn.\ref{eq:14} with the equation of continuity $\partial_t \rho = -\nabla \cdot (\rho\mathbf{v})$}

\subsubsection{\ {Mean field equations for binary mixtures and linear stability analysis}}

\ {The mean field theory for monodisperse 2D systems derived above can be easily extended to binary mixtures of two species of particles, $A$ and $B$. Assuming that the average densities of the two species, denoted by $\langle \rho_A \rangle$, and $\langle \rho_B \rangle$, respectively, are conserved, the continuum mean field equations for binary mixtures take the following form:}
\ {
\begin{gather}\label{eq:lsa1}
    \frac{\partial\rho_A}{\partial t} = -\nabla \cdot (\rho_A\mathbf{v_A})\\\label{eq:lsa2}
    \frac{\partial\rho_B}{\partial t} = -\nabla \cdot (\rho_B\mathbf{v_B})\\ \label{eq:lsa3}
    \mathbf{v_A} = g_{\parallel}^{AA}(\rho_A)\nabla\rho_A + g_{\parallel}^{AB}(\rho_A,\rho_B)\nabla\rho_B + g_{\perp}^{AA}(\rho_A) (\hat{\mathbf{e_z}}\times\nabla\rho_A) + g_{\perp}^{AB}(\rho_A,\rho_B) (\hat{\mathbf{e_z}}\times\nabla\rho_B)\\ \label{eq:lsa4}
    \mathbf{v_B} = g_{\parallel}^{BB}(\rho_B)\nabla\rho_B + g_{\parallel}^{BA}(\rho_A,\rho_B)\nabla\rho_A + g_{\perp}^{BB}(\rho_B) (\hat{\mathbf{e_z}}\times\nabla\rho_B) + g_{\perp}^{BA}(\rho_A,\rho_B) (\hat{\mathbf{e_z}}\times\nabla\rho_A)
\end{gather}}

\ {where $g_{\parallel/\perp}(\rho)$ are nonlinear, and in general, quite complicated functions of density derived from microscopic interactions. We perform linear stability analysis by perturbing the density profiles about the homogeneous steady state corresponding to $\langle \rho_A \rangle$ and $\langle \rho_B \rangle$. Since velocity depends linearly on density gradients, we have $\mathbf{v_A} = \mathbf{v_B} = 0$ in the homogeneous state. Let $\rho_{A1}$ and $\rho_{B1}$ be the perturbed density profiles. Defining $ M_{\parallel/\perp}^{ij} = g_{\parallel/\perp}^{ij}(\langle \rho_i \rangle, \langle \rho_j \rangle)$, we can linearize Eqns. \ref{eq:lsa1}-\ref{eq:lsa4} as follows:}
\ {
\begin{gather}\label{eq:lsa5}
    \frac{\partial\rho_{A1}}{\partial t} = -\langle \rho_A \rangle \nabla \cdot (\mathbf{v_{A1}})\\\label{eq:lsa6}
    \frac{\partial\rho_{B1}}{\partial t} = -\langle \rho_B \rangle \nabla \cdot (\mathbf{v_{B1}})\\ \label{eq:lsa7}
    \mathbf{v_{A1}} = M_{\parallel}^{AA}\nabla\rho_{A1} + M_{\parallel}^{AB}\nabla\rho_{B1} + M_{\perp}^{AA}(\hat{\mathbf{e_z}}\times\nabla\rho_{A1}) + M_{\perp}^{AB} (\hat{\mathbf{e_z}}\times\nabla\rho_{B1})\\ \label{eq:lsa8}
    \mathbf{v_{B1}} = M_{\parallel}^{BB}\nabla\rho_{B1} + M_{\parallel}^{BA}\nabla\rho_{A1} + M_{\perp}^{BB}(\hat{\mathbf{e_z}}\times\nabla\rho_{B1}) + M_{\perp}^{BA} (\hat{\mathbf{e_z}}\times\nabla\rho_{A1})
\end{gather}}

\ {As transverse density gradients do not contribute to divergence of the velocity, Eqns. \ref{eq:lsa5}-\ref{eq:lsa8} result in the following simplified equations for perturbed density profiles}
\ {
\begin{gather}\label{eq:lsa9}
    \frac{\partial\rho_{A1}}{\partial t} = -\langle \rho_A \rangle \left(M_{\parallel}^{AA}\nabla^2\rho_{A1} + M_{\parallel}^{AB}\nabla^2\rho_{B1}\right)\\\label{eq:lsa10}
    \frac{\partial\rho_{B1}}{\partial t} = -\langle \rho_B \rangle \left(M_{\parallel}^{BB}\nabla\rho_{B1} + M_{\parallel}^{BA}\nabla\rho_{A1}\right)
\end{gather}}

\ {
For an infinite system, we can Fourier transform in space to get the following system of ordinary differential equations for each wave vector $\mathbf{q}$:}
\ {
\begin{equation} \label{eq:lsa11}
\frac{\partial}{\partial t}
    \begin{bmatrix}
    \widetilde{\rho}_{A1} \\ \widetilde{\rho}_{B1} 
    \end{bmatrix}
    = q^2
    \begin{bmatrix}
        M_{\parallel}^{AA}\langle \rho_A \rangle & M_{\parallel}^{AB}\langle \rho_A \rangle \\
        M_{\parallel}^{BA}\langle \rho_B \rangle & M_{\parallel}^{BB}\langle \rho_B \rangle
    \end{bmatrix}
    \begin{bmatrix}
    \widetilde{\rho}_{A1} \\ \widetilde{\rho}_{B1} 
    \end{bmatrix}
\end{equation} }

\ {The eigenvalues of the matrix of coefficients are given by }
\ {
\begin{equation} \label{eq:lsa12}
    \lambda_{\pm} = \frac{M_{\parallel}^{AA}\langle\rho_A\rangle + M_{\parallel}^{BB}\langle \rho_B\rangle \pm \sqrt{\left(M_{\parallel}^{AA}\langle\rho_A\rangle - M_{\parallel}^{BB}\langle \rho_B\rangle\right)^2 + 4M_{\parallel}^{AB}M_{\parallel}^{BA}\langle \rho_A \rangle\langle \rho_B \rangle}}{2}q^2
\end{equation} }

\ { Eqn. \ref{eq:lsa12} shows that the stability of the system is determined by the sign of the relaxation terms  $M_{\parallel}^{AA}$ and $ M_{\parallel}^{BB}$ for all wave vectors. However, nonreciprocity plays a crucial role in governing the evolution of density perturbations. While perturbations decay exponentially in stable weakly nonreciprocal systems, traveling waves can be observed for sufficiently strong nonreciprocity, where $M_{\parallel}^{AB}$ and $M_{\parallel}^{BA}$ are large in magnitude and have opposite signs. Due to the dependence of the relaxation time on $q^2$, the system can exhibit persistent traveling density waves in the long wavelength limit $q \to 0$. This behavior is identical to previous studies on nonreciprocal systems \cite{scheibner2020odd}. However, it is worth noting that in the present case, the coefficients $M_{\parallel}^{ij}$ themselves depend on $\langle \rho_A \rangle$ and $\langle \rho_B \rangle$, showing that collective dynamics are governed by density dependent nonreciprocal interactions.  }

\clearpage
%%%%%%%%%%%%%%%%%%%%%%
\section{Supplementary Videos}

\begin{itemize}[leftmargin=2cm]
\setlength\itemsep{-0cm}
\item[\textbf{Video 1:}] \textbf{A young (E1) and old (E2) pair of starfish embryos exhibits run and chase dynamics.} \ { \emph{Left:} Top view of a pair of E1 (24 hours post fertilization) embryos at the air-water interface. The embryos approach each other with equal speeds, until they form a bound pair and rotate about its centroid. \emph{Center:} Top view of a pair of E1 and E2 (48 hours post fertilization) embryos at the air-water interface. E2 is identifiable by its larger and less circular morphology. E1 and E2 approach each other with different speeds, until they form a run-and-chase bound pair; E2 runs away from E1, while E1 chases E2. \emph{Right:} Top view of a pair of E2 embryos. Like the E1 pair, the embryos approach each other with equal speeds and then, as a bound pair, rotate about its centroid. }
\item[\textbf{Video 2:}] \textbf{A nonreciprocal pair of starfish embryos exhibits asymmetric tilt precessions.} Side view of a pair of E1 (24 hours post fertilization) and E2 (48 hours post fertilization) starfish embryos swimming at the air-water interface. E2, the larger embryo, tilts further away from the vertical when it is directed away from E1. Since the embryos self-propel along their AP axes, this biased tilt precession leads to a net drift of the pair toward E2.
\item[\textbf{Video 3:}] \textbf{Binary mixture of starfish embryos exhibits a transition from a traveling to a fluctuating state.} A living mixture of E1 (24 hours post fertilization) and E2 (48 hours post fertilization) embryos self-organize into traveling clusters. For the first three hours, in this traveling state, small clusters exhibit self-propulsion, and are highly dynamic, repeatedly merging and breaking apart. After three hours, the system transitions to a fluctuating state, characterized by a large, connected crystal in which embryos fluctuate about ordered positions.  
\item[\textbf{Video 4:}] \textbf{Binary mixture of starfish embryos with 2-min trajectories overlaid.} Same as supplementary video 3, but with 2-minute trajectories overlaid. The E1 embryos are in blue and E2 embryos are in red.
\item[\textbf{Video 5:}] \textbf{E1 and E2 starfish embryos form a demixed living chiral crystal (LCC).} Without initial mixing, E2 (48 hr post fertilization) embryos tend to reach the air-water interface earlier than E1 (24 hr post fertilization) embryos. As a result, E2 embryos form a nearly homogenous LCC at early times. Later on, E1 embryos swim to the air-water interface and attach to the boundary of the E2-dominated LCC. As time progresses, E2 embryos recede from the air-water interface, leading to an E1-dominated LCC, facilitated by internal rearrangements.  
\item[\textbf{Video 6:}] \textbf{E1 and E2 starfish embryos form an orbiting mixture.} At large densities, a living mixture of E1 (24 hours post fertilization) and E2 (48 hours post fertilization) embryos self-organize into a system-spanning crystal that persistently orbits along the circular boundary of the well plate. Internal rearrangements between embryos are frequent within the cluster.
\item[\textbf{Video 7:}] \textbf{Simulations tune pairwise nonreciprocal interactions, revealing diverse collective states.} Our inference-based model exhibits rich collective behavior as we tune the nonreciprocity $\mathcal{N}$ between two types of particles. Reciprocal ($\mathcal{N}=0$) mixtures lead to a crystal- a single cluster that rotates. At low ($0<\mathcal{N}<1$) nonreciprocities, the system emerges into a self-propelled crystal with translational motion. $\mathcal{N}=1$ exhibits \ {flock-like dynamics}, with small clusters that merge and break apart. Higher nonreciprocity ($\mathcal{N}>1$) leads to fragmentation without stable clusters.
\item[\textbf{Video 8:}] \textbf{Spatial distribution of the topological metric $\mathbf{d}_{\mathrm{hex}}$ in the model simulations. } In the weak nonreciprocity case ($\mathcal{N}<1$), increasing nonreciprocity reduces the number of motifs with higher $d_\mathrm{hex}$, indicative of nonreciprocal \ {error correction}. Once nonreciprocity becomes stronger ($\mathcal{N} \geq 1$), increasing nonreciprocity breaks down crystal structures, resulting in motifs with higher $d_\mathrm{hex}$.
\item[\textbf{Video 9:}] \textbf{Spatial distribution of the top five most probable topological motifs from $\mathbf{\mathcal{N} = 1}$ ($\mathbf{M}_1$-$\mathbf{M}_5$) in the model simulations.} The colors for the motifs are as follows: $\textrm{M}_1$-red, $\textrm{M}_2$-blue, $\textrm{M}_3$-green, $\textrm{M}_4$-purple, $\textrm{M}_5$-orange. Edges reflect nearest neighbor embryos.

\item[\textbf{Video 10:}] \textbf{Spatial distribution of the topological metric $\mathbf{d}_\mathrm{hex}$ in the experiment.} In the traveling state, a few hexagonal motifs ($d_\mathrm{hex} = 0$) are present within the bulk of each \ {flock-like} cluster, surrounded by a boundary of motifs with increasing $d_\mathrm{hex}$, the topological distance to a hexagonal motif. After 3 hours, in the fluctuating state, the hexagon ($d_\mathrm{hex} = 0$) becomes the dominant motif, with larger $d_\mathrm{hex}$ values on the perimeter of the crystal and around internal gaps.

\item [\textbf{Video 11:}] \textbf{Spatial distribution of the top five most probable topological motifs in the experiment ($\mathbf{M}_1$-$\mathbf{M}_5$).} These top five motifs $\textrm{M}_1$-$\textrm{M}_5$ are identical to the top five motifs in $\mathcal{N} = 1$ model simulation. The colors for the motifs are as follows: $\textrm{M}_1$-red, $\textrm{M}_2$-blue, $\textrm{M}_3$-green, $\textrm{M}_4$-purple, $\textrm{M}_5$-orange. Edges reflect nearest neighbor embryos.
\ {\item [\textbf{Video 12:}] \textbf{Nonreciprocal hysteresis emerges upon slowly decreasing or slowly increasing nonreciprocity.} Left: Using our inference-based model, we demonstrate the presence of nonreciprocal hysteresis upon slowly decreasing $\mathcal{N}$ from 1.25 to 0.75. Structures remain fragmented well past the $\mathcal{N}=1$ crossover, with a stable crystal finally forming at $\mathcal{N}=0.80$. Right: Nonreciprocal hysteresis emerges upon slowly increasing $\mathcal{N}$ from 0.75 to 1.25. Stable crystals remain up to $\mathcal{N}=1.20$, well beyond the $\mathcal{N}=1$ crossover.}

\end{itemize}

\newpage
\bibliography{SI_reference}